\documentclass[fleqn,usenatbib]{mnras}

\usepackage{newtxtext,newtxmath}
\usepackage{xcolor}

\usepackage[T1]{fontenc}
\usepackage{ae,aecompl}
\usepackage{tabularx}
\usepackage{tabularx,ragged2e,booktabs}

\usepackage{graphicx}
\usepackage[graphicx]{realboxes}
\usepackage{amsmath}	
\usepackage{natbib}
\usepackage{hyperref}
\usepackage[font=scriptsize]{caption}
\usepackage{times}
\usepackage[utf8]{inputenc}
\usepackage{amsmath,amstext}
\usepackage[figure,figure*]{hypcap}
\usepackage{scrextend}
\usepackage{footnote}
\usepackage{hyperref}
\usepackage{subcaption}
\usepackage{array}
\usepackage{epsfig}
\usepackage{natbib}
\usepackage[english]{babel}
\usepackage{tabularx}
\usepackage{rotating}
\usepackage{float}
\usepackage{amsmath}
\usepackage{tabularx,ragged2e,booktabs}
\usepackage[font=scriptsize]{caption}
\usepackage{times}
\usepackage{newtxtext,newtxmath}
\usepackage{ae,aecompl}
\usepackage{pdflscape}
\usepackage{afterpage}
\usepackage{lastpage}
\usepackage{adjustbox}

\newcolumntype{P}[1]{>{\centering\arraybackslash}p{#1}}

\renewcommand{\arraystretch}{4}

\def\kms{km~s$^{\rm -1}$}
\def\micron{$\mu$m}

\def\Vlsr{\hbox{V$_{\rm LSR}$} }
\def\arraystretch{0.7}
\def\kms{km\,s$^{-1}$}
\def\twelveCO{$^{12}$CO(2-1)}
\def\thirteenCO{$^{13}$CO(2-1)}
\def\CeighteenO{C$^{18}$O(2-1)}
\def\HtwoCO{H$_{2}$CO}

\title[CMZoom III: Spectral Line Data Release]{CMZoom III: Spectral Line Data Release}

\author[D. Callanan et al.]{Daniel Callanan,$^{1,2}$\thanks{E-mail: daniel.s.callanan@gmail.com}
Steven N. Longmore,$^{1}$
Cara Battersby,$^{2,3}$
H Perry Hatchfield,$^{3}$
\newauthor
Daniel L. Walker,$^{3,4}$
Jonathan Henshaw,$^{1,5}$
Eric Keto,$^{2}$
Ashley Barnes,$^{1,6,7}$
Adam Ginsburg,$^{8}$
\newauthor
Jens Kauffmann,$^{9}$
J. M. Diederik Kruijssen,$^{10}$
Xing Lu,$^{11}$
Elisabeth A. C. Mills,$^{12}$
\newauthor
Thushara Pillai,$^{13}$
Qizhou Zhang,$^{2}$
John Bally,$^{14}$
Natalie Butterfield,$^{15}$
Yanett A. Contreras,$^{16}$
\newauthor
Luis C. Ho,$^{17,18}$
Katharina Immer,$^{19}$
Katharine G. Johnston,$^{20}$
Juergen Ott,$^{21}$
\newauthor
Nimesh Patel$^{2}$
and Volker Tolls$^{2}$
\\
\\
$^{1}$Astrophysics Research Institute, Liverpool John Moores University, 146 Brownlow Hill, Liverpool L3 5RF, UK\\
\\
$^{2}$Harvard-Smithsonian Center for Astrophysics, MS-78, 60 Garden St., Cambridge, MA 02138 USA\\
\\
$^{3}$University of Connecticut, Department of Physics, 196 Auditorium Road, Unit 3046, Storrs, CT 06269 USA\\
\\
$^{4}$UK ALMA Regional Centre Node, Jodrell Bank Centre for Astrophysics, The University of Manchester, Manchester M13 9PL, UK\\
\\
$^{5}$Max-Planck-Institute for Astronomy, Koenigstuhl 17, 69117 Heidelberg, Germany\\
\\
$^{6}$Max-Planck-Institut f{\"u}r extraterrestrische Physik, Gie{\ss}enbachstrae 1, 85748 Garching, Germany\\
\\
$^{7}$Institut f{\"u}r theoretische Astrophysik, Zentrum f{\"u}r Astronomie der Universit{\"a}t Heidelberg, Albert-Ueberle Str. 2, D-69120 Heidelberg, Germany\\
\\
$^{8}$Department of Astronomy, University of Florida, PO Box 112055, USA\\
\\
$^{9}$Haystack Observatory, Massachusetts Institute of Technology, 99 Millstone Road, Westford, MA 01886, USA\\
\\
$^{10}$Astronomisches Rechen-Institut, Zentrum f{\"u}r Astronomie der Universit{\"a}t Heidelberg, M{\"o}nchhofstra{\ss}e 12-14, D-69120 Heidelberg, Germany\\
\\
$^{11}$Shanghai Astronomical Observatory, Chinese Academy of Sciences, 80 Nandan Road, Shanghai 200030, People’s Republic of China\\
\\
$^{12}$Department of Physics and Astronomy, University of Kansas, 1251 Wescoe Hall Dr., Lawrence, KS 66045, USA\\
\\
$^{13}$Boston University Astronomy Department, 725 Commonwealth Avenue, Boston, MA 02215, USA\\
\\
$^{14}$CASA, University of Colorado, 389-UCB, Boulder, CO 80309\\
\\
$^{15}$National Radio Astronomy Observatory, 520 Edgemont Road, Charlottesville, VA 22903, USA\\
\\
$^{16}$Leiden Observatory, Leiden University, PO Box 9513, NL 2300 RA Leiden, the Netherlands\\
\\
$^{17}$Kavli Institute for Astronomy and Astrophysics, Peking University, Beijing 100871, China\\
\\
$^{18}$Department of Astronomy, School of Physics, Peking University, Beijing 100871, China\\
\\
$^{19}$Joint Institute for VLBI ERIC, Oude Hoogeveensedijk 4 7991 PD, Dwingeloo, The Netherlands\\
\\
$^{20}$School of Physics \& Astronomy, E.C. Stoner Building, The University
of Leeds, Leeds LS2 9JT, UK\\
\\
$^{21}$National Radio Astronomy Observatory, 1003 Lopezville Rd., Socorro, NM 87801, USA
}

\date{Accepted XXX. Received YYY; in original form ZZZ}

\pubyear{2022}

\begin{document}
\label{firstpage}
\pagerange{\pageref{firstpage}--\pageref{lastpage}}
\maketitle

\begin{abstract}
We present an overview and data release of the spectral line component of the SMA Large Program, \textit{CMZoom}. \textit{CMZoom} observed \twelveCO, \thirteenCO\ and \CeighteenO, three transitions of H$_{2}$CO, several transitions of CH$_{3}$OH, two transitions of OCS and single transitions of SiO and SO, within gas above a column density of N(H$_2$)$\ge 10^{23}$\,cm$^{-2}$ in the Central Molecular Zone (CMZ; inner few hundred pc of the Galaxy). We extract spectra from all compact 1.3\,mm \emph{CMZoom} continuum sources and fit line profiles to the spectra. We use the fit results from the H$_{2}$CO 3(0,3)-2(0,2) transition to determine the source kinematic properties. We find $\sim 90$\% of the total mass of \emph{CMZoom} sources have reliable kinematics. Only four compact continuum sources are formally self-gravitating. The remainder are consistent with being in hydrostatic equilibrium assuming that they are confined by the high external pressure in the CMZ. Based on the mass and density of virially bound sources, and assuming star formation occurs within one free-fall time with a star formation efficiency of $10\% - 75\%$, we place a lower limit on the future embedded star-formation rate of $0.008 - 0.06$\,M$_{\odot}$\,yr$^{-1}$. We find only two convincing proto-stellar outflows, ruling out a previously undetected population of very massive, actively accreting YSOs with strong outflows. Finally, despite having sufficient sensitivity and resolution to detect high-velocity compact clouds (HVCCs), which have been claimed as evidence for intermediate mass black holes interacting with molecular gas clouds, we find no such objects across the large survey area.
\end{abstract}


\section{Introduction}
The central $\sim$500\,pc of our Galaxy -- the `Central Molecular Zone' (CMZ) -- provides a unique insight into the environmental dependence of the processes that govern star formation \citep{morris_serabyn_96, longmore13,kruijssen14, Henshaw2022}. The conditions found within the CMZ - in particular the Mach number, densities and temperatures of the gas, as well as the thermal and turbulent gas pressures - are far more extreme than those found in the Galactic disk, more closely resembling high redshift galaxies \citep{KL13}. The dense molecular gas in the CMZ, from which stars are expected to form, has been extensively studied both as part of large-scale Galactic plane surveys \citep[e.g.][]{Dame2001,Jackson2013,longmore17}, as well as more targeted observations \citep[e.g.][]{Rodriguez2004,Oka2007,Bally2010,Molinari2011,Jones2012,Mills2013,Lu2015,rath15,Krieger2017,Mills2017,Kauffmann2017,Kauffmann2017b,lu17,ginsburg18,Mills2018,pound_yz18,Walker2018,Lu2019}. 

The {\emph {CMZoom}} survey \citep{Battersby2020} has aimed to fill a key unexplored part of observational parameter space by
providing the first sub-pc spatial resolution survey of the CMZ at sub-millimetre wavelengths, targeting all dense gas above a column density of N(H$_{2}$) $\geq 10^{23}$\,cm$^{-2}$. The survey goals are to provide (i) a complete census of the most massive and dense cloud sources; (ii) the location, strength and nature of strong shocks; (iii) the relationship of star formation to environmental conditions such as density, shocks, and large-scale flows.

A detailed overview of the \emph {CMZoom} survey and the continuum data release was provided by \citet[][hereafter called `Paper I']{Battersby2020}. Paper I found that while the CMZ has a larger average column density than the Galactic disk, the compact dense gas fraction (CDGF) is significantly lower. This is a measure of the fraction of a cloud that is contained within the compact substructures (i.e. overdensities) that may form or are currently forming stars. Paper I concludes that identifying and understanding the processes that inhibit the formation of compact substructures is vital in explaining the current dearth of star formation within the CMZ \citep{longmore13,kruijssen14,2017MNRAS.469.2263B, Henshaw2022}.

The complete catalog of compact ($< 10\arcsec$) continuum sources was derived using dendrogram analysis and was presented in \citet[][hereafter called `Paper II']{Hatchfield2020}. Two versions of this catalog were produced: a robust catalog that contains only sources detected with high confidence - i.e only sources with a peak flux and a mean flux that are 6$\sigma$ and 2$\sigma$ above the local RMS estimates of each mosaic respectively - which was found to be 95$\%$ complete at masses of 80 M$_{\odot}$ at a temperature of 20 K; and a second catalog focusing on completeness across the CMZ. This second `high-completeness' catalog was 95\% complete at masses of 50 M$_{\odot}$ at 20 K. The catalogs contain 285 and 816 sources, respectively. These sources have typical sizes of $0.04-0.4$\,pc and are potential sites for ongoing and future star formation. Using this catalog, Paper II estimates a maximum star forming potential in the CMZ of $0.08 - 2.2$ M$_\odot$\,yr$^{-1}$, though this drops to $0.04 - 0.47$ M$_\odot$ yr$^{-1}$ when Sagittarius B2 -- the dominant site of active star formation in the CMZ -- is excluded.

In addition to the 230\,GHz continuum data, the \emph{CMZoom} survey also observed spectral line emission with an 8 GHz bandwidth using the {\sc {ASIC}} correlator, and an additional 16 GHz using the {\sc {SWARM}} correlator during later stages of the survey. In this paper, we give an overview of the spectral line data of the \emph{CMZoom} survey, and present the full spectral data cubes where available, and cubes targeting specific transitions otherwise. The spectral set-up (detailed in Paper I) targeted a number of dense gas tracers (CO isotopologues, multiple H$_{2}$CO transitions), as well as key shock tracers (SiO, SO, OCS) and compact hot core tracers (CH$_{3}$OH, CH$_{3}$CN). An overview of the targeted lines is given in Table \ref{tab:key_transitions}.

This paper is organised as follows. Section~\ref{sec:imaging} details the additional steps required for the imaging pipeline for the spectral line data beyond that described for the continuum data in Paper I. Section~\ref{sec:mmsf} outlines the generation and fitting of spectra and the production of moment maps. Section~\ref{sec:datapres} describes the data across the whole survey region and then describes the data quality and summarises the line detections on a per region basis.  Section~\ref{sec:line_brightness_variation} uses the integrated intensity maps of all detected spectral lines to explore the relative variation in line emission across the survey as a rough indicator of variations in conditions throughout the CMZ. Section~\ref{sec:cont_sources} examines the line properties of the \emph{CMZoom} continuum sources identified in Paper II. By comparing the brightness, line fitting results and detection statistics of different transitions, we aim to identify a primary kinematic tracer to describe the gas motions in the compact continuum sources. In Section~\ref{sec:analysis}, we use the results of the line fitting and conclusions in Section~\ref{sec:cont_sources} to determine the likely virial state of the continuum sources, and search for signs of proto-stellar outflows and intermediate-mass black holes in the CMZoom line data.

\section{Observations and Imaging}\label{sec:imaging}
Here we summarize the source selection, spectral setup, configurations, observing strategy and data calibration, all of which discussed in more detail in Paper I and Paper II. In this section we detail the pipeline beyond these aspects, how this pipeline differs from that of continuum imaging, and the complexities and non-uniformities that arose during this process.

\subsection{Observations and Spectral Setup}
Given the \textit{CMZoom} survey's key goal of surveying the high mass star formation across the entire CMZ, targets were selected to nearly completely include all regions of high column density (N(H$_2$)>10$^{23}$ cm$^{-2}$), with one small exception detailed in Paper I. Additionally, several regions of interest with lower column density were selected, including the ``far-side candidate'' clouds and isolated high-mass star forming region candidate clouds. A complete summary of source selection can be found in section 2.1 of Paper I, and a region file with the mosaic of the survey's pointings is published in the Dataverse at \hyperlink{https://dataverse. harvard.edu/dataverse/cmzoom}{https://dataverse.harvard.edu/dataverse/cmzoom}. 

Over the course of the program's observation, the SMA transitioned from the ASIC correlator to the SWARM correlator \citep{pri16}, and the extent of each sideband in any given observation varies depending on the date of the observation. The early ASIC observations had a lower sideband covering 216.9–220.9 GHz and an upper sideband spanning 228.9–232.9 GHz, while the widest coverage in later SWARM observations spans 211.5–219.5 GHz in the lower sideband and 227.5–235.5 GHz in the upper sideband, with the majority of observations being intermediate to these two extremes. The spectral resolution is held consistent across all published observations at about 0.812 MHz (or about 1.1 km s$^{-1}$).

\subsection{Imaging Pipeline}
Given the size of the survey both spatially and spectrally, a pipeline was developed to take the data from post-calibration to final imaging steps.  We used the software package {\sc CASA}\footnote{https://casa.nrao.edu/} to ensure a consistent approach to data imaging across the whole survey, using both compact and subcompact SMA antenna configurations. In this section, we describe the stages of this pipeline.

The input for the pipeline is the source name (variable `sourcename') and the file paths corresponding to the relevant calibrated datasets in {\sc MIR}\footnote{https://lweb.cfa.harvard.edu/$\sim$cqi/mircook.html } format. Each of these datasets are called into {\sc MIR}, which we use to determine the associated correlator (or combination of correlators for observations taken within the middle of the observing period). Once this is determined, we use {\sc IDL2MIRIAD} to convert the data from {\sc MIR} to {\sc MIRIAD} format. We split the dataset into chunks, with the number of chunks depending on the correlator, before we flag the data. We enforced an 8 channel and 100 channel flag for each chunk of data from the {\sc ASIC} and {\sc SWARM} correlators, respectively, to remove noisy  channels from both edges of the bandpass. We then convert these flagged data into \emph{uvfits} format using {\sc MIRIAD}'s \emph{fits} command with \emph{line} set to \emph{channel}. 

These \emph{uvfits} files are then loaded into {\sc CASA} and converted into a readable format using the \emph{importuvfits} task in frequency mode with an LSRK outframe. They are then concatenated into full upper and lower sidebands for each correlator using \emph{concat}. These sidebands are then continuum subtracted individually, using \emph{uvcontsub}. We do this by estimating the baseline for all channels, excluding those surrounding the brightest line within each sideband, which in this case we took to be the $^{12}$CO and $^{13}$CO transitions for the upper and lower sidebands, respectively.

To image these continuum-subtracted datasets, we first generate a `dirty' image cube to determine the appropriate R.M.S. noise level for the cleaning process. To do this, we run {\sc CASA}'s \emph{tclean} task with 0 iterations over a patch of size 100 x 100 pixels around the phase center. We also perform this over a 100 channel sub-chunk of the whole frequency space to minimise the time taken. This channel range has been predetermined to be line-free by eye in all cubes. We then use \emph{imstat} to calculate the average R.M.S. noise level throughout this cube.

Given the large variety of mosaic sizes and limited computing power, we implemented two separate methods to produce cleaned images. These methods are separated by image size, with a cut at 1000 pixels per spatial axis. For images smaller than this, we simply pass the full 4 GHz cube into a \emph{tclean} task. We set the pixel size to 0.5\arcsec{}, corresponding to 6-8 pixels per roughly 3-4\arcsec{} beam. We used a \emph{multiscale} deconvolver with scales equal to 0\arcsec{}, 3\arcsec{}, 9\arcsec{} and 27\arcsec{}  to recover both large and small scale structures. A channel width of 0.8 MHz, or 1.1 km s$^{-1}$ was enforced to ensure consistency between ASIC and SWARM datasets. The weighting for each image was set to \emph{briggs}, with a \emph{robust} parameter of 0.5. The \emph{threshold} is set to 5$\sigma$ where possible, with $\sigma$ calculated from the dirty cube previously discussed, with an arbitrarily high number (10$^8$) of \emph{iterations}  to ensure we reach this threshold. For some clouds, this 5$\sigma$ threshold led to severe imaging artifacts so the threshold for these clouds were manually modified to remove them. We make use of the \emph{chanchunks} parameter for these cleans, setting it to -1 to allow for the number of chunks that the datacube is split up into to be determined based on the available memory. We do not utilise the \emph{auto-multithresh} parameter as used for the continuum images at this stage due to the significant increase in computational time of the pipeline that it leads to.

For images larger than the 1000 pixel cut described above, we instead clean separate sub-cubes surrounding a number of key spectral lines that the CMZoom survey targeted (see Table~\ref{tab:key_transitions} for details). For the upper sideband, this is \twelveCO and OCS, and for the lower sideband we include three transitions of H$_{2}$CO in the range of 218 - 219 GHz, \thirteenCO, \CeighteenO, SiO, OCS and SO. Each of these cubes is $0.3$ GHz wide, centred on the rest frequency of the corresponding transition, which is passed into the task within the \emph{restfreq} parameter to allow for easy estimation of the velocity. All other parameters in these \emph{tclean} tasks are the same as the smaller cubes.

Each output image is then primary beam corrected by dividing the image by the corresponding .pb file, which is generated by \emph{tclean}, using {\sc CASA}'s \emph{immath} task.

\subsection{Catalog of Continuum Sources}
The spectral fitting and subsequent analysis used in this work makes use of the high-robustness version of the \textit{CMZoom} catalog, described in detail in Paper II. In this section, we provide a brief description of the source identification procedure and completeness properties. 

The \textit{CMZoom} catalogs are constructed using a pruned dendrogram. The dendrogram algorithm {\sc{astrodendro}} is used to generate a hierarchical segmentation of the 1.3mm dust continuum maps. Within this tree-like hierarchical representation, the highest level structures are defined as ``leaves'', which correspond to compact dust continuum sources cataloged in Paper II. The cataloged leaves are uniquely determined by the choice of three initial dendrogram parameters: the dendrogram minimum value, the minimum significance parameter, and the minimum number of pixels to define a unique structure. The minimum significance and minimum value are both defined in reference to a global noise estimate, and the minimum number of pixels is selected relative to the typical beam of the SMA continuum observations. Because of the high variability in noise properties across 1.3mm continuum within the CMZoom field, this initial dendrogram is overpopulated, particularly in regions with extreme local noise levels. A local estimate of the RMS noise is determined from the 1.3mm continuum residuals, and is used to prune the dendrogram, removing sources with low local signal-to-noise ratios. The sources that remain in the high-robustness catalog are dendrogram leaves that satisfy 6$\sigma$ peak flux and 2$\sigma$ mean flux minimum criteria relative to the local noise. The completeness of the catalog is determined using simulated observations of the SMA's interferometric setup, resulting in 95\% completeness to compact sources with masses above 80 M$_\odot$, assuming a dust temperature of 20K. The final robust catalog contains 285 compact sources, with effective radii between 0.04 and 0.4 pc, making them the potential progenitors of star clusters. In this work, we report on the spectral line properties of these 285 compact sources in the robust catalog.
A full description of the cataloging procedure is presented in Paper II.

\section{Spectral line fitting and moment map generation}\label{sec:mmsf}

In this section, we first describe the process  used to identify and fit spectral line emission from the compact continuum sources identified in Paper II. We then describe the process used to create moment maps to show the spatial variation in line emission across the region.

Spectra for each compact continuum source identified in Paper II were produced by averaging all emission per channel over the mask produced for that leaf within the robust dendrogram catalog in Paper II. These spectra were then fit using \emph{ScousePy}'s\footnote{https://github.com/jdhenshaw/scousepy} \citep{Henshaw2016a, Henshaw2019} stand-alone fitter functionality \citep[see also][]{Barnes2021}. We use a fiducial signal-to-noise ratio (SNR) of 5 to determine the initial threshold at which fits are accepted. The default kernel was set to 5, which smooths the spectrum by averaging every 5 channels. By-eye inspection showed that this produced reliable results for the majority of spectra. Approximately $\sim5\%$ of spectra required manual fitting as the interactive \emph{scousepy} fitter was unable to find a combination of SNR threshold and smoothing kernel to fit these spectra. 

Before analysing these fits, we enforced a series of cuts to the data that by-eye inspection showed reliably removed bad fits. We enforced a cut on the velocity dispersion, $\sigma$, and centroid velocity, $V_{\rm LSR}$, uncertainties to only keep fits with uncertainties smaller than 1.5 km\,s$^{-1}$, and only allowed for a maximum uncertainty on the amplitude of 0.5 Jy\,beam$^{-1}$ (1.3 K). To mitigate any issues with fitting multiple peaks as one single peak, we also cut out any fits that had velocity dispersions larger than 20 km\,s$^{-1}$, and removed peaks narrower than 0.5 km s$^{-1}$. Despite this check, a manual assessment confirmed no spectral components that exceeded this upper velocity dispersion threshold. Due to a combination of imaging artefacts caused by spatial filtering, and inherently more complex spectra, the $^{12}$CO and $^{13}$CO spectral line fits were both deemed too unreliable throughout most of the survey and so were removed from this process. 

\begin{figure*}
\begin{center}
    \includegraphics[width=0.9\textwidth]{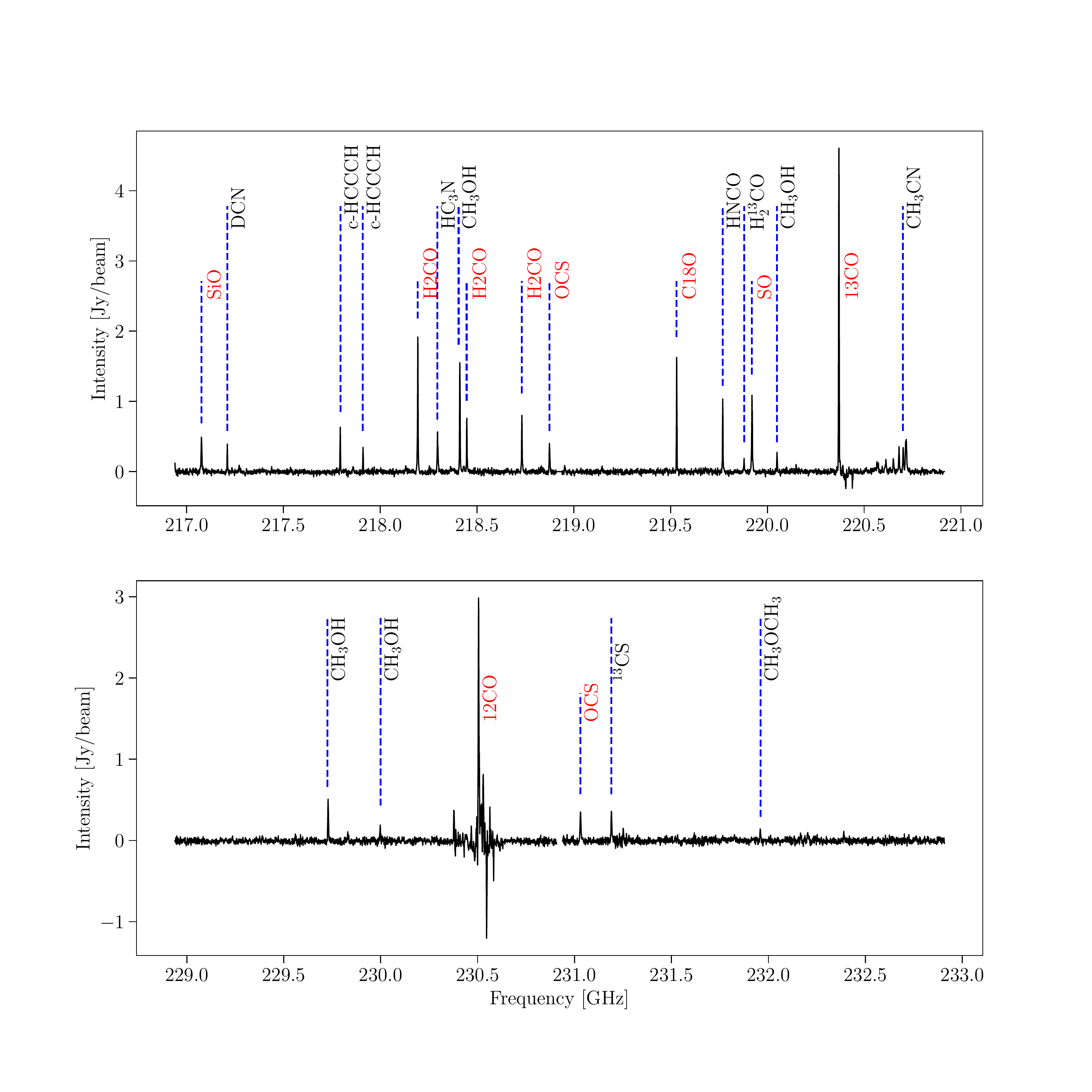}
    \caption{Complete spectra for the lower (top) and upper (bottom) sidebands for the region G0.380$+$0.050, colloquially referred to as `cloud C'. Red labels indicates the 10 transitions targeted by the CMZoom survey, with over a dozen additional lines labeled in black. Assuming a beam size of 3$\arcsec \times 3\arcsec$, at a frequency of 230 GHz, 1 Jy/beam = 2.57 K.}
    \label{fig:lines}
\end{center}
\end{figure*}

The spectra show emission from a number of lines beyond the 10 key lines targeted by the survey (see Table~\ref{tab:key_transitions}). Figure~\ref{fig:lines} shows the potential chemical complexity within a compact source in the CMZoom catalogue, using G0.380$+$0.040, or `dust ridge cloud c', as an example. To identify these lines, a single \Vlsr\ was determined for every compact source using the weighted average \Vlsr\ of all detected lines. Any lines with a centroid velocity that differed by this \Vlsr\ by more than $\pm 20$ km s$^{-1}$ were flagged as unidentified. These lines had their frequency calculated and then passed through Splatalogue\footnote{https://splatalogue.online/} with a search range of $\pm 0.04$ GHz with an upper energy limit of 100\,K. While this potentially misses some of the more high-excitation lines that may be present in the CMZ, this limit is simply a starting point to manually identify a first guess for each transition based on an assessment of the Einstein coefficient and upper energy level.

Once additional lines were assigned a most likely transition, we explored the quality of all the data by assessing the line of sight velocities, velocity dispersions, peak intensities and root-mean-square (RMS) of each compact source in the survey.

Moment maps were then produced over a velocity range of $\pm 20$ km s$^{-1}$ surrounding all dendrogram sources within a region. To generate these moment maps, an RMS map was first produced by measuring the RMS per pixel and then cutting anything over a threshold as determined by the number of channels in each pixel. This robust RMS map was used to enforce a 10$\sigma$ cut in order to identify the most significant emission within a region. This mask was then grown outwards, with \textit{scipy}'s \textit{binary dilation} task, with a lower SNR cut, down to 5$\sigma$ in order to detect low level extended emission surrounding the most robust emission. Not all clouds have emission at the 10$\sigma$ level, so this process was repeated with an iteratively lower SNR threshold until some emission was detected. If no emission was detected down to 5$\sigma$, the region was flagged as having no emission. Examples of these moment maps can be found in Appendix~\ref{sec:appendix}, which has been made available online.

\section{Data presentation}
\label{sec:datapres}

Below we present the spectral line data cubes of the 10 main molecular line transitions covered in the CMZoom spectral setup. Table~\ref{tab:key_transitions} lists these transitions and their relevant properties. 

\begin{table*}
\centering
\begin{tabularx}{0.95\textwidth}{|c|c|c|c|c|c|}
\hline
Molecule & Rest Frequency (GHz) & Quantum Number & Upper Energy Level (K) & Tracer & Detection Percentage \\
\hline
$^{12}$CO & 230.53800000 & J=2-1 & 16.59608 & Dense Gas & 96\\
$^{13}$CO & 220.39868420 & J=2-1 & 15.86618 & Dense Gas & 96 \\
C$^{18}$O & 219.56035410 & J=2-1 & 15.8058 & Dense Gas & 58 \\
H$_{2}$CO & 218.22219200 & 3(0,3)-2(0,2) & 20.9564 & Dense Gas & 82 \\
H$_{2}$CO & 218.47563200 & 3(2,2)-2(2,1) & 68.0937 & Dense Gas & 36 \\
H$_{2}$CO & 218.76006600 & 3(2,1)-2(2,0) & 68.11081 & Dense Gas & 39 \\
\hline
SiO       & 217.10498000 & 5-4 & 31.25889  & Protostellar outflows \& shocks & 39 \\
\hline
OCS       & 218.90335550 & 18-17 & 99.81016 & Shocks & 15 \\
OCS       & 231.06099340 & 19-18 & 110.89923 & Shocks & 13 \\
SO        & 219.94944200 & 6-5 & 34.9847 & Shocks & 60 \\
\hline
\end{tabularx}
\caption{Summary of 10 key transitions targeted by the CMZoom survey with the percentage of sources investigated in this paper that show emission in that transition.}
\label{tab:key_transitions}
\end{table*}

We start by providing a summary of the general emission and absorption characteristics for each transition across the full survey region, focusing on comparing the spatial extent and velocity range of the emission for the different transitions and also with the 230\,GHz continuum emission reported in Papers I and II. Our goal here is to provide the reader with a qualitative idea of the quality and the breadth of the data across the whole survey and on a per region basis.

Table~\ref{tab:line_flags} provides a description of the data quality for each of the 10 key transitions per region, and also highlights any issues which may affect the robustness and reliability of the images for analysis. We find that the $^{12}$CO and $^{13}$CO emission is detected in 100\% and 90\% of the clouds, respectively. In nearly all clouds, the emission is spatially extended across a large fraction of the survey area. There is little correspondence between the $^{12}$CO and $^{13}$CO integrated intensity emission and the 230\,GHz continuum emission. However, the $^{12}$CO and $^{13}$CO emission often suffers from severe imaging artefacts due to missing flux problems and also absorption from foreground gas clouds along the line of sight. For that reason we urge caution in interpreting the integrated intensity and moment maps from these transitions, and more generally, in blindly using the $^{12}$CO and $^{13}$CO data without the addition of zero-spacing information. Similarly, we have opted to not use these data products during the analysis until these imaging artefacts are resolved in a future paper unless there are particular aspects of the data which are relevant to highlight.


\begin{table*}
\begin{center}
 \caption{Summary of conditions of data cubes for all clouds and across 9 key molecular lines as a check of robustness and reliability for science. Each cube has been checked for a number of flags depending on extracted spectra and a visual inspection of the cubes. The flags are given as acronyms: multiple velocity components (MVC), imaging artefacts (IA), missing channels (MC), broad lines (GC) or narrow lines (N), line-wings (LW), non-detection (ND) and contamination of other spectral lines (C).}

\begin{adjustbox}{angle=90}
\begin{tabularx}{1.247\textwidth}{|l||l|c|c|c|>{\centering}p{1.6cm}|>{\centering}p{1.7cm}|>{\centering}p{1.7cm}|>{\centering}p{1.2cm}|>{\centering}p{1.2cm}|c|} 
 \hline
 \hline
    Sourcename & Colloquial Name & $^{13}$CO & C$^{18}$O & H2CO & H2CO & H2CO & OCS & OCS & SiO & SO \tabularnewline
        & & & (218.2 GHz) & (218.5 GHz) & (218.8GHz) & (218.9 GHz) & (231.1 GHz) & & & \tabularnewline
 \hline
    G0.001-0.058 & 50 km s$^{-1}$ Cloud & IA & MVC & MVC & MVC &  & MC & ND & MVC & MVC \tabularnewline
    G0.014+0.021 & Arches e1 & & ND & ND & MC & MC & MC & MC & ND & ND \tabularnewline
    G0.0.68-0.075 & Three Little Pigs: Stone Cloud & IA & MVC & GC, MVC & MVC, C & MVC, GC & MC & ND & ND & ND \tabularnewline
    G0.070-0.035 & Apex H$_{2}$CO bridge & & & & & & & & & \tabularnewline
   G0.106-0.082 & Three Little Pigs: Sticks Cloud & IA & MVC & MVC, C & MVC & MC & ND & GC, LW & LW & \tabularnewline
   G0.145-0.086 & Three Little Pigs: Straw Cloud & IA & MVC & MVC & ND & MC & MC &  & ND & ND \tabularnewline
   G0.212-0.001 & isolated HMSF candidate & IA & & & MVC & & MC & MC & MC & ND \tabularnewline
   G0.316-0.201 & isolated HMSF candidate & & & C & C &  & MC & MC & & ND \tabularnewline
   G0.326-0.085 & far-side stream candidate & IA & ND & ND & ND & ND & MC & MC & ND & ND \tabularnewline
   G0.340+0.055 & Dust Ridge: Cloud b & IA & ND & & ND & ND & MC & MC & ND & ND \tabularnewline
   G0.380+0.050 & Dust Ridge: Cloud c & MVC & C & C & C & C & MC & MVC, MC & C & MVC, C \tabularnewline
   G0.393-0.034 & isolated HMSF candidate & & MVC & MVC & ND & ND & MC & MC & ND & ND  \tabularnewline
   G0.412+0.052 & Dust Ridge: Cloud d & IA & & & & ND & MC & MC, ND & ND & ND \tabularnewline
   G0.489+0.010 & Dust Ridge: Clouds e+f & & & & & & & & & \tabularnewline
   G1.085-0.027 & 1.1$^{\circ}$ cloud & & & ND & ND & MC & MC, ND &  & ND & ND \tabularnewline
   G1.602+0.018 & 1.6$^{\circ}$ cloud & & ND & C & C & MC, ND & MC &  & & \tabularnewline
   G1.651-0.050 & 1.6$^{\circ}$ cloud & MVC & & & C & MC & MC & ND & ND & ND \tabularnewline
   G1.670-0.130 & 1.6$^{\circ}$ cloud & ND & ND & ND & MC & MC & MC & MC & ND & ND \tabularnewline
   G1.683-0.089 & 1.6$^{\circ}$ cloud & ND & ND & ND & MC & MC & MC & MC & MC & MC \tabularnewline
   G359.137+0.031 & isolated HMSF candidate & C & & C & C & MC & MC &  & N, GC & MVC, C \tabularnewline
   G359.484-0.132 & Sgr C & IA & & & & & MC & MC &  &  \tabularnewline
   G359.611+0.018 & far-side stream candidate & & ND & ND & ND & ND & MC & MC & ND & ND \tabularnewline
   G359.615-0.243 & isolated HMSF candidate & IA & & C & C & C & C MC & MC & MC & MVC, C \tabularnewline
   G359.734+0.002 & far-side stream candidate & IA & & C & C & C & MC, C & MC, C & MC & C \tabularnewline
   G359.865+0.022 & far-side stream candidate & & & & & & & & & \tabularnewline
   G359.889-0.093 & 20 km s$^{-1}$ Cloud & IA & & MC & ND & ND & ND & MC & ND & ND \tabularnewline
   G359.948-0.052 & Circumnuclear Disk & & & & & MC & MC & MC & & \tabularnewline
 \hline
 \end{tabularx}
 \end{adjustbox}
 \label{tab:line_flags}
\end{center}
\end{table*}

C$^{18}$O is detected towards 60\% of the clouds. The imaging artefacts are much less severe for C$^{18}$O than for the other CO transitions. The emission generally does appear spatially associated with the 230\,GHz continuum emission.

SO and SiO are detected towards 7 (20\%) and 5 (15\%) clouds, respectively, and are mostly well correlated --  all clouds with detection SiO emission are also detected in SO. This is perhaps unsurprising given they are both species thought to trace shocks. We explore the correlation between different tracers more fully in $\S$~\ref{sec:cont_sources}.

As expected, the three H$_2$CO transitions show a very good correspondence, both spatially and in velocity. At least one transition of H$_2$CO was observed towards 50\% of clouds. In the spectra containing the H$_{2}$CO 3(2,2)-3(2,1) transition, there is often an apparent `additional' velocity component offset by 50\,\kms\ from the main velocity component that actually corresponds to CH$_{3}$OH-e (4(2) - 3(1)) with a rest frequency of 218.4401 GHz.

A discussion of each of the CMZoom clouds in turn can be found in Appendix~\ref{sec:region}, focusing on notable characteristics of the emission and specific issues with the data. The emission characteristics and issues for all clouds are summarised in Table~\ref{tab:line_flags}. Through visual inspection of the spectral line data cubes and integrated intensity maps, we found that except where specifically mentioned, there is significant emission in all $^{12}$CO and $^{13}$CO cubes, often with strong emission and absorption over a \Vlsr\ range of $\pm$100\, \kms. However, there are severe imaging artefacts, including strong negative bowls due to missing extended structure, making these cubes unreliable.
\section{Spatial variation in line emission across the CMZ}
\label{sec:line_brightness_variation}

With a fairly uniform sensitivity across the CMZ and a homogeneous analysis of the emission, CMZoom is well suited to investigating changes in line brightness on sub-pc scales as a function of location \citep{Battersby2020}. Detailed modelling of this line emission is required to fully understand the excitation conditions, opacity and chemistry to derive accurate physical properties of the gas. Such detailed modelling is beyond the scope of this paper. Instead, in this section we search for large differences in line strength ratios between clouds as a rough indicator of variations in conditions as a function of position throughout the CMZ.
\begin{figure*}
\begin{center}
    \includegraphics[width=0.95\textwidth]{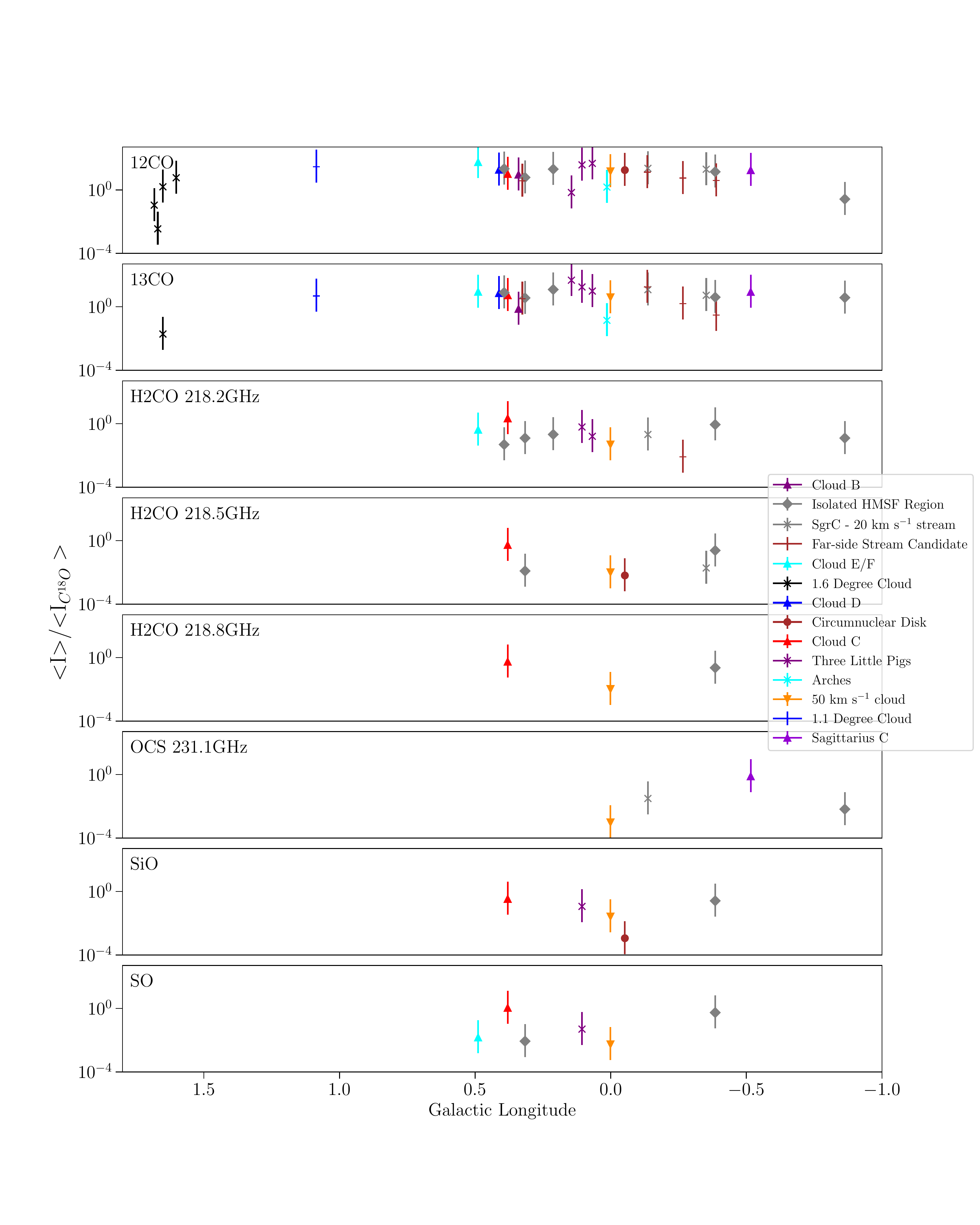}
    \caption{Normalized integrated intensity ratios in each region normalised by the integrated intensity of the C$^{18}$O emission in that region. Representative uncertainties of $\pm 1$ dex are shown, as these integrated intensity ratios likely suffer from both observational and physical uncertainties due to spatial filtering, optical depth effects, etc. OCS (218.9 GHz) has been removed from both this Figure and Figure 3 as it only has a single data point.}
    \label{fig:C18O_line_ratio}
\end{center}
\end{figure*}
\begin{figure*}
\begin{center}
    \includegraphics[width=.95\textwidth]{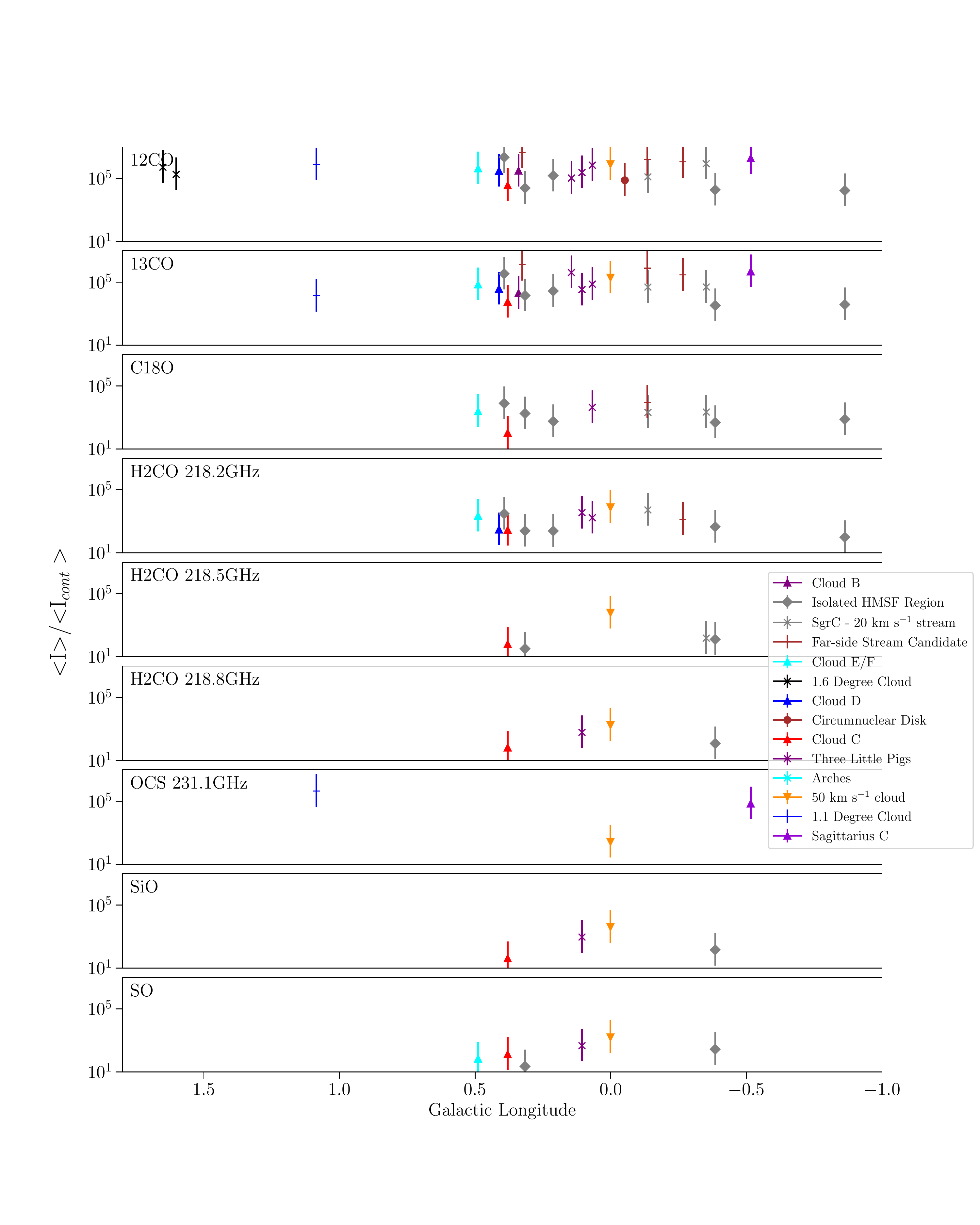}
    \caption{Normalized integrated intensity ratios in each region compared to the 230 GHz continuum. Representative uncertainties of $\pm 1$ dex are shown, as these integrated intensity ratios likely suffer from both observational and physical uncertainties due to spatial filtering, optical depth effects, etc.}
    \label{fig:cont_ratio}
\end{center}
\end{figure*}

For every region, if a transition was detected, all unmasked pixels in the moment map (see $\S$~\ref{sec:mmsf}) were summed and compared to the total integrated intensity of C$^{18}$O and the 230 GHz continuum emission. Figures~\ref{fig:C18O_line_ratio} and~\ref{fig:cont_ratio} show the distribution of these ratios as a function of Galactic longitude. Note that the Sgr B2 region (between $0.50^\circ < l < 0.72^\circ $) and the circumnuclear disk are not included on these figures due to the imaging difficulties described in $\S$~\ref{sec:region}.

Comparing the longitude range of the different transitions, $^{12}$CO and $^{13}$CO are detected across the full survey extent. With the exception of G1.085$-$0.027, which has a strong OCS (231.1 GHz) detection, the ratios for all other transitions are confined to $|l|<0.5^\circ$.  

As expected for a first look for general trends which does not solve for excitation, opacity, chemistry, etc., there is a large (order of magnitude) scatter in the line brightness ratios between clouds. Nevertheless, there are several interesting aspects of these figures, which we discuss below.

Firstly, we find that $^{12}$CO and $^{13}$CO have the highest ratios and are detected within the most clouds, followed by C$^{18}$O, and then the lowest energy transition of H$_2$CO. This simple trend is, of course, expected given that these lines are the brightest and most extended across the cloud sample.

Secondly, the integrated intensity ratios with respect to dust emission of SO, SiO, and the two upper energy levels of H$_2$CO all increase by several orders of magnitude towards the Galactic Centre (i.e., as $|l| \rightarrow 0^\circ$). Detailed modelling is required to understand the origin of this, but it is interesting to note that the highest excitation lines and shock tracers all increase in the same way, as may be expected due to changing physical conditions (e.g. increased shocks in the gas). This substantiates previous observations from \citet{Mills2017} who found a similar trend towards the Galactic Centre in a number of molecular species, a trend that was further supported by HC$_{3}$N observations by \citet{Mills2018} who found an increase in the dense gas fraction inwards of R $\lesssim 140$ pc.

Finally, we can compare the integrated intensity ratios of the CMZoom sources (all points apart from the grey diamonds in Figures~\ref{fig:C18O_line_ratio} and~\ref{fig:cont_ratio}) in the Galactic Centre with the isolated high mass star-forming (HMSF) regions in the survey. These lie along our line of sight towards the CMZ but are actually located in the disk, providing a useful control sample. 

The scatter of line brightness ratios of the isolated HMSF regions are consistent from the Galactic Centre sources in Figures~\ref{fig:C18O_line_ratio} and~\ref{fig:cont_ratio}. This is in direct contrast to observations of clouds in the Galactic Centre and the Galactic disk on $\gtrsim$\,pc scales, which show very different emission integrated intensity ratios. Molecular line observations of clouds in the Galactic Centre on $\gtrsim$\,pc scales show that bright emission from dense gas tracers (e.g. NH$_3$, N$_2$H$^+$, HCO$^+$) is extended across the entire CMZ (e.g. \citealp{Jones2012,longmore13}). However, emission from these dense gas tracers on similar scales in local clouds, such as Orion, is confined to the highest density regions of the clouds \citep[see][]{Lada2010, Pety2017,Kauffmann2017,Hacar2018}. The apparent similarity in these observed tracers (H$_2$CO, OCS, SiO, SO) may therefore indicate a difference in the chemistry between the various tracers, or it may simply be a product of observational uncertainties.

We note, however, several caveats in interpreting this at face value. Firstly, we do not observe the same lines that show these cloud-scale differences in \emph{CMZoom} and therefore cannot rule out that these differences would present themselves at the core-scale if these lines were observed. Secondly, it is not clear if the high mass star formation regions observed in the \textit{CMZoom} survey are representative of other such regions throughout the Galaxy. Thirdly, the variation in CMZ integrated intensity ratios may simply be so large that it encompasses the range in typical Galactic disk integrated intensity ratios.

\section{Line properties of 230\,GHz continuum sources}
\label{sec:cont_sources}

We now investigate the detection statistics and line properties of the CMZoom 230\,GHz continuum sources using the fits to the spectra for each of the main individual transitions targeted in the CMZoom survey (see Table~\ref{tab:key_transitions}).

\subsection{Detection statistics of brightest lines and identification of primary kinematic tracer}
Table~\ref{tab:key_transitions} also shows the detection statistics for each of the key tracers. We note here that the complete number of sources in our dataset differs substantially from the complete robust catalog presented in Paper II, as we have left several larger mosaics -- including Sagittarius B2 -- out of this analysis until additional steps can be made to suitably clean these. Of the remaining clouds, $^{12}$CO and $^{13}$CO are detected in 96$\%$ of all sources. However, all $^{12}$CO and most $^{13}$CO data suffer from image artefacts so they can not be used as reliable tracers for the kinematics of the sources. We remove these transitions in the kinematic analysis from here on.

After $^{12}$CO and $^{13}$CO, C$^{18}$O and the lowest energy H$_2$CO transition are the next most often detected, being found in 58$\%$ and 82$\%$ of all sources, respectively. As these transitions tend to be well correlated, sources with only one of these transitions are interesting targets for potential follow-up observations. As summarised in Table~\ref{tab:key_transitions}, the images of these transitions do not suffer from imaging artefacts and the line profiles are generally well fit with single or multiple Gaussian components. The emission from both of these transitions should therefore provide robust information about the compact source kinematics. Given the prevalence of the lower transition of H$_2$CO and the fewer deviations in line profiles from that well described by a single Gaussian component, we opt to use H$_2$CO as our fiducial tracer of the compact source kinematics.

\begin{figure}
    \centering
    \includegraphics[width=0.5\textwidth]{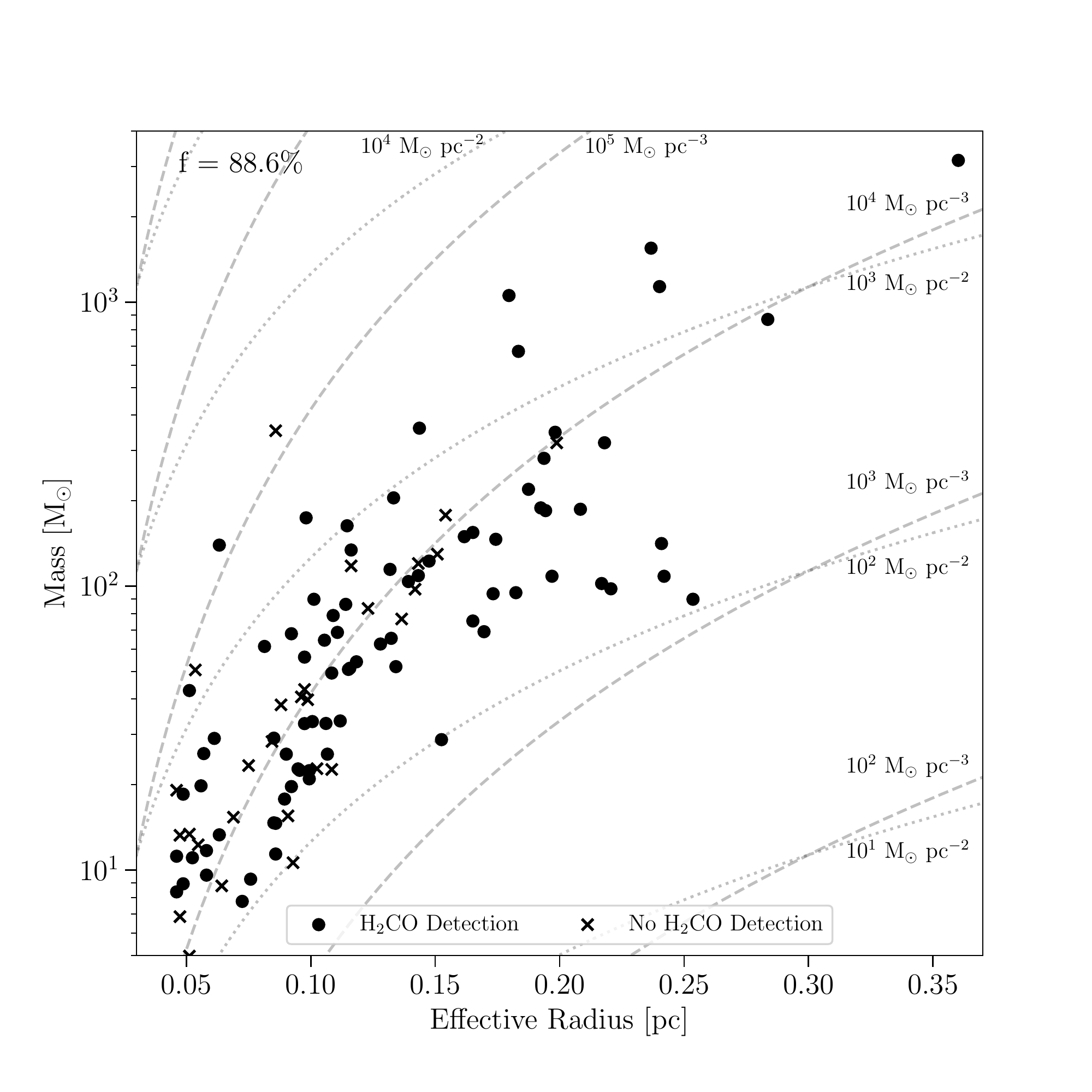}
    \caption{Mass vs effective radius relation with markers indicating sources with a H$_{2}$CO (218.2 GHz) detection. The number in the top-left corner states that sources with H$_{2}$CO (218.2 GHz) detections account for 88.8\% of the mass of all sources in this work. The dashed lines are lines of constant volume density where 10$^4$ M$_{\odot}$ pc$^{-3}$ $\sim$ 1.5x10$^5$ cm$^{-3}$ assuming a mean particle mass of 2.8 AMU. The detections lie in the range ~10$^4$-10$^6$ cm$^{-3}$. Dotted lines indicate lines of constant column density.}
    \label{fig:H2CO_mR}
\end{figure}

Figure~\ref{fig:H2CO_mR} shows the mass-radius relation for all sources included in this analysis, with circles indicating sources with a H$_{2}$CO (218.2 GHz) detection. As expected, the larger and more massive sources are more likely to be detected in H$_{2}$CO, though this transition is still detected in a majority of small, low mass sources. Overall, these sources represent 88.8$\%$ of the total mass of sources that have been included in this analysis. As such, using this transition as our fiducial tracer provides significant coverage across the whole survey.

\subsection{Analysis of compact source velocities}

\begin{figure}
    \centering
    \includegraphics[width=0.48\textwidth]{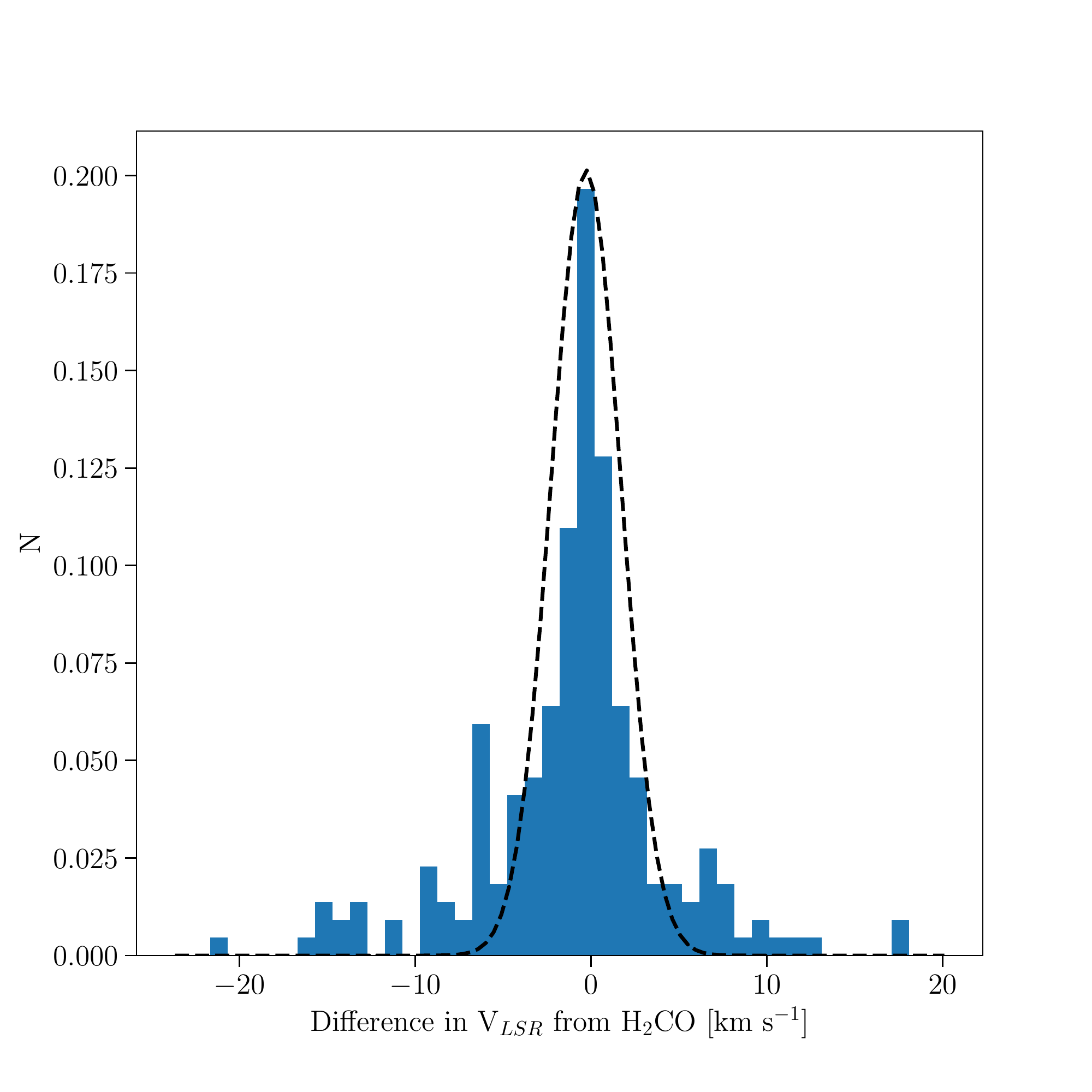}
    \caption{Histogram of the \Vlsr difference of each key transition when compared to the lower transition of \HtwoCO  for every compact source. The dashed line represents a Gaussian fit to the mean and standard deviation - ($\mu$, $\sigma$ = -0.29,1.98) - of the data.}
    \label{fig:core_vel_hist}
\end{figure}

Figure~\ref{fig:core_vel_hist} shows a histogram of the \Vlsr difference for each compact source between H$_2$CO and all other lines detected detected towards that compact source. The black dashed line shows the best-fit Gaussian to all data within a \Vlsr difference $\Delta \Vlsr \leq 5$\,\kms{}. The small mean and dispersion of $-0.29$\,\kms\ and 1.98\,\kms, respectively, gives confidence that the observed \Vlsr for sources is robust. There are 30 sources with $\Delta \Vlsr > 5$\,\kms{} which lie in 9 clouds throughout the survey. Of these 30 sources, 12 of them belong to G359.889$-$0.093, 5 to G0.001$-$0.058 and 4 to G0.068$-$0.075 -- i.e. they lie very close in projection to the Galactic centre. This is the most complicated part of position-position-velocity space, with multiple, physically distinct components along the line of sight, so these \Vlsr{} offsets are not unexpected \citep{Henshaw2016}.

We then seek to understand how these compact source \Vlsr{} values compare to the observed velocities of their parent clouds on larger scales. In order to determine a representative velocity range for each parent cloud, we use the catalogue of Walker et al. (in prep.), who extracted spatially averaged spectra for each cloud from single-dish data in the literature. To do this, they used archival data from the APEX CMZ survey at 1mm \citep{Ginsburg2016}, and the MOPRA CMZ survey at 3mm \citep{Jones2012}. The results used here are specifically from the Gaussian fits to the integrated spectra of the HNCO (4$_{0,4}$ - 3$_{0,3}$) emission.

\begin{figure}
    \centering
    \includegraphics[width=0.5\textwidth]{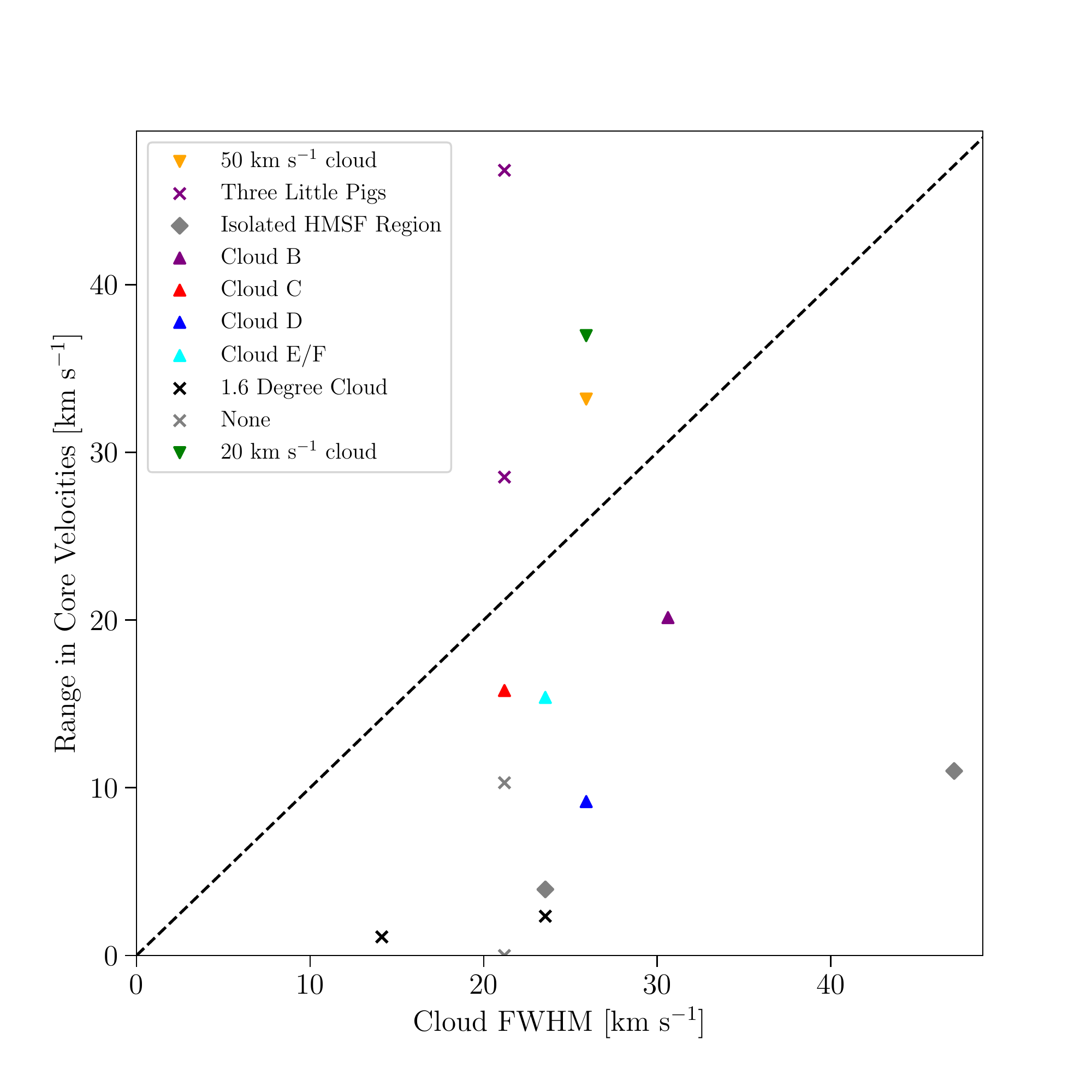}
    \caption{Comparison of the range in compact source velocities of the 10 key transitions targeted by the survey as measured by \textit{scousepy} to the observed FWHM of the cloud. The dashed line shows the one-to-one line. The majority of sources fall below the dashed line, as expected if these sources are distributed within the cloud. In the main text we discuss each of the clouds which lie above the dashed line.}
    \label{fig:cloud}
\end{figure}

Figure~\ref{fig:cloud} compares the full-width half maximum (FWHM) of the Walker et al. (in prep.) single-dish observations to the range of observed compact source velocities within the same cloud, using only the compact source velocities measured for the 10 key transitions described in Table~\ref{tab:key_transitions}. The dashed line shows the one-to-one relation between those velocities. In general,  we would expect the range of compact source velocities within a cloud to be similar to or smaller than the cloud's FWHM if the sources lie within the parent cloud, i.e. points should lie below the one-to-one line. As expected, most of the clouds satisfy this criteria. 

Two of the four clouds that do not meet this criteria are the 20- and 50- km s$^{-1}$ clouds. This is somewhat expected, firstly as these clouds are composed of large mosaics (67 and 24 pointings, respectively). Secondly, these clouds have large velocity gradients across them, causing the compact source velocities on one side of the region to differ significantly from the other side. Such velocity gradients are expected due to the evolution of gas clouds under the influence of the external gravitational potential \citep[see e.g.][]{kdl15,kruijssen19,dale19,petkova22}.

The `Three Little Pigs' clouds that lie above the one-to-one line, however, are small and do not have large velocity gradients across them. The region farthest above the one-to-one line -- `G0.068-0.075' -- contains 12 dense sources identified by Paper II. To try and understand the much larger than expected range in compact source velocities, we inspect the individual spectra for this region in detail. 

Figure~\ref{fig:0068a} shows the spectra extracted from each spectral cube of the most massive compact source (G0.068-0.075a) in which $^{13}$CO, C$^{18}$O, H$_{2}$CO (218.2 GHz) and SiO all peak at $\sim 20$\,km\,s$^{-1}$, differing from the average \Vlsr of the remaining sources within the cloud by $\sim 30$\,km\,s$^{-1}$.  Figure~\ref{fig:0068b} shows the same spectra for the second most massive compact source in the cloud, in which these key transitions peak well within the shaded region indicating the cloud's velocity dispersion. Since this is the case for all sources other than `a', it suggests that this compact source may not be contained within the cloud, and instead may be unassociated with the cloud identified in Walker et al (in prep.). \citet{Henshaw2016} identified a second velocity component along the same line of sight as this cloud, separated by $\sim20$ km s$^{-1}$, which could potentially be the source of these additional features. However, further work is required to understand the nature and location of compact source 'a'.

The fourth cloud above the dashed line, `G0.106-0.082', contains multiple, broad velocity components in the spectra (Figure~\ref{fig:0106a}). The peak of the CMZoom emission sits within the shaded region showing the cloud's velocity dispersion. However, additional velocity components in most of the  transitions lie outside this range. It seems likely that the Walker et al. (in prep.) catalogue only derived the cloud velocity and velocity dispersion from one of these two velocity components.

\subsection{Compact source velocity dispersions} 

\begin{figure}
    \centering
    \includegraphics[width=0.48\textwidth]{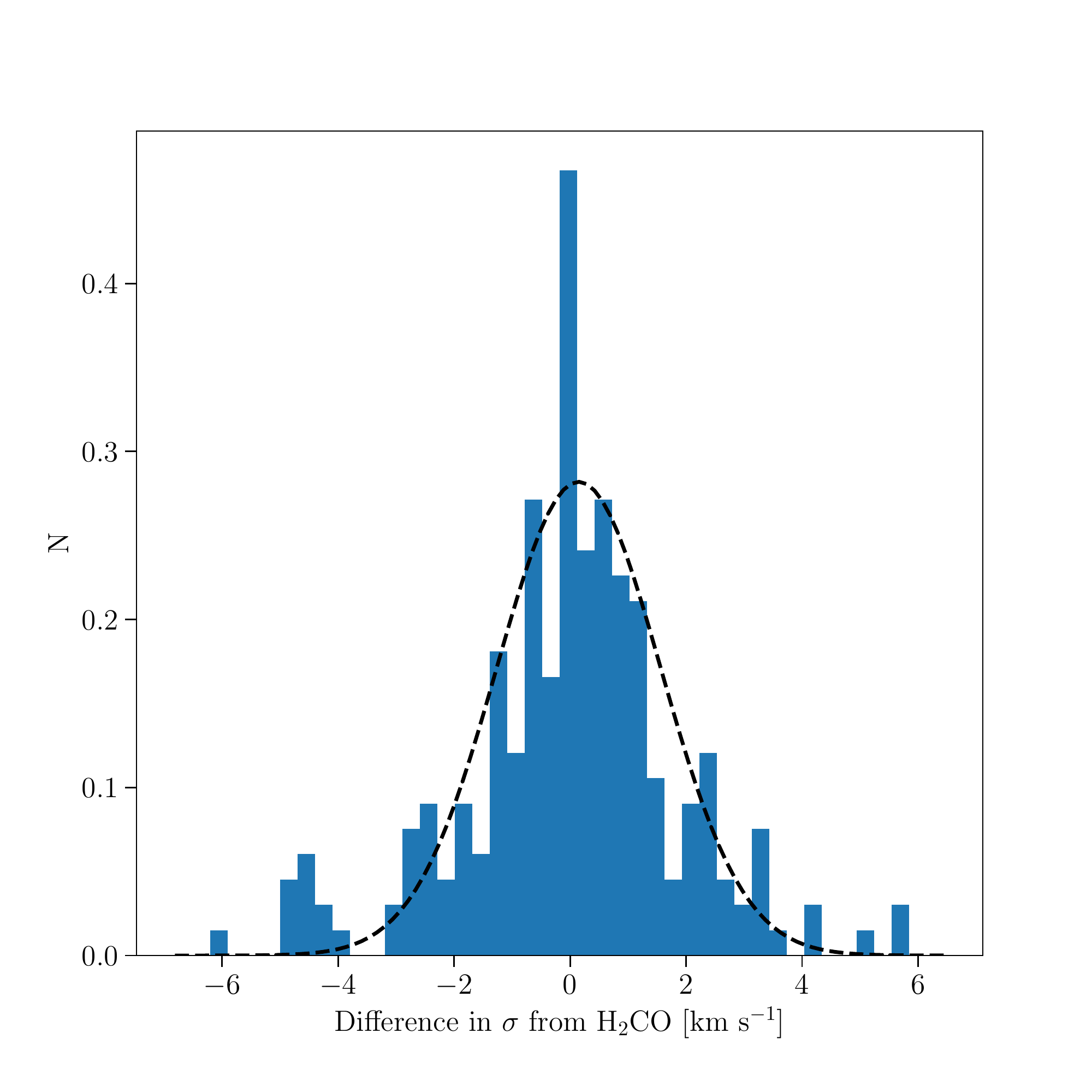}
    \caption{Histogram of the velocity dispersion difference of each key transition when compared to the lower transition of \HtwoCO for every compact source. The dashed line represents a Gaussian fit to the mean and standard deviation - ($\mu$, $\sigma$ = 0.15,1.41) - of the data.}
    \label{fig:core_disp_hist}
\end{figure}

Figure~\ref{fig:core_disp_hist} shows a histogram of the velocity dispersion difference for each compact source between H$_2$CO and all other lines detected towards that compact source. The black dashed line shows the best-fit Gaussian to all data within $\Delta \sigma \leq 4$\,\kms{}. The small mean and dispersion of 0.15\,\kms\ and 1.41\,\kms{}, respectively, gives confidence that the observed velocity dispersion for the sources are robust. There are 10 sources with $\Delta \sigma > 4$\,\kms{} from 4 different clouds. Of these 10 sources, 3 belong to G0.001$-$0.058, 3 to G0.068$-$0.075, 2 to G0.106$-$0.082 and 2 to G359.889$-$0.093. We note that most sources with $\Delta \sigma > 4$\,\kms{} also have $\Delta \Vlsr > 5$\,\kms{}, likely a result of either multiple velocity components being averaged together or poorer fit results from lower signal-to-noise spectra. 

\subsection{Number of lines detected per compact source} 

\begin{figure}
    \centering
    \includegraphics[width=0.5\textwidth]{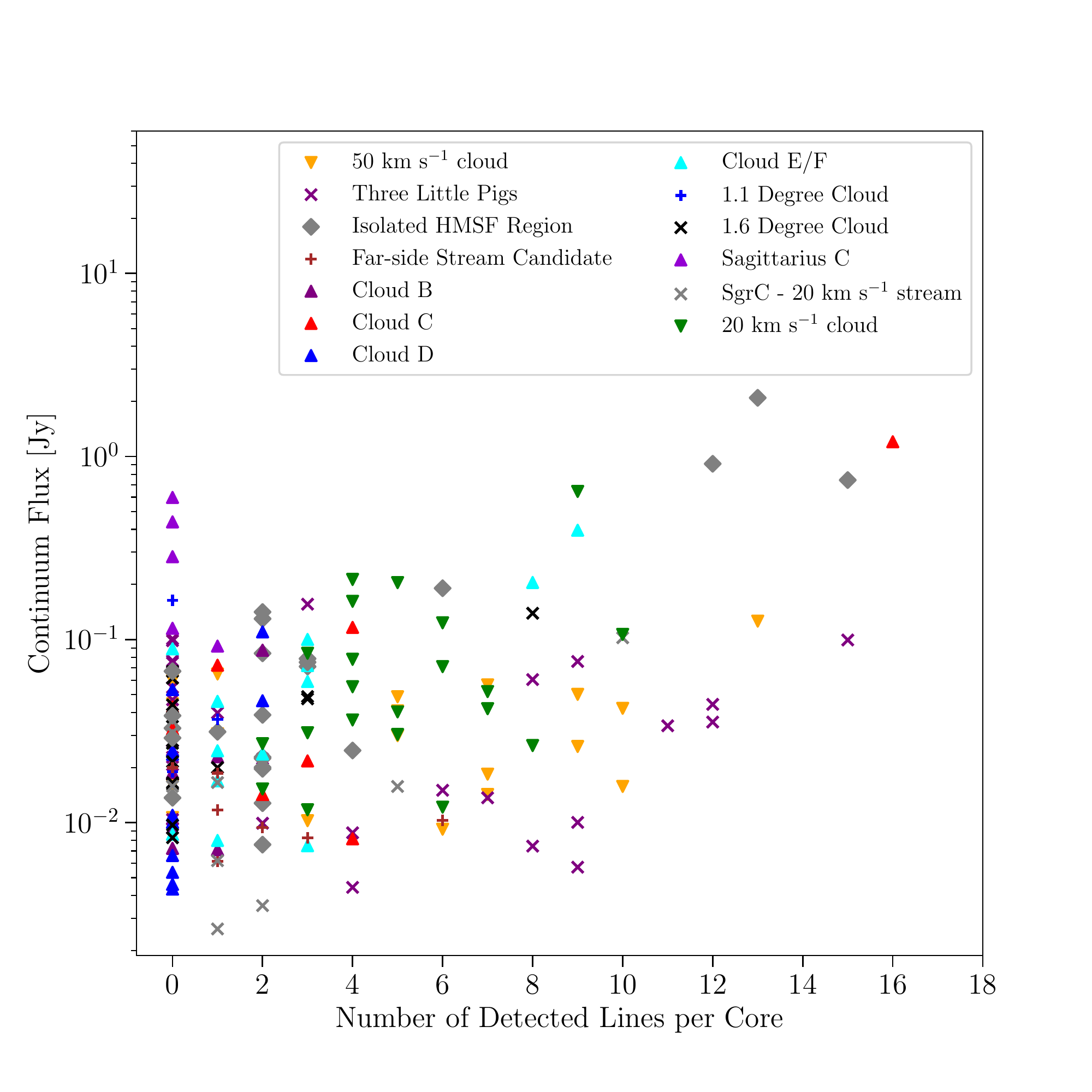}
    \caption{Comparison of the total continuum flux of each dense compact source to the number of total detected lines within the compact source. A general upwards trend implying that the brighter sources tend to have more line complexity.}
    \label{fig:contvn}
\end{figure}

Figure~\ref{fig:contvn} shows the relation between the observed continuum flux of each compact source and the number of spectral lines detected. There is a slightly upward trend showing that the brighter sources tend to have more lines detected. Three of the six observed dense sources within cloud `b' have no detected emission lines despite having continuum fluxes of $\gtrsim$0.2\,Jy. All other sources with such high continuum fluxes have $\ge$9 detected lines. These `line-deficient, continuum-bright' sources are interesting to followup as potential precursors to totally metal stars that have been predicted to exist \citep{hopkins_tot_metal14}. A source with bright continuum flux and no line emission suggests that either the gas to dust ratio is very low or the line abundances are very low. Very low gas to dust ratios are predicted by the `totally metal' star scenario, while the latter may highlight sources with interesting chemical or excitation regimes.

Conversely, sources in the `Three Little Pigs' clouds, and to a lesser extent the 50\,\kms{} cloud, stand out as having a large number of lines detected at low continuum flux levels. We note that in the right panel of Figure~\ref{fig:virial_pressure}, the sources in both of these clouds lie in the same portion of external pressure vs gas surface density space, and have a similar (low) fraction of star forming sources, with only one or two ambigious tracers of star formation activity. We speculate that the large number of lines detected in sources at low continuum flux levels in the `Three Little Pigs' clouds and 50\,\kms{} cloud may be the result of shocks in the high pressure gas beginning to compress the gas and instigate star formation. Further work is needed to test this hypothesis.

\subsection{Correlations between the emission from different transitions}

\begin{figure}
    \centering
    \includegraphics[trim={0 0 5cm 0},clip,width=1\columnwidth]{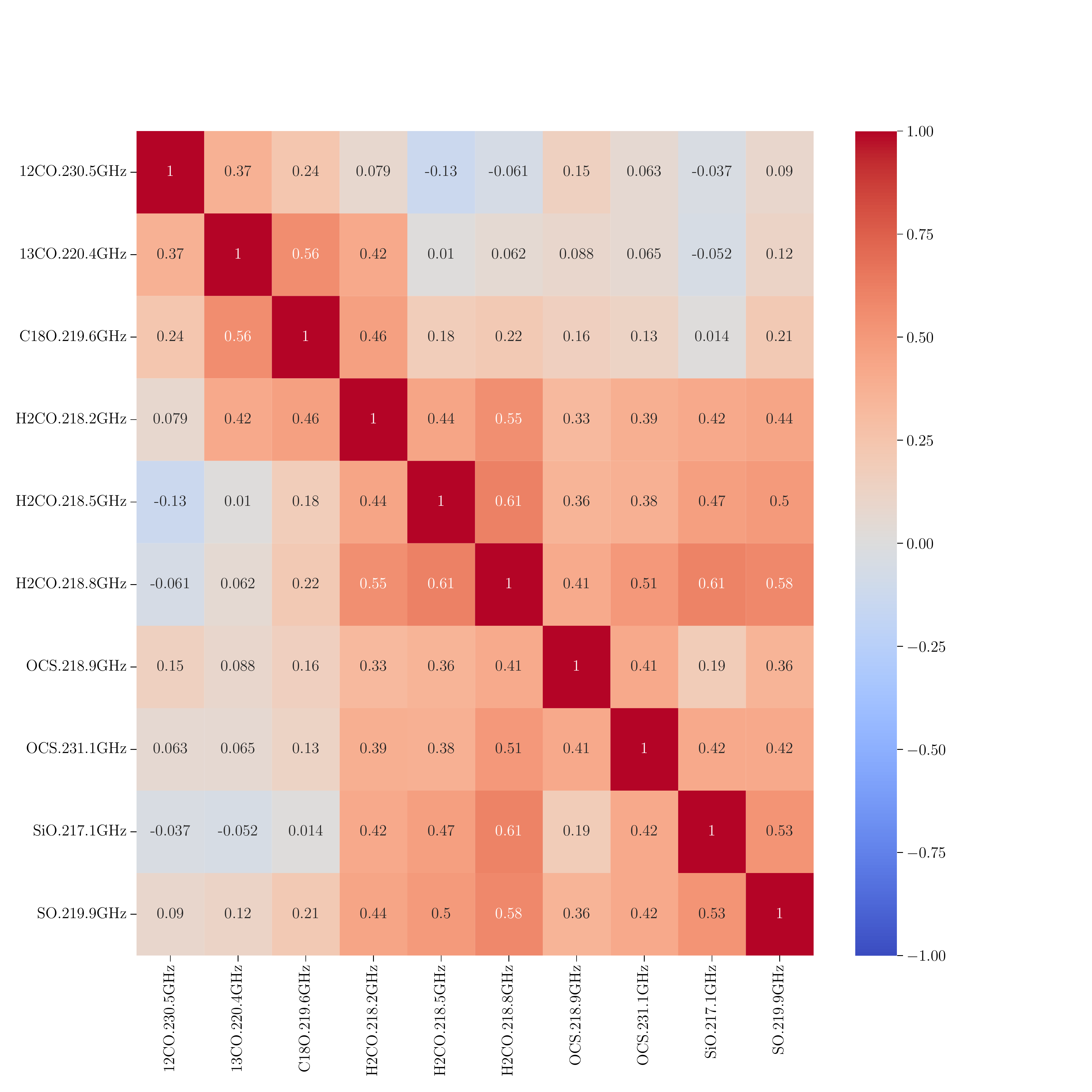}
    \caption{Correlation matrix showing the correlation coefficients between the amplitude of Gaussian peaks fit by \textit{scousepy} for the 10 key transitions targeted by the survey.}
    \label{fig:data_corr}
\end{figure}

We now investigate how well the emission from the 10 key different transitions correlate with each other. Figure~\ref{fig:data_corr} shows the correlation matrix for the measured amplitudes of the detected emission from these lines. The larger the correlation coefficient shown in each grid cell, the stronger the correlation between the two lines in that row and column. Negative values indicate the emission in the lines is anti-correlated. The correlation coefficient of $1.0$ along the diagonal of the matrix shows the auto-correlation of the emission from each line with itself. 

We begin by looking at the correlations between the three main `groups' of transitions --  the CO isotopologues, the H$_2$CO transitions tracing dense gas, and the shock tracers -- before investigating the correlations between transitions in different groups.

Unsurprisingly, emission from the three CO isotopologues are well correlated. The imaging artefacts in the $^{12}$CO and $^{13}$CO datacubes may well contribute to a lower correlation coefficient between these transitions than may have been expected. Again unsurprisingly, the three H$_2$CO transitions are also positively correlated, with the highest two energy levels having the highest correlation coefficient of all line pairs. Emission from the SiO, SO and OCS transitions are all well correlated too. As these transitions trace emission from shocks, these correlation makes sense. 

We then turn to comparing correlations between transitions in different groups. The emission from $^{12}$CO and $^{13}$CO is almost completely uncorrelated (and sometimes even slightly anti-correlated) with the emission from all the other transitions. The only stark exception to this is that emission from $^{13}$CO is well correlated with emission from the lowest energy level of H$_2$CO. 

The C$^{18}$O emission only shows a very weak correlation with most of the other non-CO transitions. Again the notable exception to this is that the C$^{18}$O emission is well correlated with the lower energy transition of H$_2$CO. The increasing correlation between the CO isotopologues with the lower energy transition of H$_2$CO, from $^{12}$CO to $^{13}$CO to C$^{18}$O, suggests that these transitions are increasingly better tracers of denser gas, as expected given their relative abundances.

Comparing the H$_2$CO transitions with the shock tracers, there is an apparent increase in correlation with increasing H$_2$CO transition energy for all shock tracers. This suggests there is a relation between clouds containing dense gas with higher excitation conditions and the prevalence and strength of shocks \citep{Turner1991,Lu2021}. Such clouds might be expected where there are the convergent points of large-scale, supersonic, colliding gas flows or increased star formation activity. It is interesting that while the 218.5\,GHz and 218.8\,GHz transitions of H$_{2}$CO have nearly identical upper state energies, the 218.8\,GHz transition correlates much better with SiO than the other. This apparent trend could be the result of large correlation uncertainties and these correlations are in fact statistically equivalent. If this is not the case, then it is highlighting a potential problem in interpreting the difference between these lines, as the two upper transitions of H$_{2}$CO have the same upper state energy levels and excitation properties and should therefore be correlated to other transitions by the same amount.

Summarising the results of the correlation matrix analysis, we conclude that: (i) $^{12}$CO (and to a lesser extent $^{13}$CO) is a poor tracer of the dense gas; (ii) the C$^{18}$O and lowest energy H$_2$CO transition are the most robust tracers of the dense gas; (iii) the higher energy H$_2$CO transitions and the shock tracers are all consistently pinpointing regions with elevated shocks and/or star formation activity.

\section{Analysis}
\label{sec:analysis}

In this section we use the results of the line fitting and conclusions in Section~\ref{sec:cont_sources} to determine the likely virial state of the continuum sources ($\S$~\ref{sub:core_kinematics}) and its relation to their star forming potential  ($\S$~\ref{sub:core_kin_sf_pot}), then search for signs of proto-stellar outflows ($\S$~\ref{sub:search_outflows}) and intermediate-mass black holes ($\S$~\ref{sub:IMBH}) in the CMZoom line data.

\subsection{Determining the virial state of the compact continuum sources }
\label{sub:core_kinematics}

As described above, H$_{2}$CO (218.2 GHz) was used to determine the kinematic properties for the sources within Paper II's dendrogram catalog due to its prevalence throughout the survey and typically being a bright line with a Gaussian profile and a single velocity component. Using the line fit parameters for this transition, we calculated the virial parameter, $\alpha$, for every source with a H$_{2}$CO (218.2 GHz) detection using the observed velocity dispersion ($\sigma_{obs}$), by considering a compact source's kinetic energy support against its own self gravity through,
\begin{equation*}
    \alpha = \frac{5 \sigma^{2} R}{G M},
\end{equation*}
\citep{McKee2003} where $\sigma$ is the velocity dispersion, $R$ and $M$ are the radius and mass of the dendrogram compact source derived in Paper II, and $G$ is the gravitational constant. The constant `5' comes from the simplistic assumption that these sources are uniform spheres, which may not be the case for all sources in the survey.

\begin{figure*}
\begin{center}
    \includegraphics[width=1\textwidth]{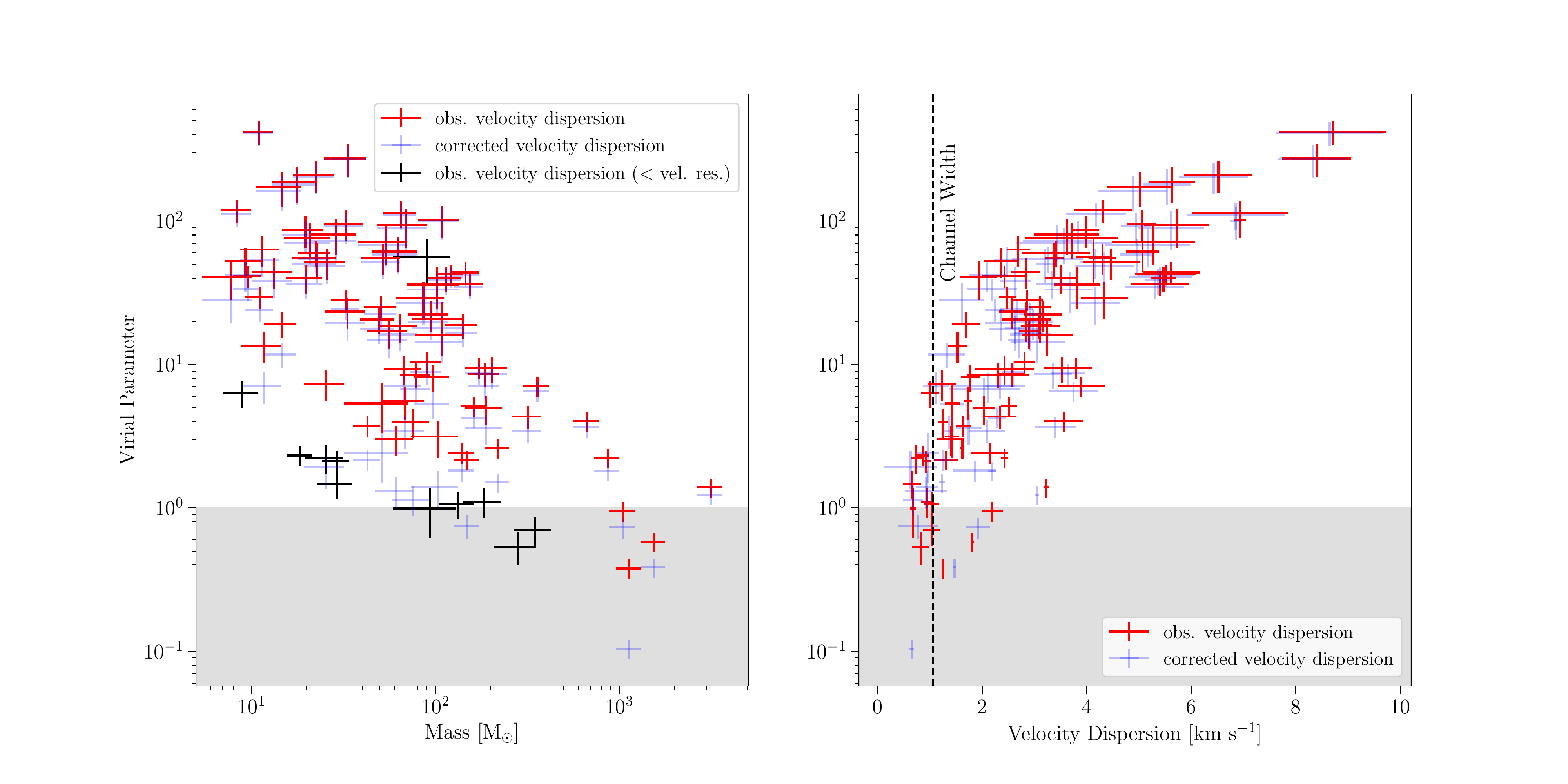}
    \caption{Virial parameters as a function of dendrogram compact source mass [left] 
    and observed velocity dispersion [right]. The red crosses show the observed velocity dispersion, the blue crosses show the velocity dispersion corrected for the instrumental velocity resolution. The black crosses in the left panel show those measurements for which the fit result for the velocity dispersion is lower than the channel width, and thus cannot be corrected for the instrumental velocity resolution. The vertical dashed line in the right panel indicates the channel width of the {\sc ASIC} data. The shaded grey region represents the condition a compact source must meet to be virially bound. These plots show that when only considering the support from gas kinetic energy against self-gravity, most of the sources are not gravitationally bound. The fact that the measured linewidths for most sources are larger than the channel width demonstrates that this result is not affected by the velocity resolution of the observations. }
    \label{fig:simple_virial}
\end{center}
\end{figure*}

Figure~\ref{fig:simple_virial} shows the distribution of virial parameters as a function of compact source mass and compact source velocity dispersion. Using this form of the virial analysis, only six (out of 103) of the more high-mass sources are virially bound based on observed velocity dispersions, and four are virially bound based on the corrected velocity dispersion. $94 - 96 \%$ of sources in the survey are gravitationally unbound when only considering a compact source's kinetic energy support against its own self gravity. Similar results have been observed in the past by various dynamical studies, with \citet[][and references therein]{Singh2021} finding there are a number of systematic errors that can affect virial ratio measurements.

To explore if this is a physical representation of the compact source population within the CMZ or a result of the limited velocity resolution of the survey, we first repeated the analysis in Figure~\ref{fig:simple_virial} after correcting for the instrumental velocity resolution (blue crosses in Fig~\ref{fig:simple_virial}). We calculated the virial parameter using the corrected velocity dispersion ($\sigma_{int}$) by subtracting the channel width ($\sigma_{inst}$) in quadrature from the observed velocity dispersion, $\sigma_{int} = \sqrt{\sigma_{obs}^{2} - \sigma_{inst}^{2}}$. The velocity dispersion of most sources are significantly larger than the channel width, so the virial ratios $>$1 for the majority ($\sim 78 \%$) of sources are not affected by the instrumental velocity resolution. 

\begin{figure}
    \includegraphics[width=0.5\textwidth]{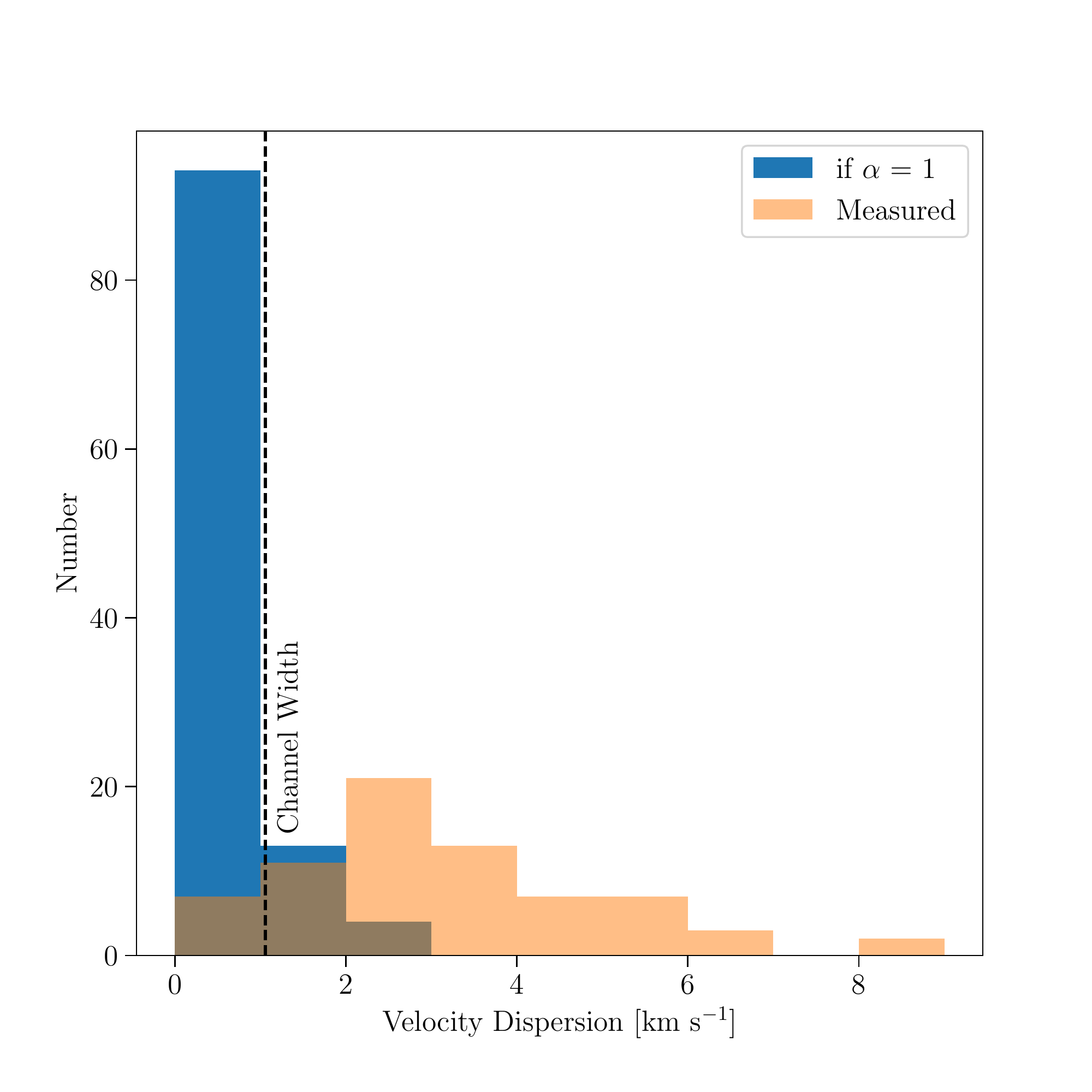}
    \caption{Histogram of measured velocity dispersions (orange) compared to the velocity dispersion required for every compact source to be virially bound (blue). The fact that the observed velocity dispersion is larger than the channel width for most of the sources suggests that the CMZoom velocity resolution is not biasing the virial analysis for most sources. A velocity resolution of $\sim$0.1\,\kms\ would be required to determine if the small number of sources with linewidths comparable to the CMZ velocity resolution are gravitationally bound.}
    \label{fig:disp_hist}
\end{figure}

We then determined what velocity dispersion each compact source would need to have for it to be gravitationally bound, i.e. to have $\alpha = 1$. Figure~\ref{fig:disp_hist} shows a histogram of these `$\alpha = 1$' velocity dispersions compared to the measured velocity dispersions of the sources. This shows that in order to unambiguously determine the virial state of those sources with $\sigma$ close to the channel width of 1.1 km s$^{-1}$ requires re-observing them with an instrumental velocity resolution of $\sim$0.1\,\kms\ to resolve the smallest plausible sound speed of $\sim$0.2 km s$^{-1}$. We highlight these low velocity dispersion sources as particularly interesting to follow-up in the search for potential sites of star formation activity with the CMZ.

Having concluded these high virial ratios are real for the majority of sources, we then seek to understand whether these sources are simply transient overdensities, or longer-lived structures. Previous work on the clouds within the dust ridge by \citet{Walker2018} and \citet{Barnes2019} found that while dust ridge clouds are gravitationally unbound according to virial metrics comparing the gravitational potential and kinetic energies, the intense pressure inferred within the CMZ is sufficient to keep these sources in hydrostatic equilibrium. 

\begin{figure*}
    \centering
    \begin{tabular}{c c}
        \includegraphics[width=0.49\textwidth]{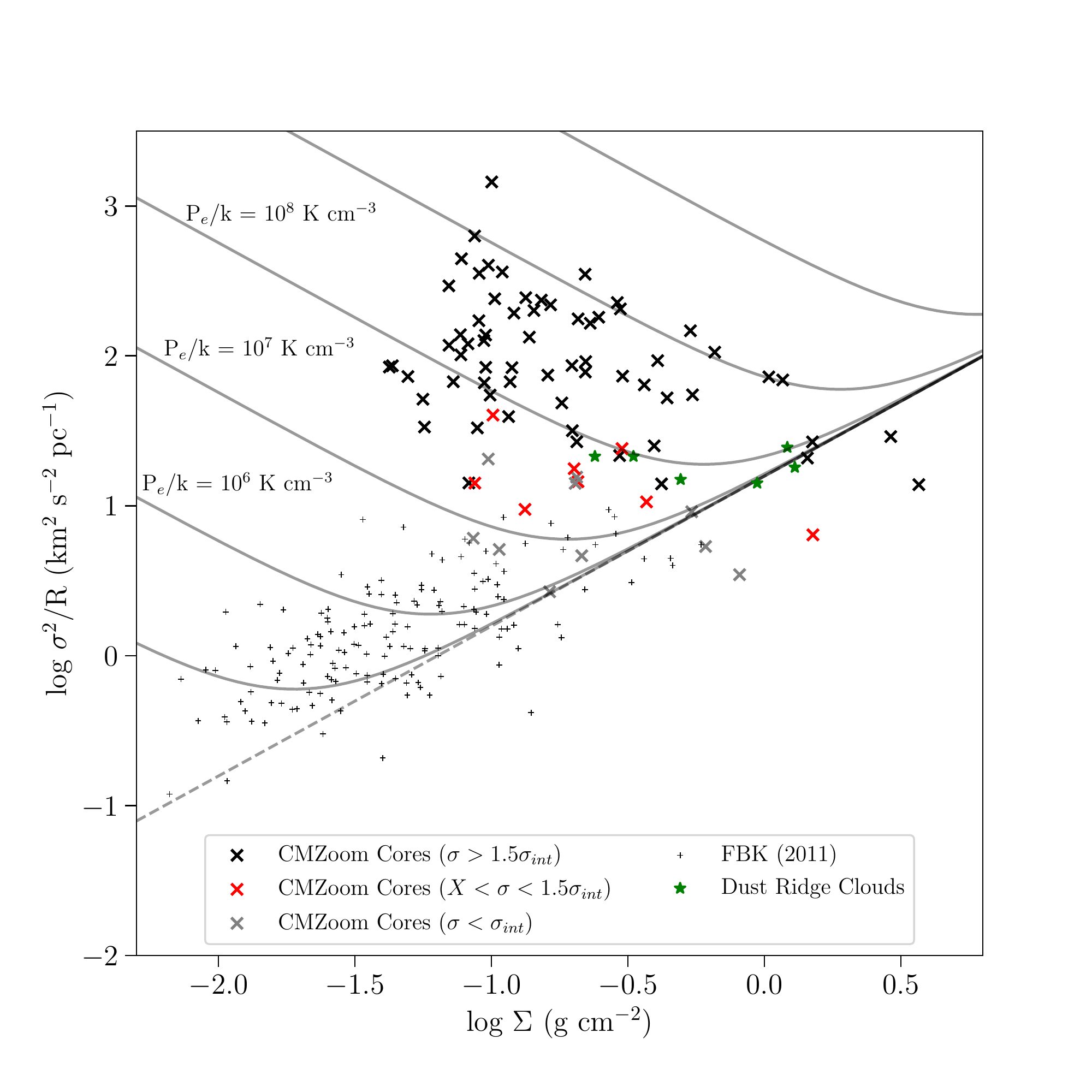} & \includegraphics[width=0.49\textwidth]{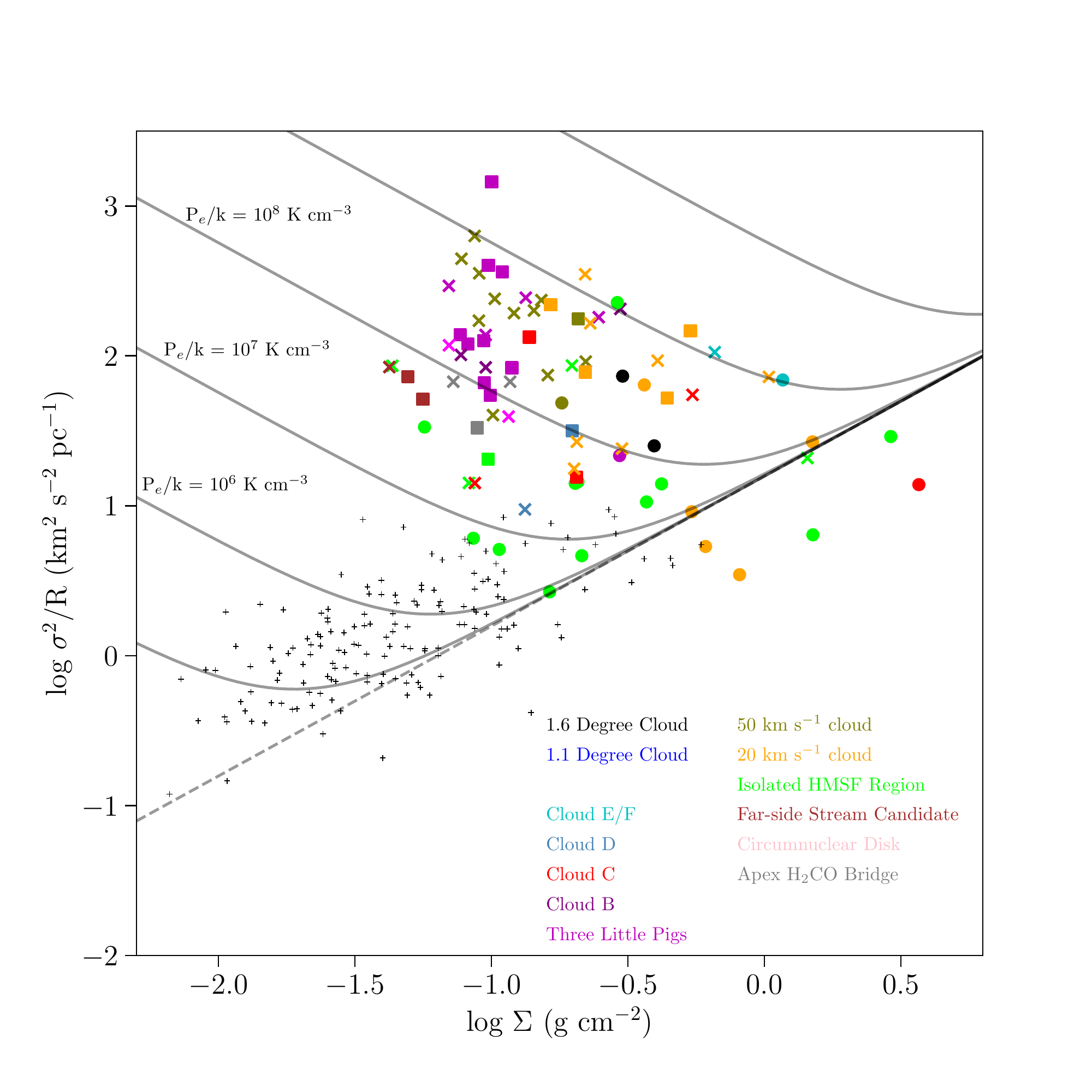} \\
    \end{tabular}
    \caption{Left: Comparison of the CMZoom sources shown by crosses, to Galactic Ring Survey clouds \citep{Field2011} shown by black plus symbols. Grey crosses indicate sources with a velocity dispersion less than the channel width ($\sigma_{int} = 1.2$\,\kms{}) of the survey. Red crosses indicate sources with only slightly resolved velocity dispersions between 1 and 1.5 times the channel width. The black crosses indicate lines with a velocity disperion  more than 1.5 times the channel width, so are well resolved. Overlaid are green star markers corresponding to \citet{Walker2018}'s measurements of dust ridge clouds. The dashed line represents virial equilibrium with P$_{\text{e}} = 0$ and the curved lines represent objects in hydrostatic equilibrium at the stated pressure. While few of the CMZoom sources would be self-gravitating with P$_{\text{e}} = 0$, at pressures of P$_{\text{e}} = 10^{8}$ K km$^{-1}$ the majoriry of these sources would be in hydrostatic equilibrium. Right: Left panel with marker colors indicating different key clouds throughout the CMZ. Circles indicate sources that have associated star formation tracers according to Hatchfield et. al. (in prep) or \citet{Lu2015}, squares indicate sources with potential star formation tracer association according to Hatchfield et. al. (in prep) and crosses indicate sources with no star formation tracer association. All sources, except for one isolated HMSF core, below or close to P$_{e} = 0$ (shown by the dashed line) are found to be star forming, while the fraction of sources that are star forming drops off quickly against increased pressure or distance above this line.}
    \label{fig:virial_pressure}
\end{figure*}

In Figure~\ref{fig:virial_pressure}, we replicate the Figure 4 of \citet{Walker2018} -- which in turn replicated Figure~3 of \citet{Field2011} -- for all sources in the CMZoom survey with a detected \HtwoCO  (218.2\,GHz) transition. The black curved lines show where sources would be in hydrostatic equilibrium if confined by external pressures described by,

\begin{equation}
    V_{0}^{2} = \frac{\sigma^2}{R} = \frac{1}{3}\left(\pi \Gamma G \Sigma + \frac{4 P_{e}}{\Sigma}\right),
\end{equation}

where $V_{0}$ is the linewidth-size scaling relation, $\sigma$ and $R$ are the velocity dispersion and radius of the compact source, $\Gamma$ is a form factor related to the density structure \citep[as described by ][]{Elmegreen1989} and here we adopt $\Gamma = 0.73$ which describes an isothermal sphere at critical mass, $\Sigma$ is the mass surface density, $G$ is the gravitational constant and $P_{e}$ is the external pressure. The black dashed line represents the simple virial condition of P$_{e} = 0$ as shown in Figure~\ref{fig:simple_virial}. 

Given the gas pressure in the CMZ of 10$^{7-9}$\,K\,cm$^{-3}$ calculated by \citep{kruijssen14} based on observations by \citet{Bally1988}, Figure~\ref{fig:virial_pressure} further enforces the conclusion of \citet{Walker2018} that while only a small number of these sources are gravitationally bound according to simply virial analysis, the intense pressures found within the CMZ are capable of keeping a large fraction of these sources in hydrostatic equilibrium, so they may still be long-lived structures.

\subsection{The relation of compact source gas kinematics to a compact source's star forming properties}
\label{sub:core_kin_sf_pot}

We then seek to understand what role, if any, the kinematic state of the gas plays in setting the star formation potential of the sources. The right panel of Figure~\ref{fig:virial_pressure} repeats the left, but with marker colours representing a number of key structures throughout the CMZ. Hatchfield et. al. (in prep) use a number of standard high-mass star formation tracer catalogs including methanol masers \citep{Caswell2010}, water masers \citep{Walsh2014}, 24\micron{} point sources \citep{Gutermuth2015} and 70\micron{} point source \citep{Molinari2016} catalogs to identify which dense sources within Paper II's catalog may be associated with ongoing star formation. They defined three categories: sources definitely associated with these high-mass star formation tracers, sources definitely not associated with these star formation tracers, and an ``ambiguously star-forming'' category for sources where it was difficult to determine whether the observed star formation tracer was associated with that compact source or not.  We combined these star formation tracer activities with targeted observations of the 20\,km\,s$^{-1}$ cloud from \citet{Lu2015}, who detected a number of deeply embedded H$_{2}$O masers towards this cloud. In the right panel of Figure~\ref{fig:virial_pressure}, sources with robust associated star formation tracers are marked with a filled circle. Ambiguously star-forming sources are marked with a square. Sources with a robust non-detection of any star formation tracers are marked with crosses.

We find that all CMZ sources below the P$_{e} = 0$ line, i.e. all sources with $\alpha \le 1$, are associated with a star formation tracer. As sources move upwards and to the left of the P$_{e} = 0$ line the fraction of sources with star formation tracers drops to $\sim40\%$. 

\begin{figure*}
\centering
\begin{tabular}{c}
    \includegraphics[width=1.0\textwidth]{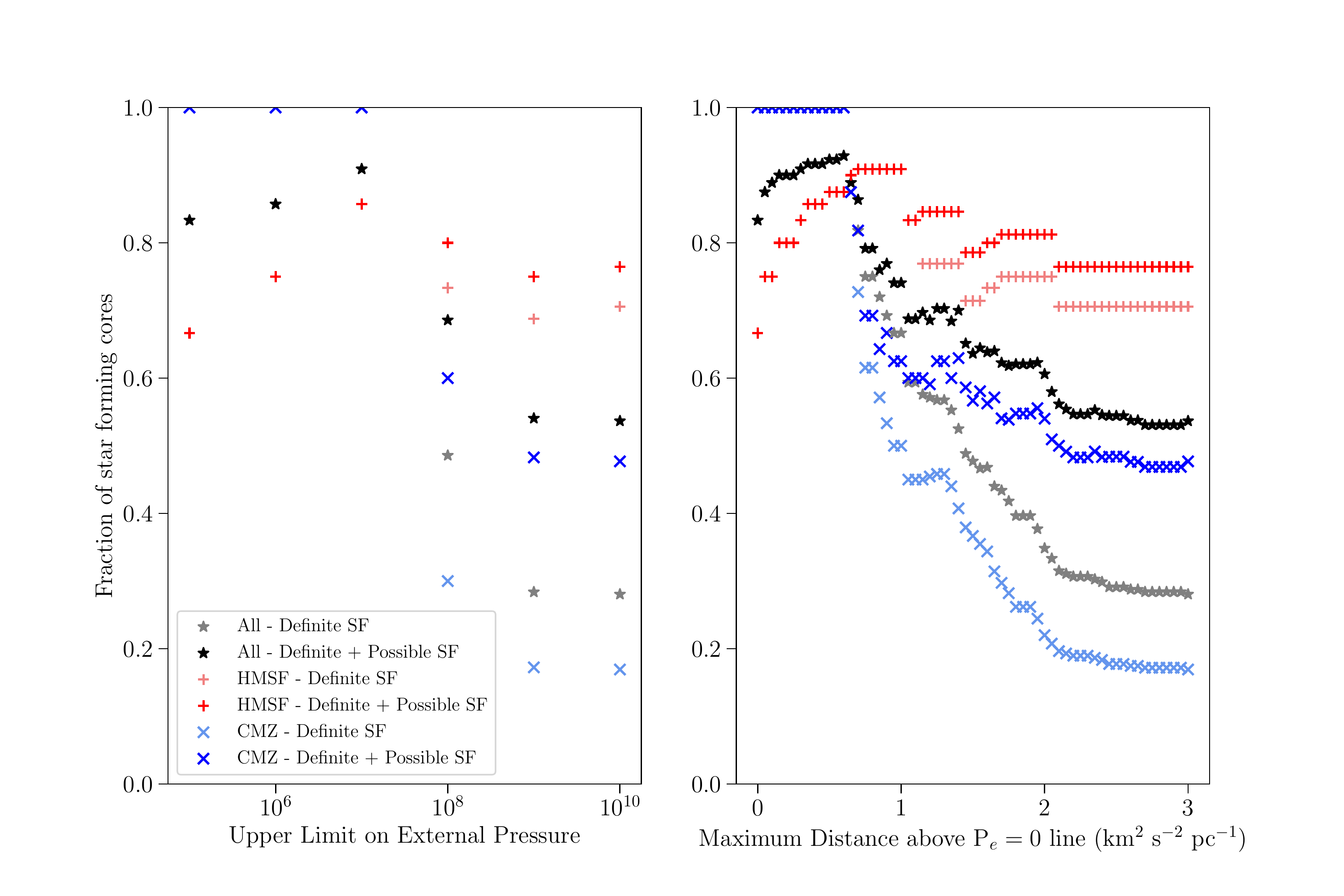}
\end{tabular}
    \caption{Fraction of sources that are star forming as a function of upper limit on the external pressure [left] or maximum distance above the P$_{e} = 0$ line [right]. Grey markers indicate all sources with definitive star formation tracers, while black markers indicate all sources with definitive or possible star formation tracers. In addition to this, light red markers indicate isolated HMSF sources with definitive star formation tracers and dark red markers indicates isolated HMSF sources with definitive or potential star formation tracers. Finally, light blue markers indicate CMZ sources that have definite star formation tracers and dark blue indicates sources with definite or possible star formation tracers.}
    \label{fig:frac_SF}
\end{figure*}

We then try to quantify if there is a combination of physical properties that can be used to determine the likelihood that a given compact source will be star forming or not. Figure~\ref{fig:frac_SF} shows the fraction of sources that are star forming below lines of constant pressure (left) or as a function of distance from the $P_{e} = 0$ line (right). We show the total population of sources in black stars, as well as breaking down the population of sources into CMZ sources (blue crosses) and isolated HMSF sources (red pluses). In addition to this breakdown, we have also split these fractions up into regions that show definite association with star formation tracers, indicated by light coloured markers, as well as sources with either definite or ambiguous star formation tracers in dark coloured markers.

All CMZ sources below a maximum external pressure of $10^{7}$\,K\,cm$^{-3}$ have associated star formation tracer activity while the isolated HMSF sources peak at $10^{7}$\,K\,cm$^{-3}$ before plateauing at $70 - 80\%$ while the CMZ sources drop to $20 - 50\%$. These isolated HMSF regions were selected due to their potential star formation activity, so it is no surprise that this population of sources differ significantly from CMZ sources. A similar trend occurs as a function of star forming sources against maximum distance from $P_{e} = 0$, though the CMZoom sources separate from the isolated HMSF regions at a faster rate than as a function of external pressure. This suggests that while the external pressure factors in to whether or not a compact source will begin to form stars, the proximity of a compact source to being virially bound provides a more accurate indication of its star formation activity. 

\subsection{Searching for proto-stellar outflows}
\label{sub:search_outflows}

The CMZoom spectral set up was specifically selected to target a number of classic outflow tracers; SiO \citep{Schilke1997,Gueth1998,Codella2007,Tafalla2015} and CO \citep{Beuther2003}. The energies involved in protostellar outflows are sufficiently high enough to vaporize SiO dust grain mantles and while CO is more prevalent and excited at lower temperatures, it has been used to observe protostellar outflows towards high-mass star forming clouds in the past \citep[e.g.][]{Beuther2003}.

\begin{figure*}
	\centering
	\includegraphics[width=1.0\textwidth]{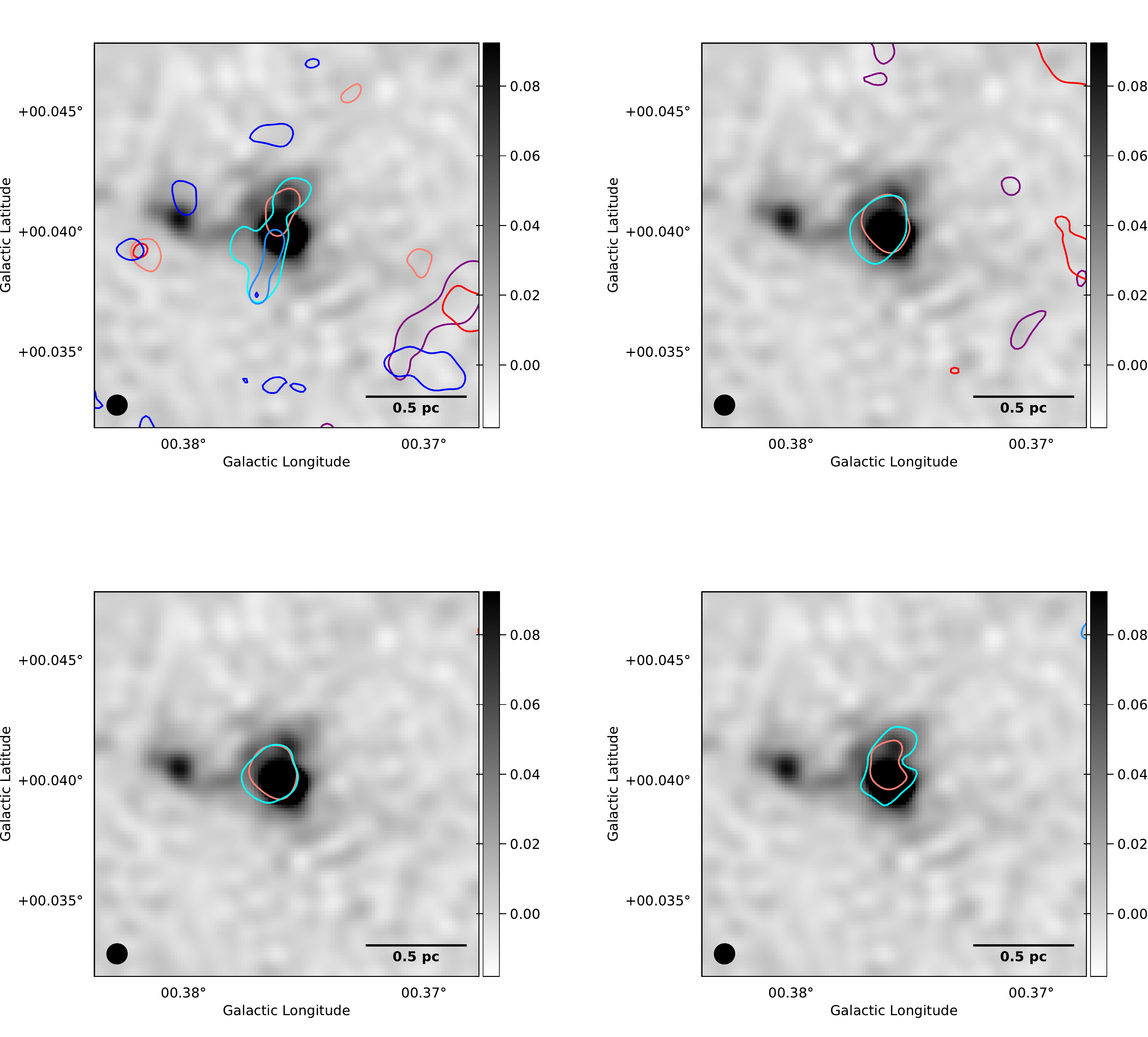}
\caption{The grey scale images show the 230 GHz continuum emission centred on the most massive compact source within G0.380+0.050. The colour bar shows the flux density in Jy. Overlaid are contours of moment maps produced over 10 km s$^{-1}$ intervals from $\pm 30$ km s$^{-1}$ from the compact source's $\Vlsr$, for $^{12}$CO (top-left), $^{13}$CO (top-right), C$^{18}$O (bottom-left) and SiO (bottom-right).}
\label{fig:0380_outflow_maps}
\end{figure*}

\begin{figure*}
	\centering
	\includegraphics[width=1.0\textwidth]{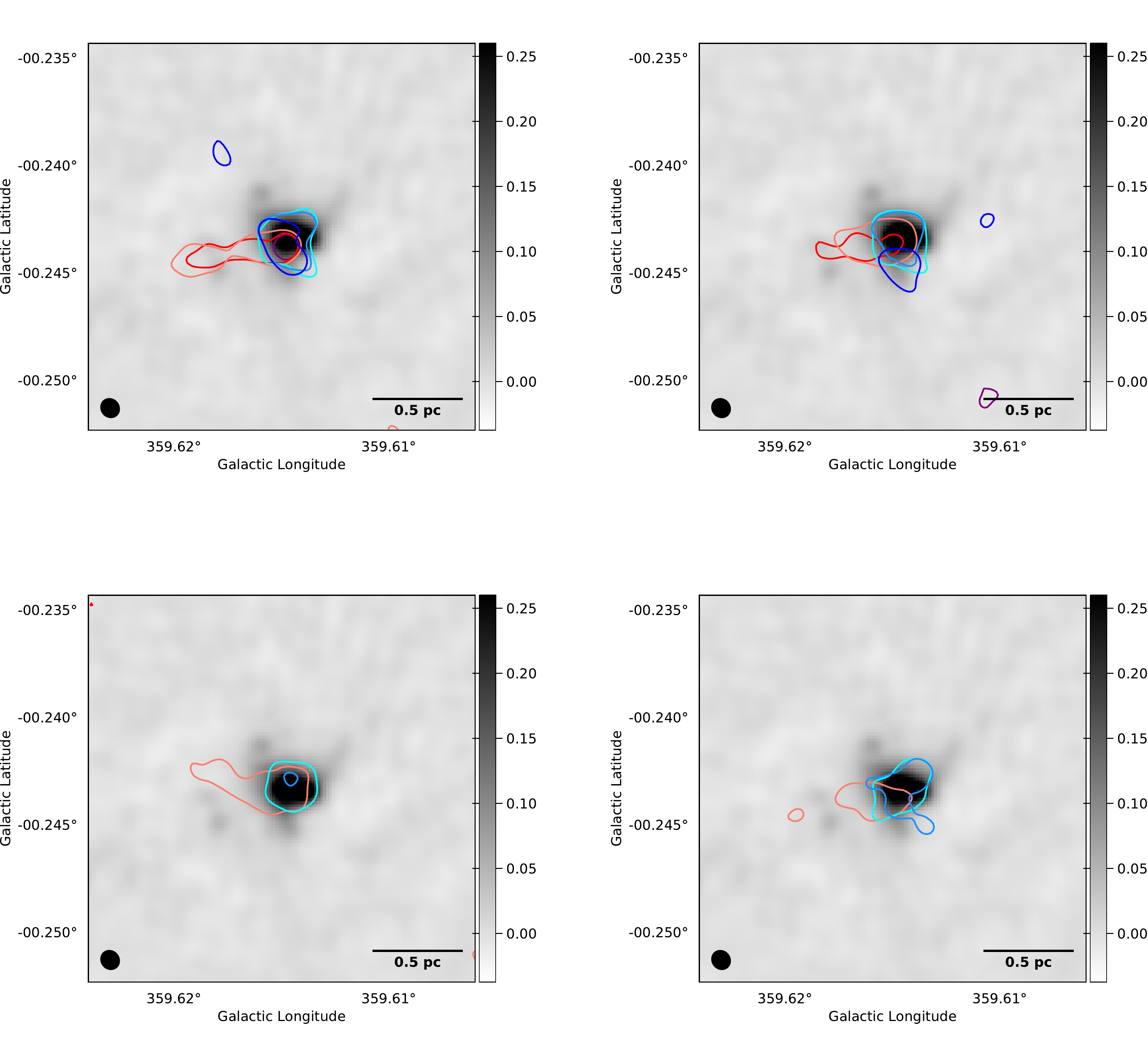}
\caption{The grey scale images show the 230 GHz continuum emission centred on the most massive compact source within G359.615+0.243. The colour bar shows the flux density in Jy. Overlaid are contours of moment maps produced over 10 km s$^{-1}$ intervals from $\pm 30$ km s$^{-1}$ from the compact source's $\Vlsr$, for $^{12}$CO (top-left), $^{13}$CO (top-right), C$^{18}$O (bottom-left) and SiO (bottom-right).}
\label{fig:359615_outflow_maps}
\end{figure*}

As the most detected transition within the quality controlled data set, and with the most reliable line profiles, we first used H$_2$CO (218.2 GHz) to provide a single \Vlsr for each compact source. Combining this with the $l$ and $b$ positions from paper II, we generated \{$l,b,\Vlsr$\} positions for a large majority of the sources within Paper II's robust catalog. These data were then overlaid on non-primary beam corrected\footnote{The increased noise at the edge of the primary beam corrected images obscured the outflow emission.} 3D cubes of SiO and the three CO isotopologues within \emph{glue}\footnote{https://glueviz.org}. Each compact source was then examined by eye to check for extended structure along the velocity axis. During this process, only two convincing outflows were detected in clouds G0.380$+$0.050 and G359.615$-$0.243 as shown in Figures~\ref{fig:0380_outflow_maps} and~\ref{fig:359615_outflow_maps}.

These two clouds were followed up by creating a series of moment maps for SiO and the three CO isotopologues over 10\,km\,s$^{-1}$ intervals across the surrounding $\pm$\,30\,km\,s$^{-1}$ from the compact source $\Vlsr$. Figures~\ref{fig:0380_outflow_maps} and~\ref{fig:359615_outflow_maps} show these moment maps as contours overlaid on the 230\,GHz continuum emission for G0.380$+$0.050 and G359.615$-$0.243. While $^{12}$CO emission shows evidence of red/blue lobes surrounding the compact source at 30\% of peak brightness, there is no sign of similar outflow morphology in any other transition, despite other work having identified an outflow at this compact source in SiO emission. However, \citet{Widmann2016} cautions the use, and in particular the absence, of SiO in interpreting outflows.

The emission in SiO and the three CO isotopologues of 359.615-0.243 all show consistent structures in the form of a significant red lobe to the left of the compact source. The lack of a strong  blue lobe on the opposite side of the compact source may be the result of sensitivity, opacity or different excitation conditions.

We also search for outflow candidates in a more automated way. For every region in the survey, a representative velocity is measured by fitting Gaussian components to a spatially-averaged spectrum of the HNCO emission from the MOPRA CMZ survey \citep{Jones2012}. Using this velocity, we then create blue and red-shifted maps of four different tracers ($^{12}$CO, $^{13}$CO, C$^{18}$O, and SiO) by integrating the emission over 10~\kms \ either side of the V$_{\textrm{lsr}}$ ($\pm$ 1~km~s$^{-1}$). The blue and red-shifted maps were then combined for each region, and inspected to search for any potential outflow candidates.

Overall, 6 candidates were identified using this method. Figures~\ref{fig:G0.316_outflows}~--~\ref{fig:G359.615_outflows} show the integrated emission for each of the 4 molecular line tracers for all 6 candidates, along with $^{12}$CO position-velocity plots taken along the candidate outflows. The PV-plots in particular reveal that only 3 of these are likely to be molecular outflows, namely those in G0.316$-$0.201, G0.380$+$0.050, and G359.615$-$0.243. The latter two of these are the same as those identified via visual inspection in \emph{glue}.

Of the 3 regions with robust outflow detections, only 1 is actually known to be in the CMZ. In Paper I, it was concluded that both G0.316$-$0.201 and G359.615$+$0.243 do not reside in the CMZ based on their kinematics and comparison with results from \citet{Reid2019}. The only molecular outflow(s) that we detect in the CMZ is therefore in G0.380$+$0.050 (aka dust ridge cloud C), which is a known high-mass star-forming region \citep{Ginsburg2015}.

Recently $\sim$ 50 molecular outflows have been detected across 4 molecular clouds in the CMZ with ALMA at 0.1\arcsec -- 0.2\arcsec resolution \citep{Lu2021, Walker2021}. All of these clouds are targeted with \textit{CMZoom}, yet we do not detect any of the outflows detected with ALMA. This is likely due due a combination of angular resolution and sensitivity of the SMA data. Indeed, many of the outflows reported are \textless \ 0.1~pc in projected length, and would not be resolved by our observations. However, some of the larger-scale outflows reported in \citet{Lu2021} are much larger than our resolution, suggesting that they are fainter than our detection limit. Given that the only CMZ-outflow detected with \textit{CMZoom} is in a high-mass star-forming region, this indicates that our observations are capable of detecting large, bright outflows from massive YSOs only.

In conclusion, \emph{CMZoom} provides the first systematic, sub-pc-scale search for high mass proto-stellar outflows within the CMZ. We detect only three outflows throughout the survey -- one in a known high mass star forming region, and two more in isolated high mass star forming regions that are likely not in the CMZ. We can therefore rule out the existence of a wide-spread population of \textit{high-mass} stars in the process of forming that has been missed by previous observations, e.g. due to having low luminosity of weak/no cm-continuum emission.

\subsection{Intermediate Mass Black Holes}
\label{sub:IMBH}

Intermediate mass black holes (IMBHs) are considered to be the missing link between stellar mass black holes and supermassive black holes (SMBHs), with multiple merging events of smaller "seed" IMBHs growing to form SMBHs \citep{Takekawa2021}. Despite this, their existence has yet to be confirmed. A number of IMBH candidates have been identified in the CMZ via the observation of `high-velocity compact clouds', or HVCCs. These are dense gas clouds ($<$ 5\,pc) with high brightness temperatures and large velocity dispersions ($\sigma > 50$\,km\,s$^{-1}$) \citep{Oka1998,Oka2012,Tokuyama2019}, and have been interpreted as the signpost of an intermediate mass black hole (IMBH) passing through a gas cloud and interacting with the gas. As the first sub-pc-scale resolution survey of the dense gas across the whole CMZ, CMZoom is ideally placed to find such HVCCs. 

To determine CMZoom's ability to detect such HVCCs we turn to the papers reporting detections of IMBHs through this method. \citet{Oka2016} reported a compact ($\le 1.6$ pc, using the NRO telescope with a half-power beamwidth of 20\arcsec) candidate IMBH detected in HCN and SiO with an extremely broad velocity width ($\lesssim 100$ km s$^{-1}$), located 0.$^{\circ}$2 southeast of Sgr C. Using the volume density of N(H$_{2}$) $\geq$ 10$^{6.5}$ cm $^{-3}$ given by \citet{Oka2016}, we estimate column densities of three of our dense gas tracers -- $^{13}$CO, C$^{18}$O and H$_{2}$CO, assuming standard abundance ratios ([$^{13}$CO]/[H$_{2}$] $= 2 \times 10^{-6}$, \citet{Pineda2008}, [C$^{18}$O]/[H$_{2}$] $= 1.7 \times 10^{-7}$ \citet{Frerking1982} and [H$_{2}$CO]/[H$_{2}$] $= 10^{-9}$, \citet{Tak2000}). Using these column densities, a kinetic temperature of 60 K and a linewidth of 20 km s$^{-1}$, we use RADEX \citep{vdT_radex_2007} to estimate a brightness temperature of between $16 - 40$K for the interacting gas around this IMBH candidate.

\begin{figure*}
    \centering
    \includegraphics[width=0.95\textwidth]{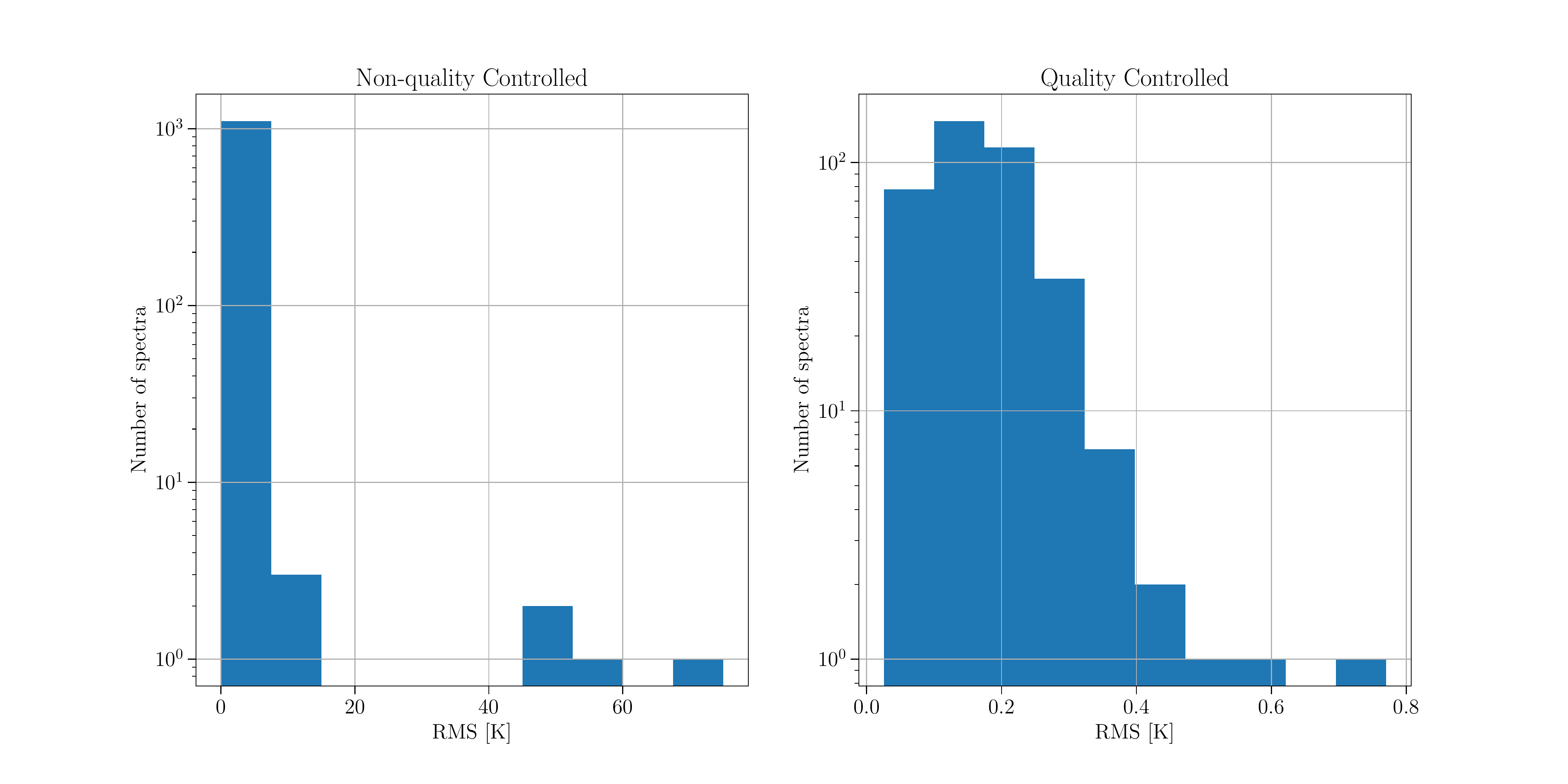}
    \caption{Histogram of the RMS of every spectra throughout the survey measured in Kelvin. The quality controlled data set peaks at $\sim 0.2$ K.}
    \label{fig:RMS_K}
\end{figure*}

Assuming a typical beam size of 3$^{\prime\prime} \times$ 3$^{\prime\prime}$ at a frequency of 230\,GHz we calculate the RMS for each spectra in K, as shown in Figure~\ref{fig:RMS_K}, which peaks at $\sim 0.2$ K. If the HVCC reported in \citet{Oka2016} is representative of IMBH candidates at these transitions in terms of brightness temperature and size we would expect to easily detect $\sim 1$ pc features using the \textit{CMZoom} survey. However, even before quality control, we find no spectral components fit with velocity dispersions $\geq 20$ km s$^{-1}$ throughout the data. The only exceptions are from protostellar outflows.

In summary, we can rule out the presence of HVCC's or IMBH's with properties like those in \citet{Oka2016} within the region covered by this work.

\section{Conclusions}

We present 217--221\,GHz and 229-233\,GHz spectral line data from the SMA's Large Program observing the Galactic Centre, \emph{CMZoom}, and the associated data release. This data extends the work of previous papers published from this survey -- the 230\,GHz dust continuum data release and a dense compact source catalog.

These data were imaged via a pipeline that is an extension to the previously developed imaging pipeline built for the 230\,GHz dust continuum data. During this process, a number of clouds -- in particular Sagittarius B2 and the Circumnuclear Disk -- were found to suffer from severe imaging issues, which prevented these clouds from being analysed. Once imaged, all data were examined by eye to identify both imaging artefacts as well as potentially interesting structures. The quality controlled data were then used to produce moment maps for each cloud, as well as spectra for most dense sources identified by Paper II.

Using \textit{scousepy} \citep{Henshaw2016a, Henshaw2019}, these spectra were fit and then quality controlled to remove spurious fit results before being used to extract kinematic information for a majority of these dense sources and also identify a number of spectral lines beyond the 10 major transitions of dense gas and shocks that were targeted by \textit{CMZoom}.

By measuring the normalized integrated intensity with respect to both C$^{18}$O and 230\,GHz dust continuum, we find that the shock tracers, SiO and SO, as well as the two higher energy \HtwoCO transitions increase by several orders of magnitude towards the Galactic Centre. We also find that the population of isolated HMSF sources that were included in the survey due to their association with star formation tracers, but which likely lie outside the Galactic Centre, have indistinguishable integrated intensity ratios from the CMZ sources. This may present an interesting avenue for follow-up studies using chemical and radiative transfer modelling to disentangle the opacity and excitation effects, and make a quantitative comparison between the physical conditions within the CMZ and the (foreground) Galactic Disk star-forming regions we have identified. Doing so could have important implications for understanding the similarities and differences in the processes controlling star formation between the two (potentially very different) environments.

We identified \HtwoCO (218.2\,GHz) as the best tracer of compact source kinematics, due both to the frequency with which it was detected in sources, but also its tendency to be fit by single Gaussian components. Using this transition, we determine a single \Vlsr and velocity dispersion for every compact source where \HtwoCO was detected and calculated a virial parameter for each compact source. Using a simple virial analysis, only four dense sources were found to be gravitationally bound. 

Expanding this analysis to factor in external pressure and compare this to sources identified as having associated star formation tracers, we find most  sources appear to be consistent with being in hydrostatic equilibrium given the high external pressure in the CMZ. All sources below a maximum external pressure of 10$^{7}$\,K\,cm$^{-3}$ have  associated star formation activity. Above this pressure, the fraction of star forming sources drops. We find that the fraction of star forming sources drops even more steeply the farther it lies from virial equilibrium. We conclude that while the external pressure plays a role in determining whether or not a compact source will begin to form stars, how close a compact source is to being gravitationally bound provides a more accurate indication of its star formation activity.

Through visual inspection of the three CO isotopologues and SiO, only two protostellar outflows (in clouds G0.380$+$0.050 and G359.614$+$0.243) were detected throughout the entire survey. We can therefore rule out a wide-spread population of high-mass stars in the process of forming that has been missed by previous observations, e.g. due to having low luminosity of weak/no cm-continuum emission

Recent observations of the CMZ have highlighted a number of high-velocity compact clouds (HVCCs) which have been interpreted as candidate intermediate mass black holes (IMBHs). Despite having the sensitivity and resolution to detect such HVCCs, we do not find any evidence for IMBHs within the \emph{CMZoom} survey spectral line data. 

\section*{Acknowledgements}
JMDK gratefully acknowledges funding from the Deutsche Forschungsgemeinschaft (DFG) in the form of an Emmy Noether Research Group (grant number KR4801/1-1), as well as from the European Research Council (ERC) under the European Union’s Horizon 2020 research and innovation programme via the ERC Starting Grant MUSTANG (grant agreement number 714907).
LCH was supported by the National Science Foundation of China (11721303, 11991052, 12011540375) and the China Manned Space Project (CMS-CSST-2021-A04).EACM gratefully acknowledges support by the National Science Foundation under grant No. AST-1813765.

\section*{Data Availability}
The data underlying this article will be made available via dataverse, at https://doi.org/10.7910/DVN/SPKG2S.
\bibliographystyle{yahapj}
\bibliography{main}
\appendix
\section{Beam Correction}\label{subsec:beam}
\begin{figure*}
    \centering
    \includegraphics[width=0.97\textwidth]{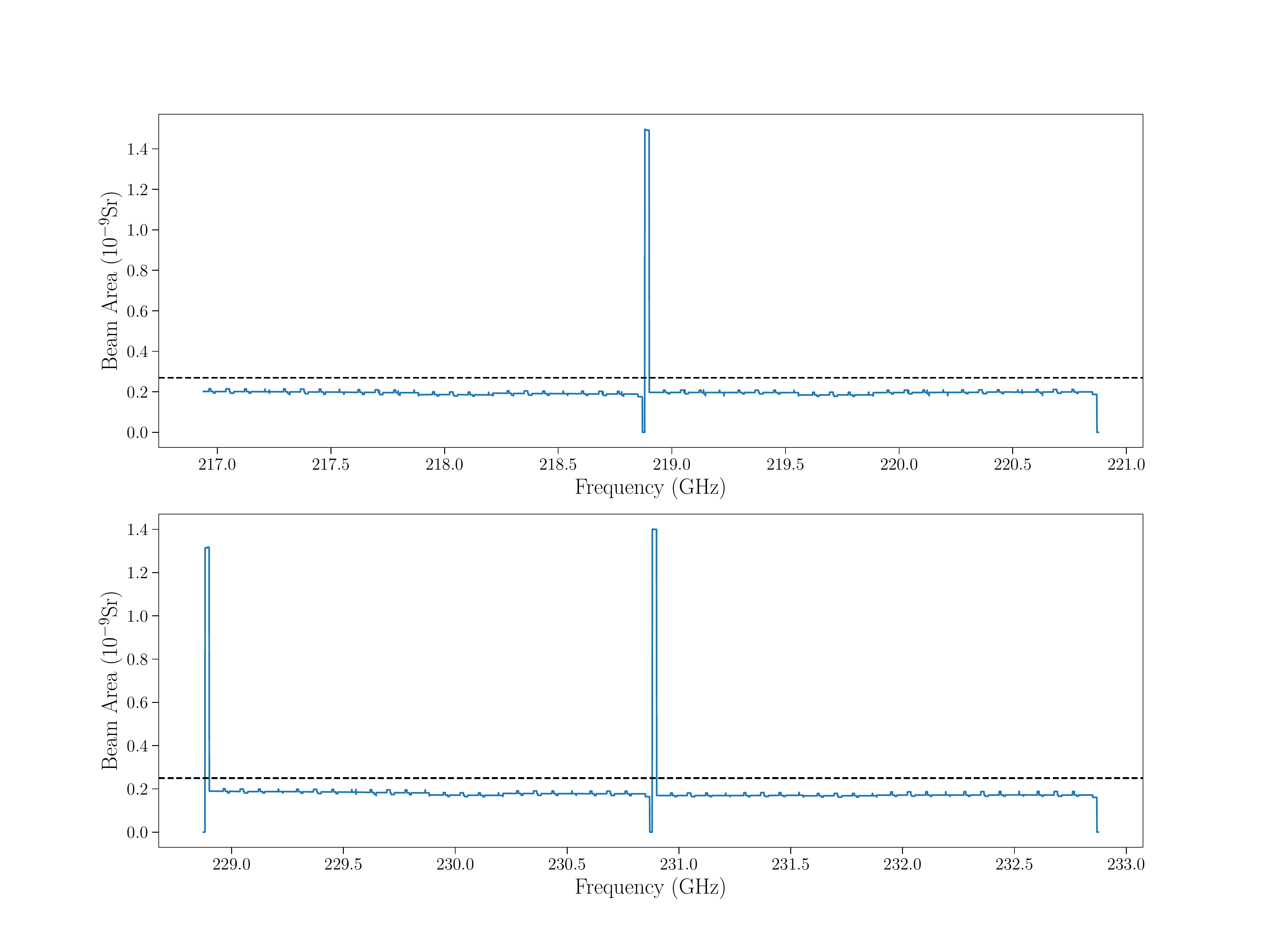}
    \caption{Beam area versus frequency for the lower (top) and upper (bottom) {\sc ASIC} sidebands for the source G0.001-0.058. The sharp peaks at the centre of both panels and the left of the bottom panel show the channels with a problematic beam. The horizontal dashed line indicates the area of the smoothed beam in the final cube.}
    \label{fig:beam_area}
\end{figure*}

A manual inspection of these cubes showed that for a number of channels the beam size increased by factor of a few, typically at the start and end of the frequency coverage, as well as the centre of the datacube, where there is a natural gap in frequency coverage. Figure~\ref{fig:beam_area} shows the variation in beam area as a function of frequency for an example region, G0.001-0.058. This is the result of a natural gap in the SMA's spectral coverage which shifts in absolute frequency depending on when the observation is taken. As these data are the combination of compact and subcompact configurations, if the frequency shift causes a channel to only have compact or subcompact data the beam will be different. This variation in beam size typically resulted in a very different noise profile within these channels in the cube, causing spikes in the spectra that could be mistaken as line emission.

To resolve this issue, we used the python package \emph{spectral cube} to identify these `bad' beams. We found that defining `bad' beams as those that vary from the median beam by 30\% either in semimajor or semiminor axis, or beam area, identified all the problem channels. The channels with beams that are caught by this flag are masked and then the rest of the cube is convolved to a beam corresponding to the smallest beam size that exceeds all unmasked beams using the function \textit{common beam} from python package \textit{radio beam}\footnote{https://radio-beam.readthedocs.io/en/latest/} with a tolerance set to $10^{-5}$.

The cubes are then reprojected into Galactic coordinates using the python package \emph{reproject}. We do this using python instead of {\sc CASA} (version 5.3.0) due to a known bug that introduces a slight offset when reprojecting within the \emph{imregrid} task\footnote{This bug has been fixed as of {\sc CASA} version 5.4.0 (see https://casa.nrao.edu/casadocs/casa-5.4.0/introduction/release-notes-540) for details}.
At this stage, the cubes are split into smaller subcubes targeting key dense gas tracers as well as star formation and shock tracers.

\section{Data Statistics}\label{sec:histograms}
\begin{figure*}
    \centering
    \includegraphics[trim={0 5cm 0 5cm},clip,width=0.9\textwidth]{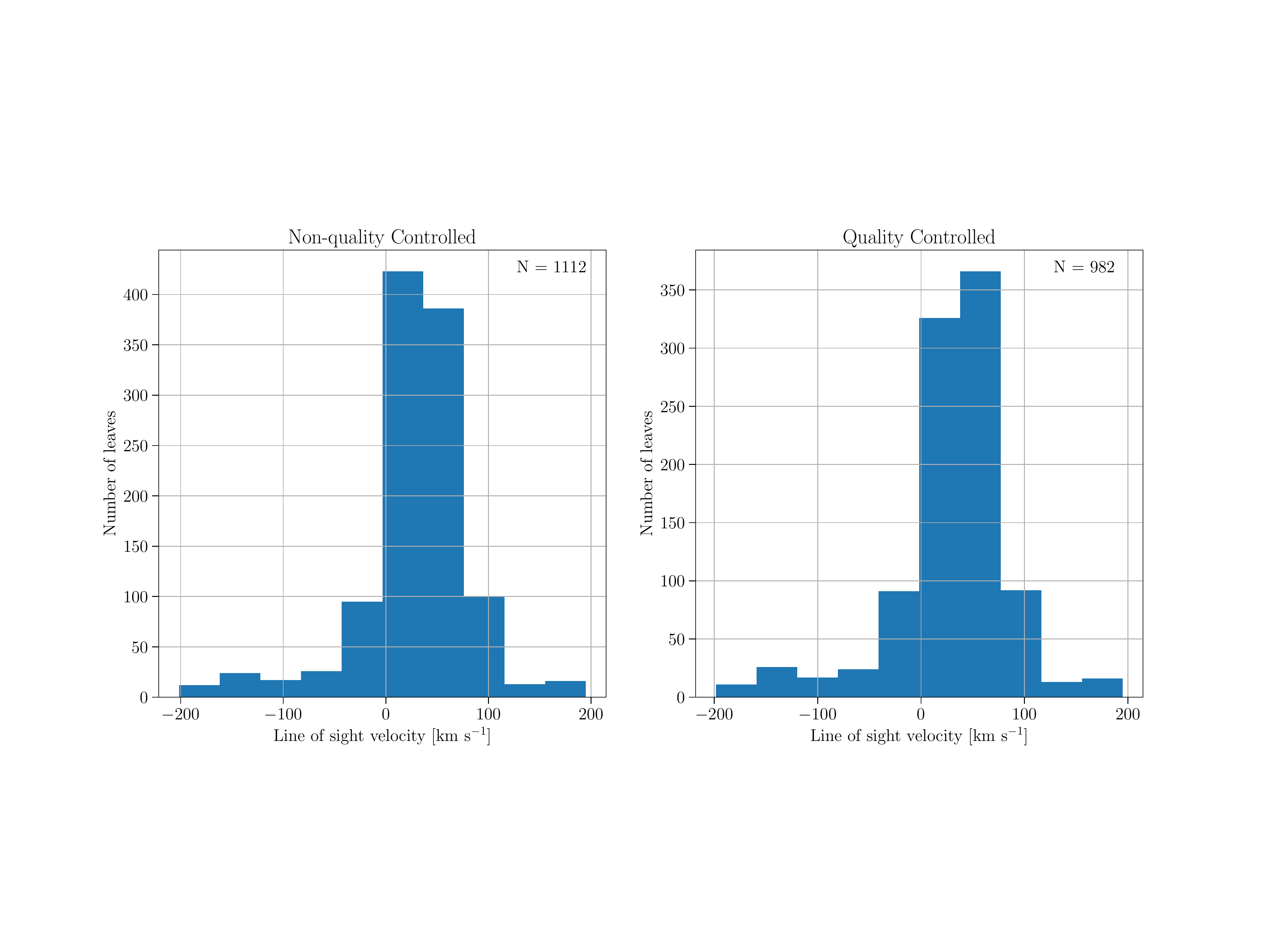}
    \caption{Histogram of all \textit{scousepy} fit \Vlsr measurements throughout the survey for the original data set [left] and the quality controlled data set [right]. A similar format is used for the figures up to Figure~\ref{fig:rms_all}. The majority of the spectral line emission observed by CMZoom lies between $0$ km s$^{-1} <$ \Vlsr $< 100$ km s$^{-1}$}
    \label{fig:vlsr}
\end{figure*}

\begin{figure*}
    \centering
    \includegraphics[trim={0 5cm 0 5cm},clip,width=0.9\textwidth]{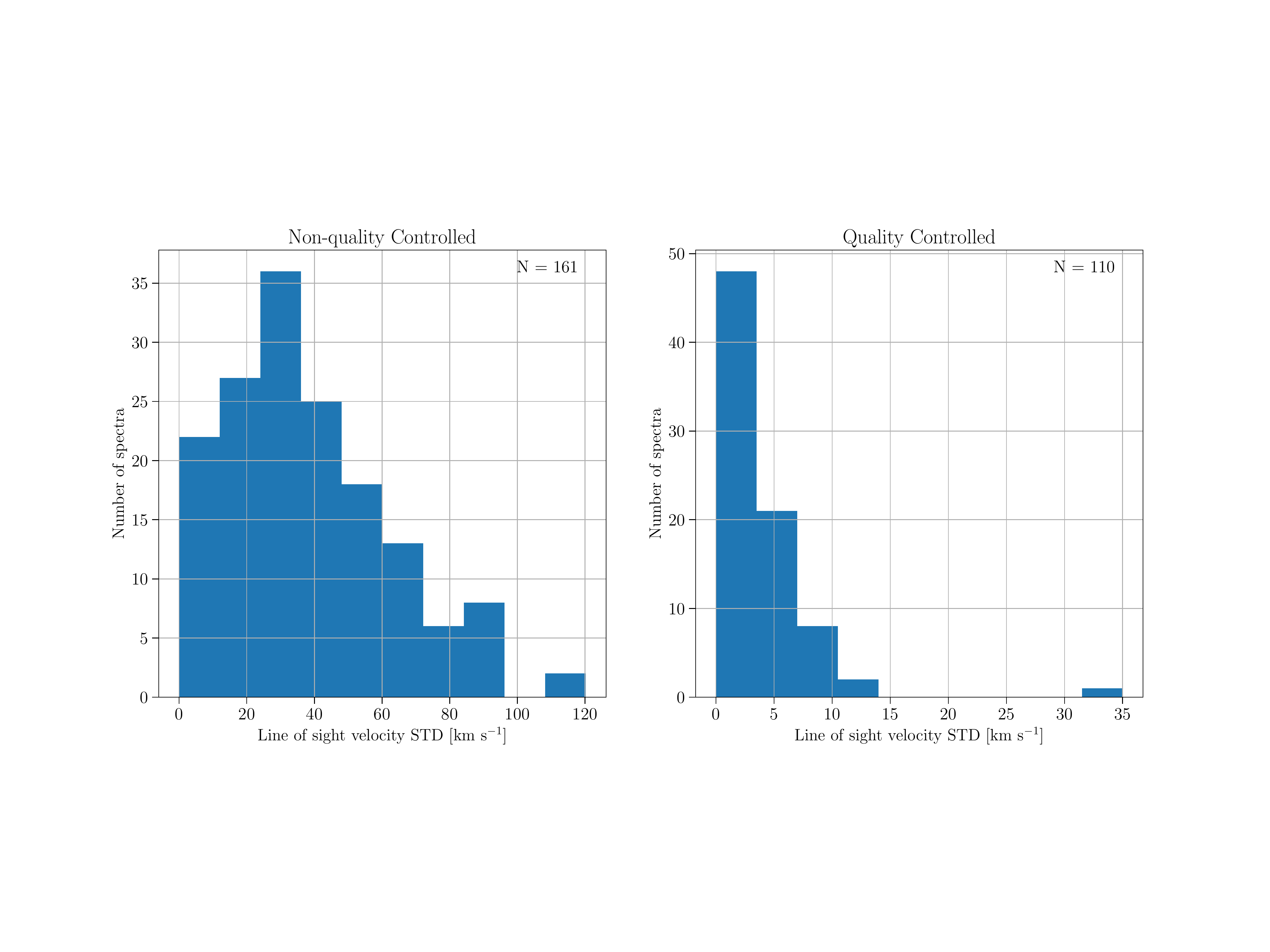}
    \caption{Histogram of the standard deviation in \textit{scousepy} fit \Vlsr measurements for each unique compact source.}
    \label{fig:vlsr_std}
\end{figure*}

Figure~\ref{fig:vlsr} shows the histogram of all \textit{scousepy} fit \Vlsr measurements across the survey, with the majority of the emission observed throughout the region lies between $0$ km s$^{-1}$ and $100$ km s$^{-1}$, as this range in \Vlsr contains most of the dense gas in the CMZ \citep{Henshaw2016}. Figure~\ref{fig:vlsr_std} shows a histogram of the standard deviation of the \Vlsr measurements for each unique compact source. While the non-quality controlled panel (left) shows a typical standard deviation of $\sim 30$\,km s$^{-1}$, this drops to $\le 5$ \,km s$^{-1}$ in the quality controlled data set, with only a single outlier at $\sim 30$ km s$^{-1}$.

\begin{figure*}
    \centering
    \includegraphics[trim={0 5cm 0 5cm},clip,width=0.9\textwidth]{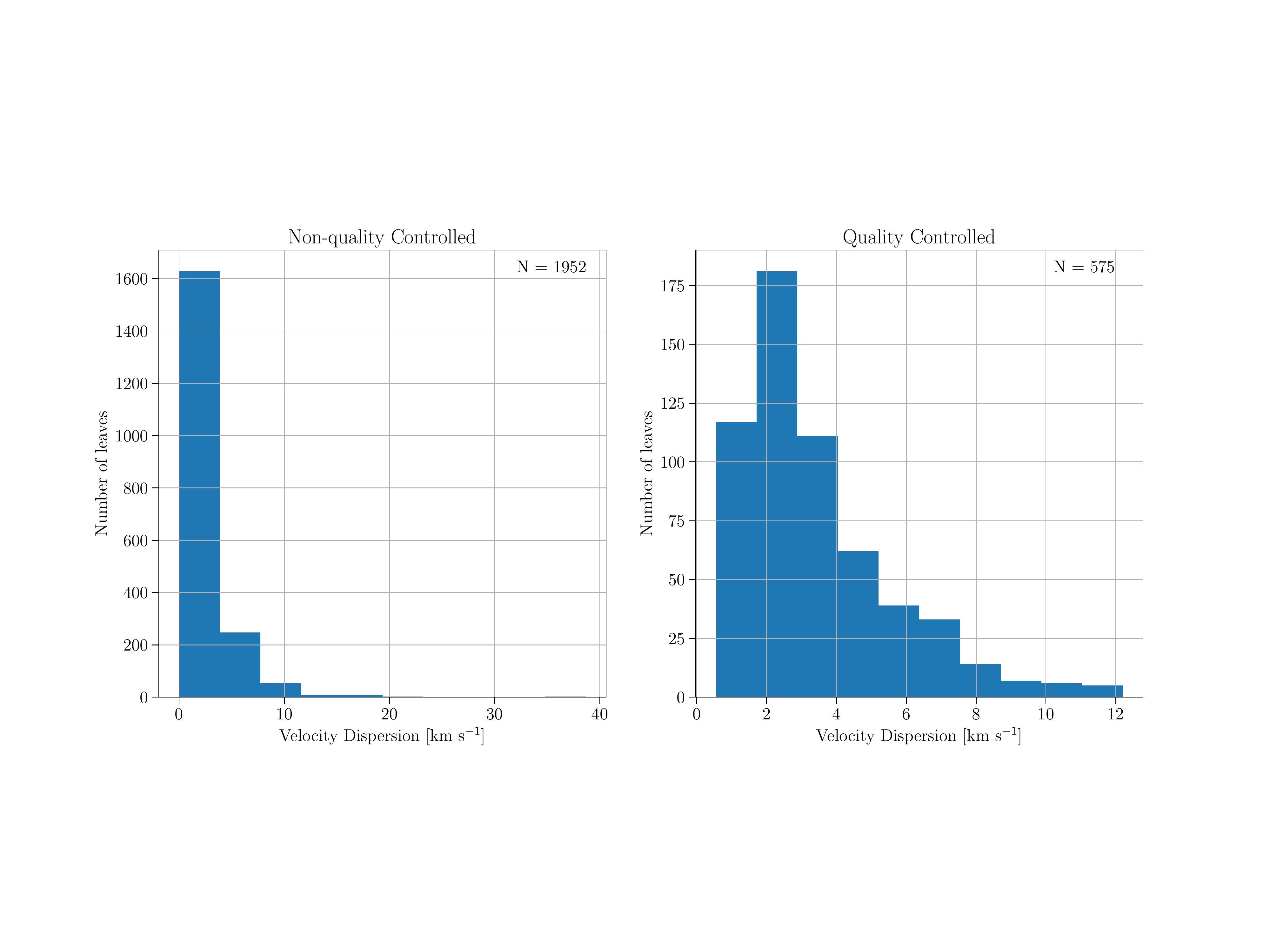}
    \caption{Histogram of \textit{scousepy} fit velocity dispersion measurements for each unique compact source.}
    \label{fig:fwhm}
\end{figure*}

\begin{figure*}
    \centering
    \includegraphics[trim={0 5cm 0 5cm},clip,width=0.9\textwidth]{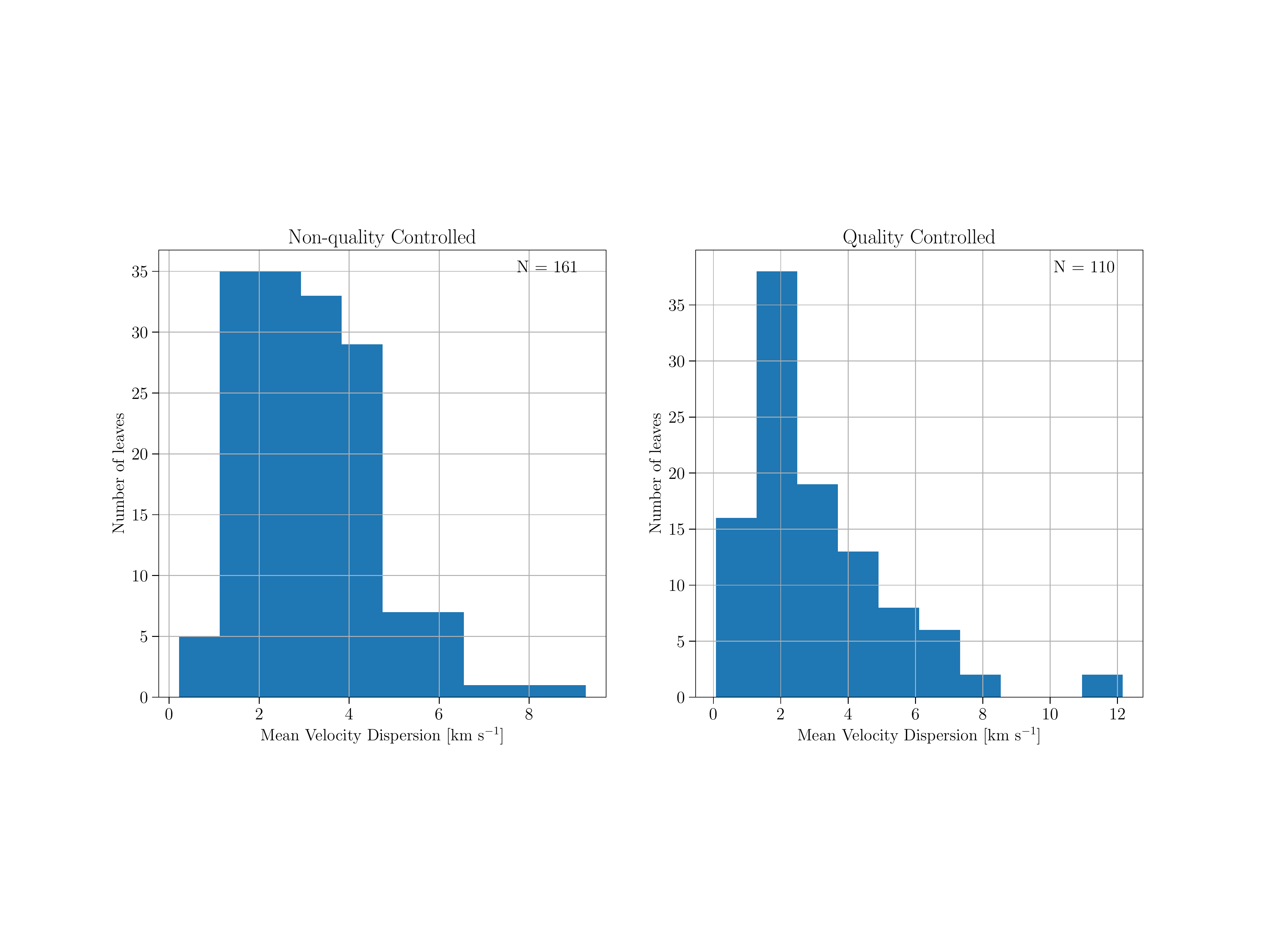}
    \caption{Histogram of the mean \textit{scousepy} fit velocity dispersion measurements for each unique compact source.}
    \label{fig:fwhm_av}
\end{figure*}

While Figure~\ref{fig:vlsr_std} shows the velocity dispersion of centroids across each core, Figure~\ref{fig:fwhm} displays the line-of-sight velocity dispersion measured directly using \textit{scousepy}. Figure~\ref{fig:fwhm_av} shows the average of these velocity dispersion measurements for each unique compact source. Figure~\ref{fig:fwhm} shows that quality control does not have a drastic impact on the typical velocity dispersion of a fit spectral peak. However, it removes several broad components. The points at $\sim 12$\,\kms{} in the right hand panel of Figure~\ref{fig:fwhm_av} belong to G0.001$-$0.058r and G0.489$+$0.010j. These are clouds with complicated velocity structure, containing multiple peaks with small velocity dispersions superimposed on a broader component. The narrow peaks were removed by the quality control conditions, leaving behind single broad peaks.

\citet{Kauffmann2013} observed a number of low density cores with linewidths $\lesssim 1$ km s$^{-1}$ on scales of 0.1 pc within G0.253+0.016. These features primarily manifested as a narrow feature superimposed on top of a broad feature, similar to what we observe in G0.001$-$0.058r and G0.489$+$0.010j. \citet{Kauffmann2017} explore this further using SMA and APEX observations of the region between Sgr C and Sgr B2. \cite{Kauffmann2017} observed narrow features ranging from 0.6 km s$^{-1}$ (in the brick) to 2.2 km s$^{-1}$ (in 20 km/s cloud). We detect similarly narrow features within these clouds when using \textit{scousepy}, ranging from 0.55 km s$^{-1}$ to 1.56 km s$^{-1}$, though we do not observe Sgr B1 off and Sgr D in the \textit{CMZoom} survey.

\begin{figure*}
    \centering
    \includegraphics[trim={0 0 0 0},clip,width=0.9\textwidth]{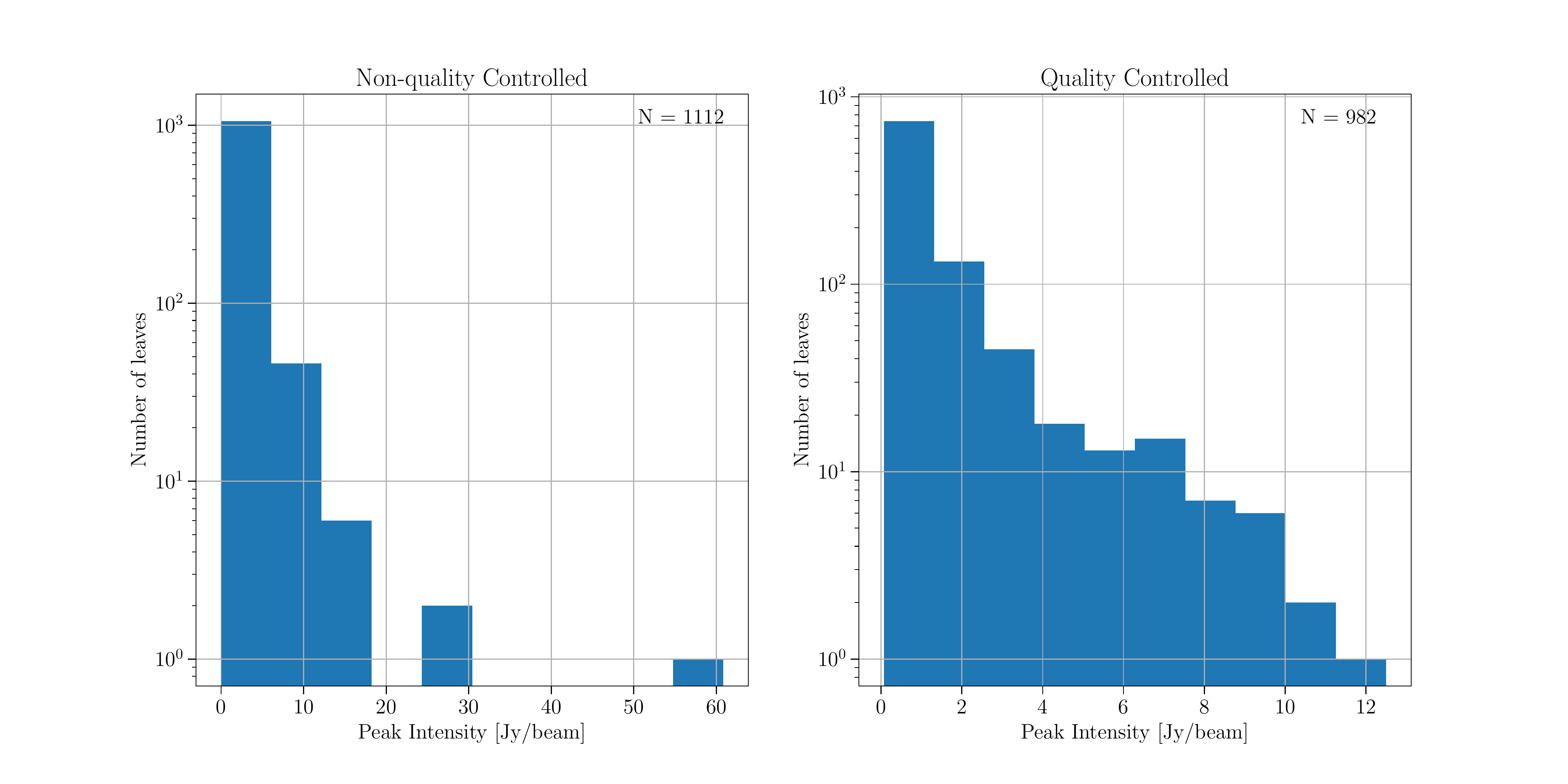}
    \caption{Histogram of all \textit{scousepy} fit peak intensities throughout the survey. Outliers in the left-hand panel are the result of poorly fit \twelveCO.}
    \label{fig:amp}
\end{figure*}

Figure~\ref{fig:amp} shows the histogram of all \textit{scousepy} fit peak intensity measurements across the survey. This shows a number of very bright peaks that are removed by the quality control conditions as they belong to $^{12}$CO, a transition that suffer from severe imaging issues. The majority of spectral peaks in both data sets have low peak intensities and are not affected by quality control.

\begin{figure*}
    \centering
    \includegraphics[width=0.9\textwidth]{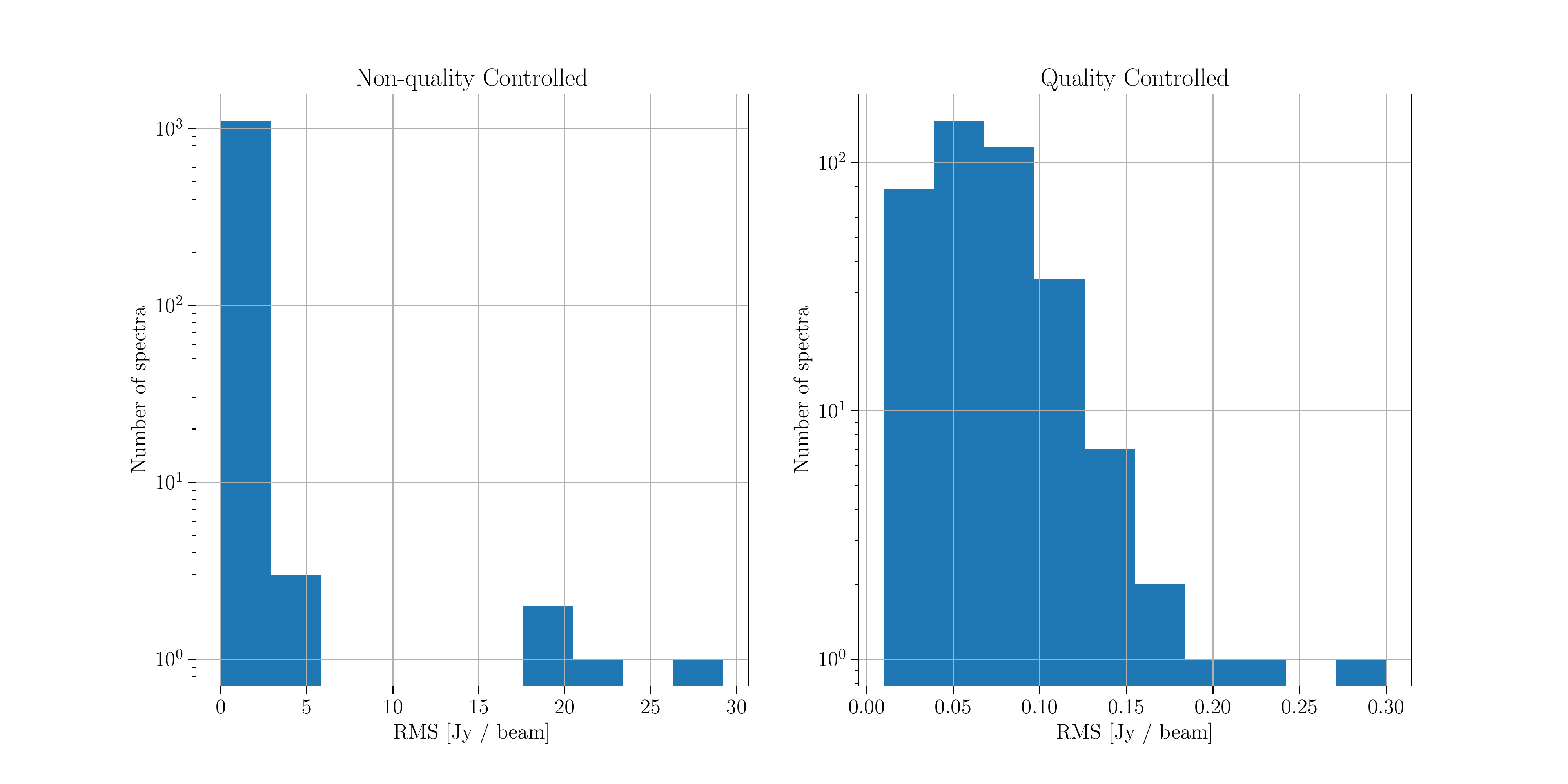}
    \caption{Histogram of the RMS for each unique compact source.}
    \label{fig:rms_all}
\end{figure*}

Figure~\ref{fig:rms_all} shows the histogram of the RMS of all spectra across the survey. While a majority of spectra in the survey have low RMS values in the left hand panel of Figure~\ref{fig:rms_all}, there are a number of very noisy spectra that were removed due to the quality control condition.

\section{Region Summary}\label{sec:region}
In this section we provide information on the transitions detected in each region and their main velocity components. Several regions have been excluded from the analysis contained in this paper due to various issues that arose during imaging and are indicated here. Complex regions like Sagittarius B2 and the circumnuclear disk would required significant larger computing power and time than was available and have also been excluded from this paper.

\subsection{G0.001$-$0.058}
\label{sub:G0.001$-$0.058}
C$^{18}$O emission is confined to two spectral components found at $-10$\,\kms\ and 30\,\kms. Two of the three transitions of H$_{2}$CO show significant emission between $30-40$\,\kms, coinciding well spatially with the continuum emission. The spectral cube centred on the middle transition of H$_{2}$CO also has a second spectral feature at 80 \kms corresponding to CH$_{3}$OH-e at 218.44006300 GHz. Weak OCS emission is detected, though only in the higher frequency transition (231.1 GHz). The OCS spatial distribution corresponds well with the continuum structure and both SiO and SO emission.

\subsection{G0.014+0.021}
\label{sub:G0.014$+$0.021}
H$_{2}$CO 3(2,1)-3(2,0) and both transitions of OCS and the velocity range from 10 to $-200$\,\kms\ of the $^{12}$CO transition was masked during the beam correction process (see $\S$~\ref{subsec:beam}). In the unmasked channels of the $^{12}$CO data cube, significant emission is detected, but there are severe image artefacts including strong negative bowls due to missing extended structure, making this cube entirely unreliable. Two spectral components were observed in the $^{13}$CO cube, with a single narrow peak at $-15$\,\kms\ and a broader component from $0-30$\,\kms{}. No emission was observed in any other line.

\subsection{G0.054+0.027}
\label{sub:G0.054$+$0.027}
This region will be included in a future publication.

\subsection{G0.068$-$0.075}
\label{sub:G0.068$-$0.075}
C$^{18}$O traces the continuum emission well, with two spectral components at 45 \kms{} and 70 \kms{}. Two of the three H$_{2}$CO transitions show strong emission around the continuum structures, with multiple peaks at 45 \kms{} and 55 \kms{}. CH$_{3}$OH-e is also detected at the same velocity. No emission was observed in any other lines.

\subsection{G0.070$-$0.035}
\label{sub:G0.070$-$0.035}
This region suffered from image artifacts and will be included in a future publication.

\subsection{G0.106$-$0.082}
\label{sub:G0.106$-$0.082}
\begin{figure*}
    \centering
    \includegraphics[angle=-90,width=0.85\textwidth]{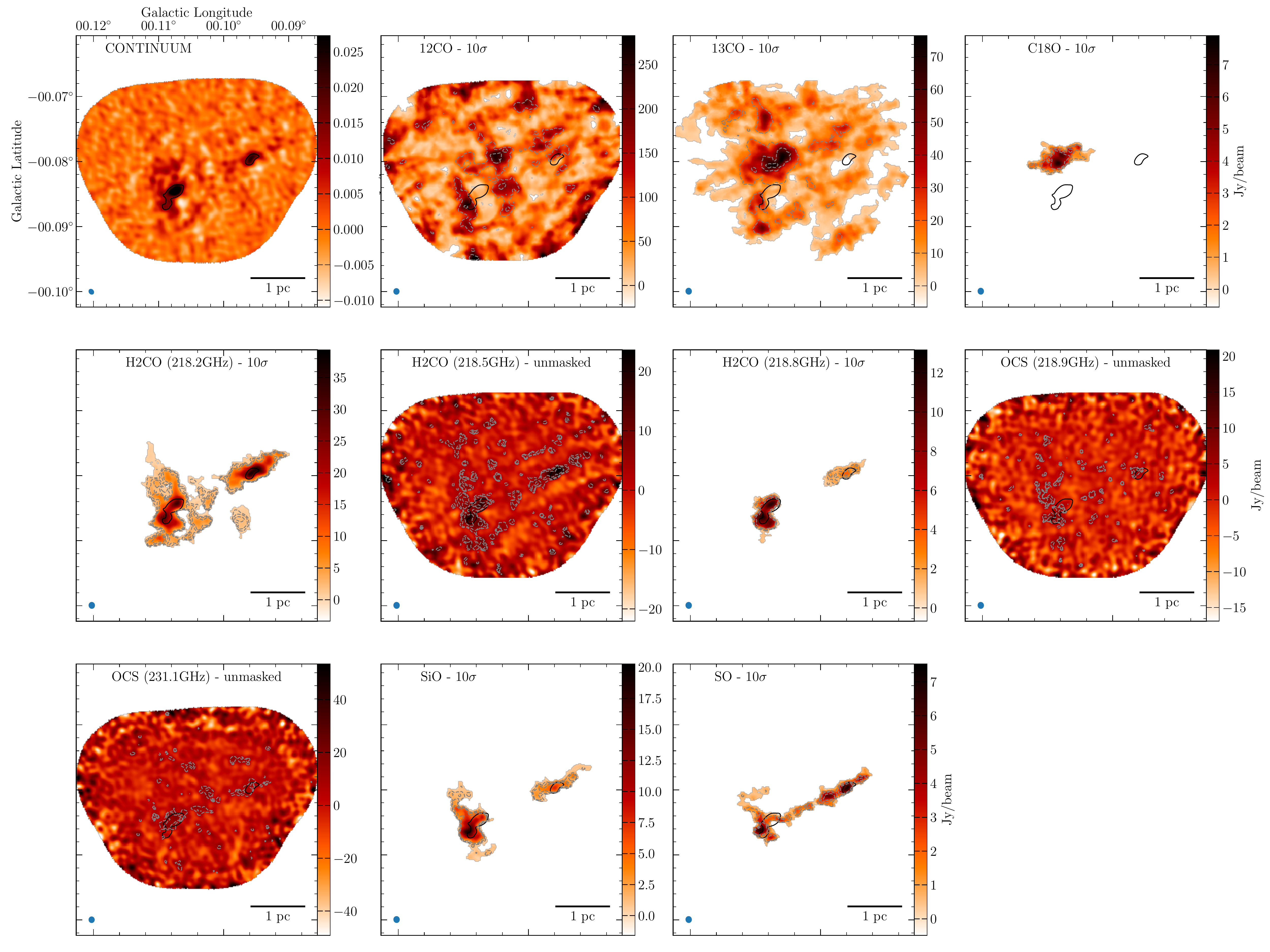}
    \caption{G0.106-0.082 integrated intensity moment maps}
    \label{fig:0106int}
\end{figure*}

\begin{figure*}
    \centering
    \includegraphics[angle=-90,width=0.83\textwidth]{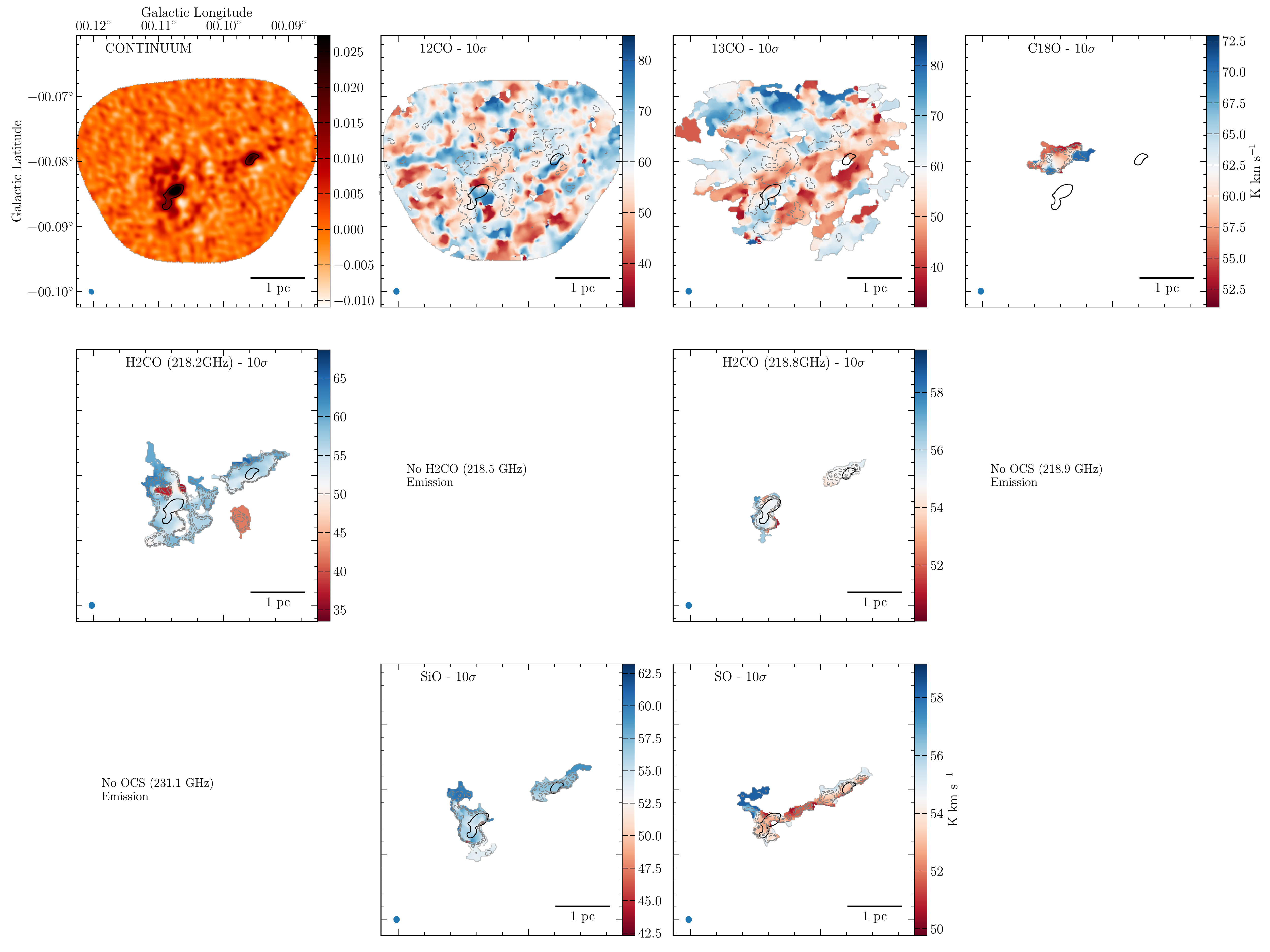}
    \caption{G0.106-0.082 \Vlsr moment maps}
    \label{fig:0106vel}
\end{figure*}

\begin{figure*}
    \centering
    \includegraphics[angle=-90,width=0.83\textwidth]{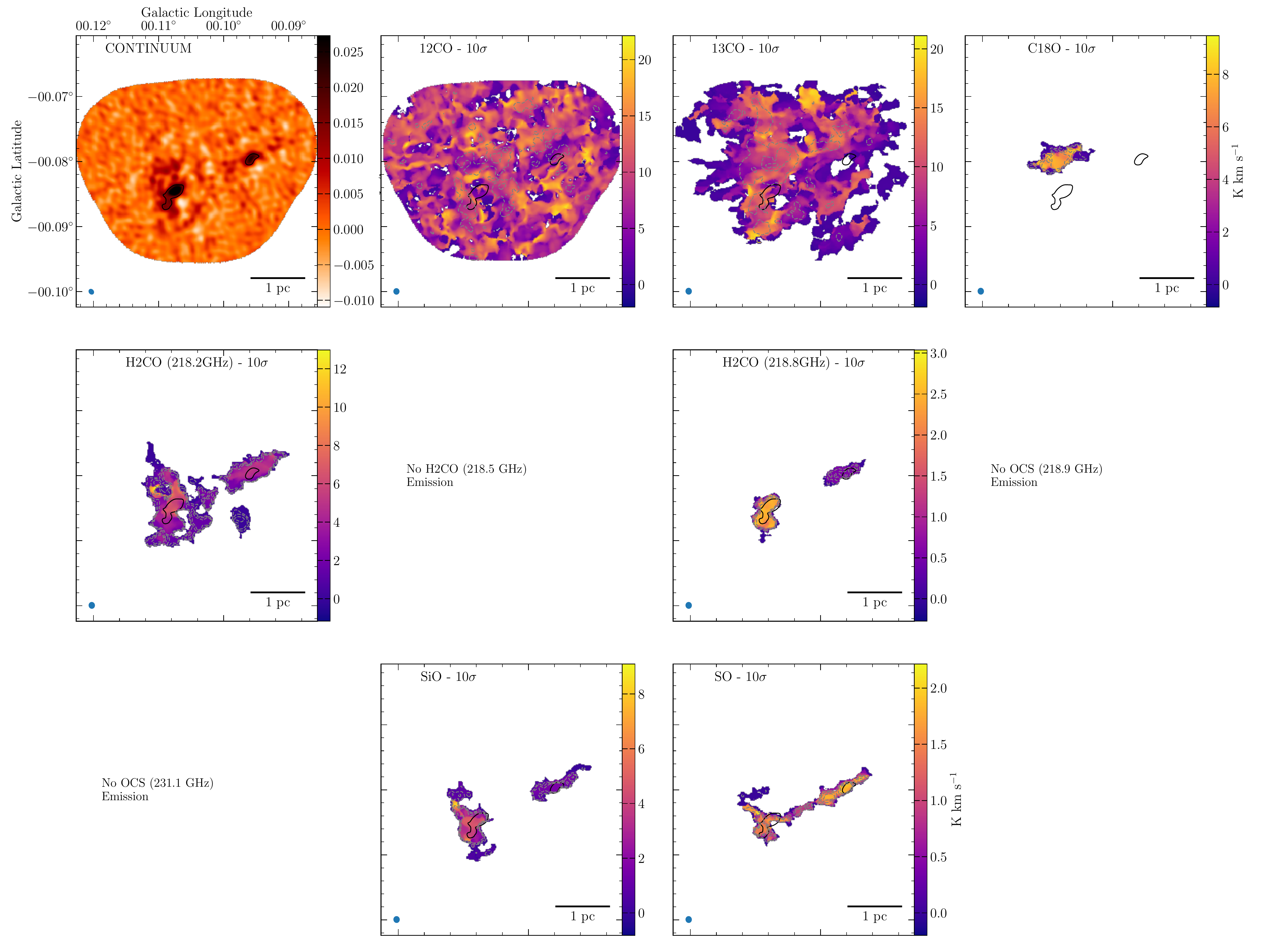}
    \caption{G0.106-0.082 velocity dispersion moment maps}
    \label{fig:0106fwhm}
\end{figure*}

C$^{18}$O emission is spatially compact, with two spectral components found at 55\,\kms and 70\,\kms. There is a spatial offset between the locations of these two components and an overall offset in the C$^{18}$O emission with respect to the continuum emission. Two of the three H$_{2}$CO transitions also show several spectral components. SiO and SO both trace the same spatial structures, and the line profile of both transitions show the same double peaked distribution. Figures~\ref{fig:0106int} to~\ref{fig:0106fwhm} in Section~\ref{sec:appendix} show the integrated moment maps, moment 1 maps and moment 2 maps and the \textit{scousepy} fit spectra for each compact source within this region.

\subsection{G0.145$-$0.086}
\label{sub:G0.145$-$0.086}
C$^{18}$O emission peaks at the lower continuum peak at $- 15$\,\kms{}. No emission is detected in any other line.

\subsection{G0.212$-$0.001}
\label{sub:G0.212$-$0.001}
C$^{18}$O and H$_{2}$CO $3_{03}-2_{02}$ both peak at 45 \kms{} coinciding very well with the continuum emission. No emission is seen in any other line.

\subsection{G0.253+0.0216}
\label{sub:G0.253$+$0.016}
This region will be included in a future publication.

\subsection{G0.316$-$0.201}
\label{sub:G0.316$-$0.201}
C$^{18}$O emission peaks at 18 \kms{} at the continuum peak. However, significant negative bowls are present within this data cube. Each of the three H$_{2}$CO transitions have emission at this same $\Vlsr$, though the intensity of emission at 218.8 GHz is too low to appear in the moment map. SO emission is also seen at 18 \kms{} at the location of the continuum peak, but no emission is seen in any other line.

\subsection{G0.326$-$0.085}
\label{sub:G0.326$-$0.085}
The current emission in the C$^{18}$O moment map is the result of a single channel peak which is likely masking the real emission seen at 15 \kms{} and causing anomalous moment 1 and 2 maps. No emission is seen in any other line.

\subsection{G0.340$-$0.055}
\label{sub:G0.340$-$0.055}
No emission was detected in any lines other than $^{12}$CO and $^{13}$CO.

\subsection{G0.380$+$0.050}
\label{sub:G0.380$+$0.050}
Including $^{13}$CO, all lines other than both OCS transitions show strong emission at 40 \kms{}. Other lines are present: in the C$^{18}$O datacube at 110\,\kms{}, in the 218.2 GHz \HtwoCO{} datacube at $-100$\,\kms{}, in the \HtwoCO{} 218.5 GHz cube at 85\,\kms{}, and in the \HtwoCO{} 218.8 GHz cube at $-160$\,\kms{}. Similarly, a second line is observed in the SiO cube at $-150$\,\kms{} and two additional lines associated with the SO cube at 95\,\kms{} and $-140$\,\kms{}. 

\subsection{G0.393$-$0.034}
\label{sub:G0.393$-$0.034}
Two spectral components are observed at 75 \kms and 92 \kms in both C$^{18}$O and the lower energy transition of \HtwoCO{}. No emission was detected in any other line.

\subsection{G0.412$+$0.052}
\label{sub:G0.412$+$0.052}
C$^{18}$O shows a single peak at 37 \kms{}, though this emission lies far from any continuum structures. Emission from the lowest energy transition of \HtwoCO{} appears associated with the central continuum peak at a \Vlsr\ of 27 \kms{}. No emission was detected in any other line.

\subsection{G0.489$+$0.010}
\label{sub:G0.489$+$0.010}
C$^{18}$O and the lower transition of \HtwoCO{} shows emission at a \Vlsr\ of 32 \kms{}, though this does not coincide well with the continuum emission. The lower continuum peak also shows SO emission at a \Vlsr\ of 29 \kms{}. No emission was seen in any other line.

\subsection{G0.619$+$0.012 and G0.699$-$0.028}
\label{sub:G0.619$+$0.012}

Due to the proximity of these clouds to Sgr B2, the pipeline was unable to suitably clean this data without the appropriate single dish data to include the zero-spacing information. For this reason, these clouds have been removed from all preceeding work.

\subsection{G0.714-0.100}
\label{sub:G0.054$+$0.075}
This region suffered from image artifacts and will be included in a future publication

\subsection{G0.891$-$0.048 and G1.038$-$0.074}

These two clouds, both associated with the 1.1$^\circ$ cloud, suffered from significant imaging problems and have not been included in this work.

\subsection{G1.085$-$0.027}
\label{sub:G1.085$-$0.027}
Significant emission is detected throughout the $^{13}$CO cube, the bulk of which occurs at 28 \kms{}. Emission in C$^{18}$O, the upper transition of \HtwoCO{}, and the upper transition of OCS is detected in a single channel and is therefore unreliable. No emission is seen in any other line.

\subsection{G1.602$+$0.018}
\label{sub:G1.602$+$0.018}
No emission is seen in any lines other than $^{12}$CO and $^{13}$CO.

\subsection{G1.651$-$0.050}
\label{sub:G1.651$-$0.050}
Two spectral components are seen in $^{13}$CO at $-35$ \kms and 55 \kms. These components are separated spatially from the continuum emission but coincide well with each other. No emission is seen in any other line.

\subsection{G1.670$-$0.130 and G1.683$-$0.089}
\label{sub:G1.670$-$0.130}
Half of the $^{12}$CO, H$_{2}$CO 3(2,1)-3(2,0) and both transitions of OCS were entirely masked during the beam correction process described in Section~\ref{subsec:beam}. This prevents a reasonable production of the moment map, as the unmasked half contains mostly emission and not enough emission free channels to accurately measure the rms. As such this moment map should not be considered reliable. No emission is seen in any other line.

\subsection{G359.137$+$0.031}
\label{sub:G359.137$+$0.031}
$^{13}$CO shows two structures separated both spatially and kinematically, with a peak at the continuum emission at a \Vlsr\ of 0 \kms{}, and a secondary peak at $-40$\,\kms{} which lies south of the continuum peak. C$^{18}$O and the lower transition of \HtwoCO{} peak at 0 \kms, with a peak at this \Vlsr\ in the middle transition of \HtwoCO{} that is too weak to be included in the moment map. The baseline within the upper transition of OCS is offset from 0, and as such the moment map should not be considered reliable. No emission was seen in any other line.

\subsection{G359.484$-$0.132}
\label{sub:G359.484$-$0.132}
Emission is detected in C$^{18}$O, the lower two transitions of \HtwoCO{}, both transitions of OCS, as well as SiO. There appears to be no consistent position or \Vlsr\ for the emission between the transitions. No emission was seen in any other line.

\subsection{G359.611$+$0.018}
\label{sub:G359.611$+$0.018}
No emission is seen in any line other than $^{12}$CO and $^{13}$CO.

\subsection{G359.615$-$0.243}
\label{sub:G359.615$-$0.243}
C$^{18}$O and all three \HtwoCO{} transitions show emission at a \Vlsr\ of 20 \kms{}. A peak at 70 \kms{} in the cube of the middle \HtwoCO{} transition is likely produced by CH$_{3}$OH-e. A peak at $-120$\,\kms{} was also detected in both the 218.2 GHz and 218.8 GHz \HtwoCO{} transitions, with an additional peak in this latter cube at $-175$\,\kms{}. The emission seen in the moment map of OCS (218.9 GHz) is detected in a single channel, leading to anomalous moment 1 and 2 maps. Emission is also detected at a \Vlsr\ of 20 \kms{} in SO. All of these lines coincide well within the single continuum peak. No emission is seen in any other line.

\subsection{G359.865$+$0.022}
\label{sub:G359.865$+$0.022}
$^{13}$CO shows multiple velocity components at $-40$, 10 and 60\,\kms{}, with C$^{18}$O also peaking at a \Vlsr\ of $-4$\,\kms{}. No emission is seen in any other line.

\subsection{G359.889$-$0.093}
\label{sub:G359.889$-$0.093}
C$^{18}$O, three transitions of \HtwoCO{}, SiO and SO all show strong emission at a \Vlsr\ or $\sim$15 km s$^{-1}$ coinciding strongly with the continuum emission, with numerous other peaks throughout the region within the range of $\pm 50$\,\kms{}, likely the result of contamination from other transitions. The OCS transition in the lower sideband shows weak emission at $\sim$15 km s$^{-1}$, however channels from $-30 - 10$\,\kms{} were masked during the beam correction process described in Section~\ref{subsec:beam}. The OCS transition in the upper sideband shows emission from $\pm 50$\,\kms{}, but with no coincidence with the continuum emission. 

\subsection{G359.948$-$0.052}
\label{sub:G359.948$-$0.052}
This region suffered from image artifacts and will be included in a future publication

\clearpage
\section{G0.106-0.082 Moment Maps and Spectral Fits}\label{sec:appendix}

\begin{figure*}
    \includegraphics[width=1\textwidth]{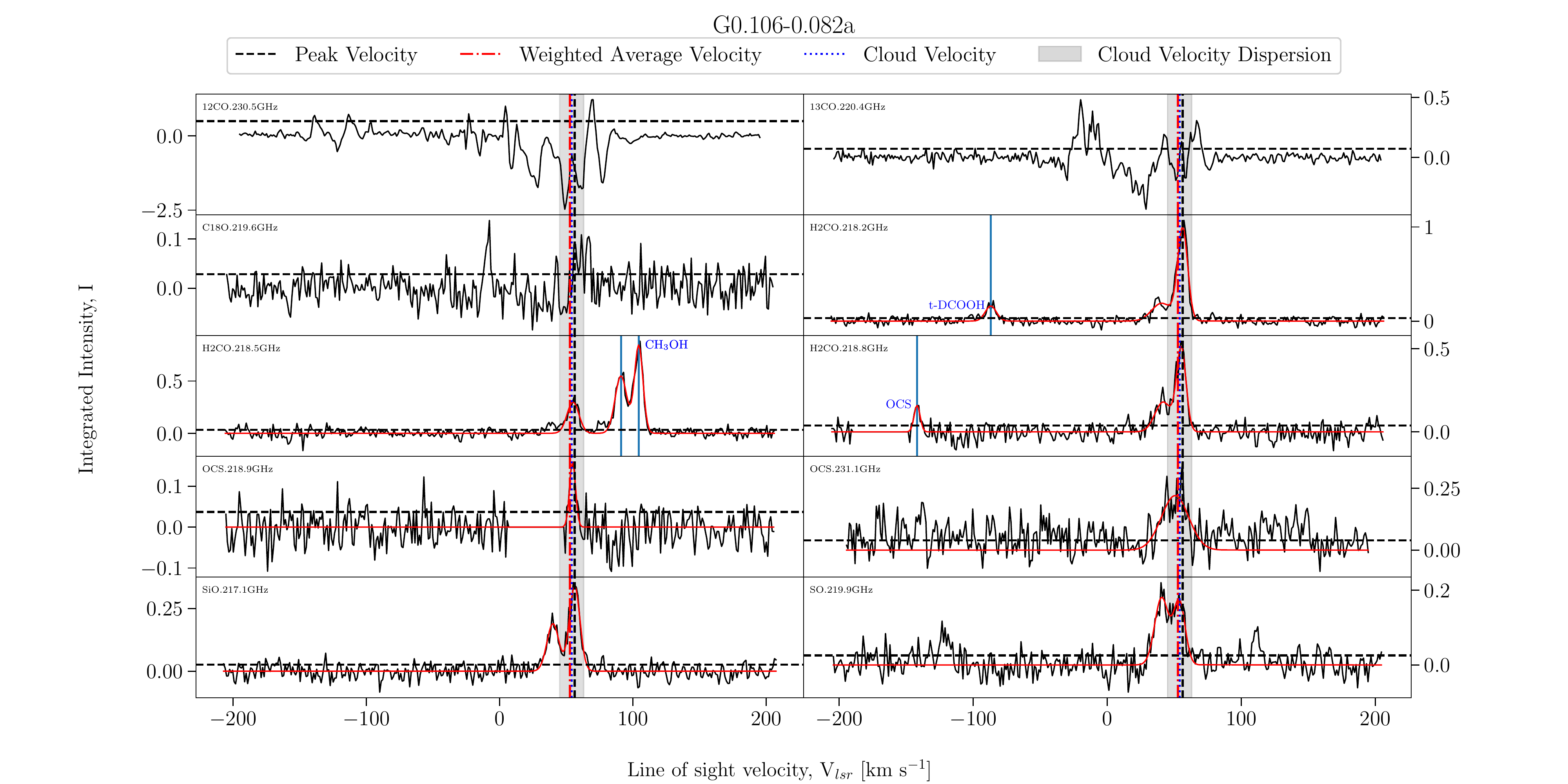}
    \caption{Fitted spectra for dendrogram leaf G0.106-0.082a, with scouse fits overlaid in red.}
    \label{fig:0106a}
\end{figure*}

\begin{figure*}
    \includegraphics[width=1\textwidth]{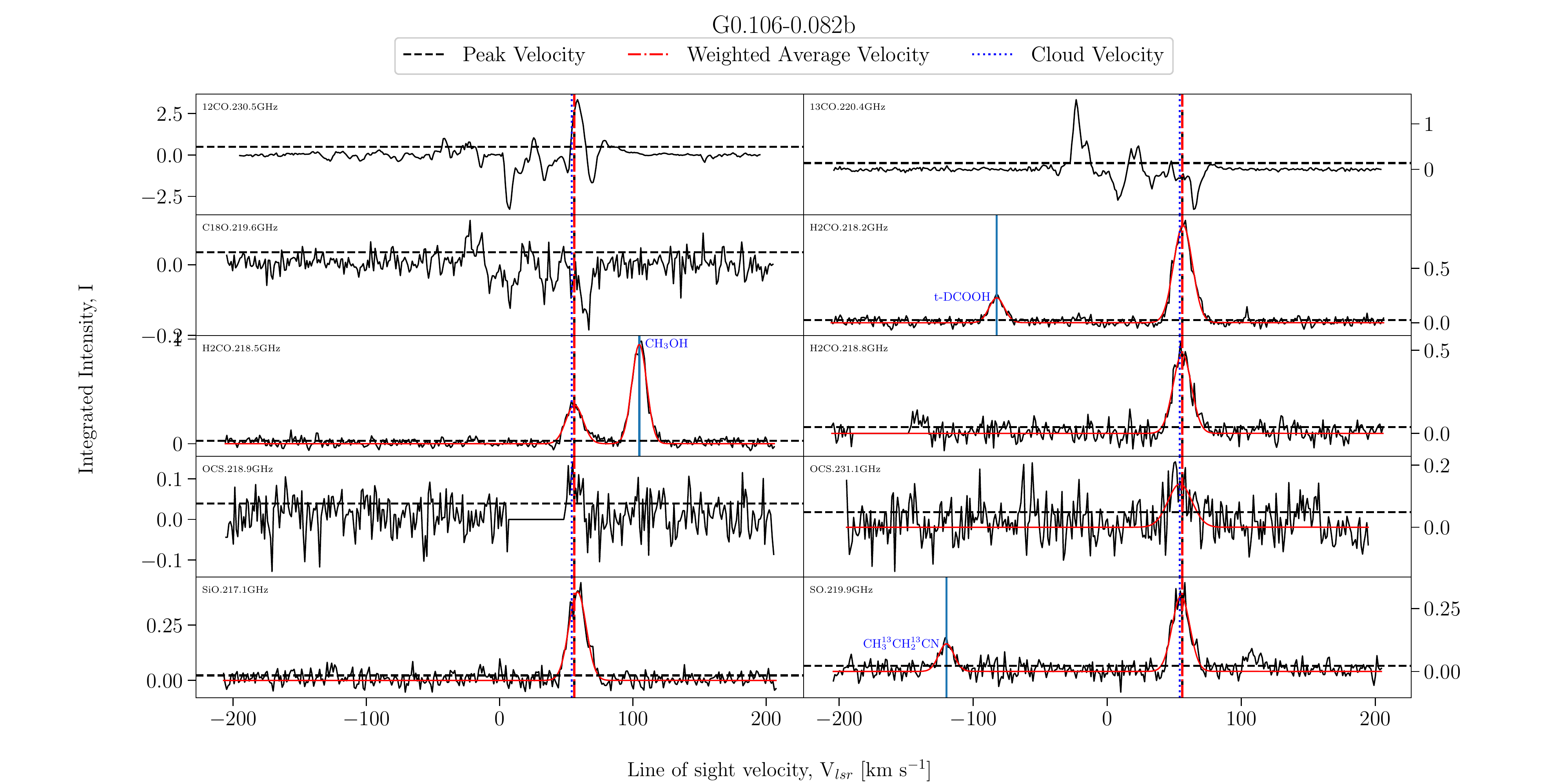}
    \caption{Fitted spectra for dendrogram leaf G0.106-0.082b, with scouse fits overlaid in red.}
    \label{fig:0106b}
\end{figure*}

\begin{figure*}
    \includegraphics[width=1\textwidth]{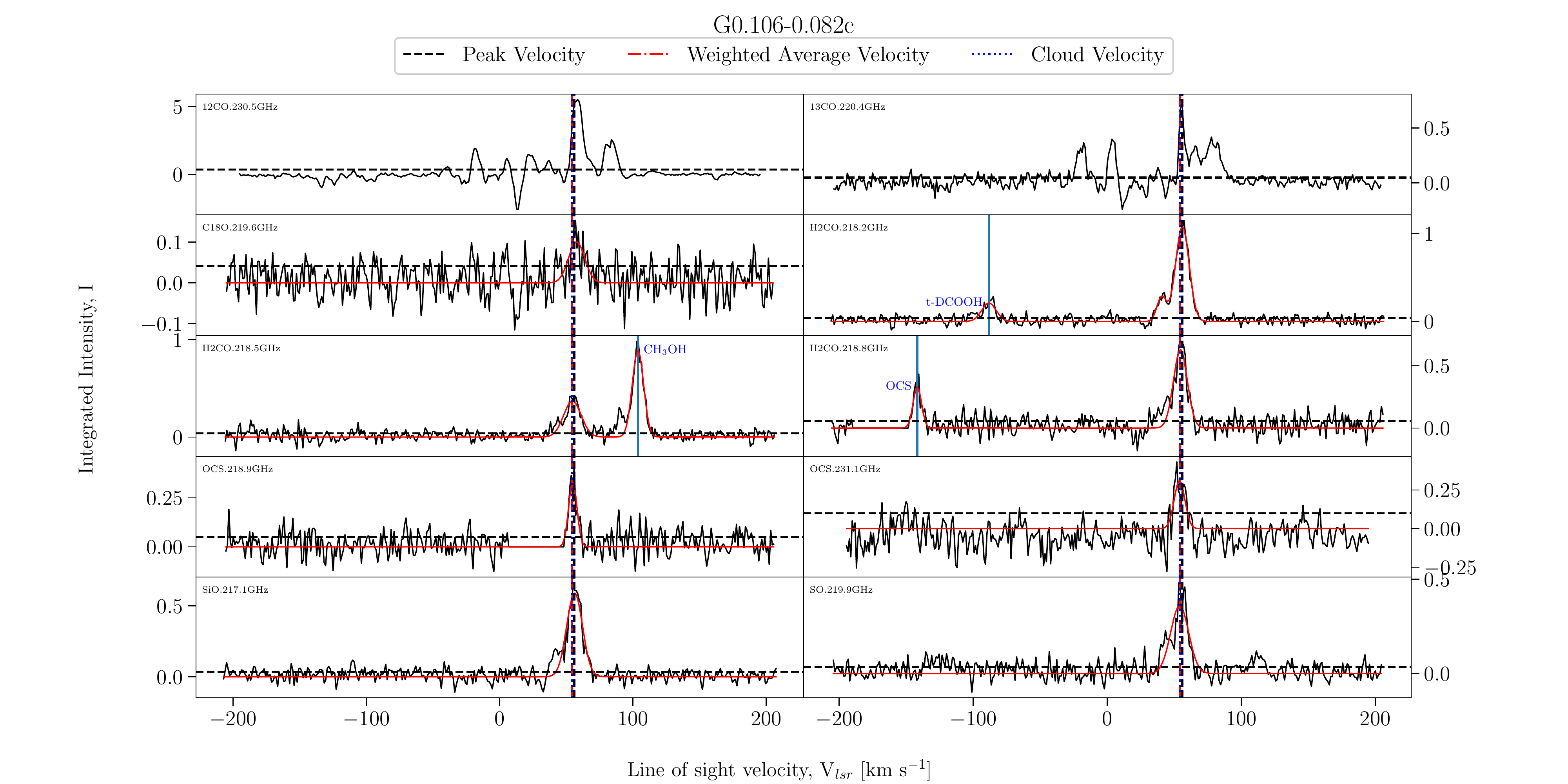}
    \caption{Fitted spectra for dendrogram leaf G0.106-0.082c, with scouse fits overlaid in red.}
    \label{fig:0106c}
\end{figure*}

\begin{figure*}
    \includegraphics[width=1\textwidth]{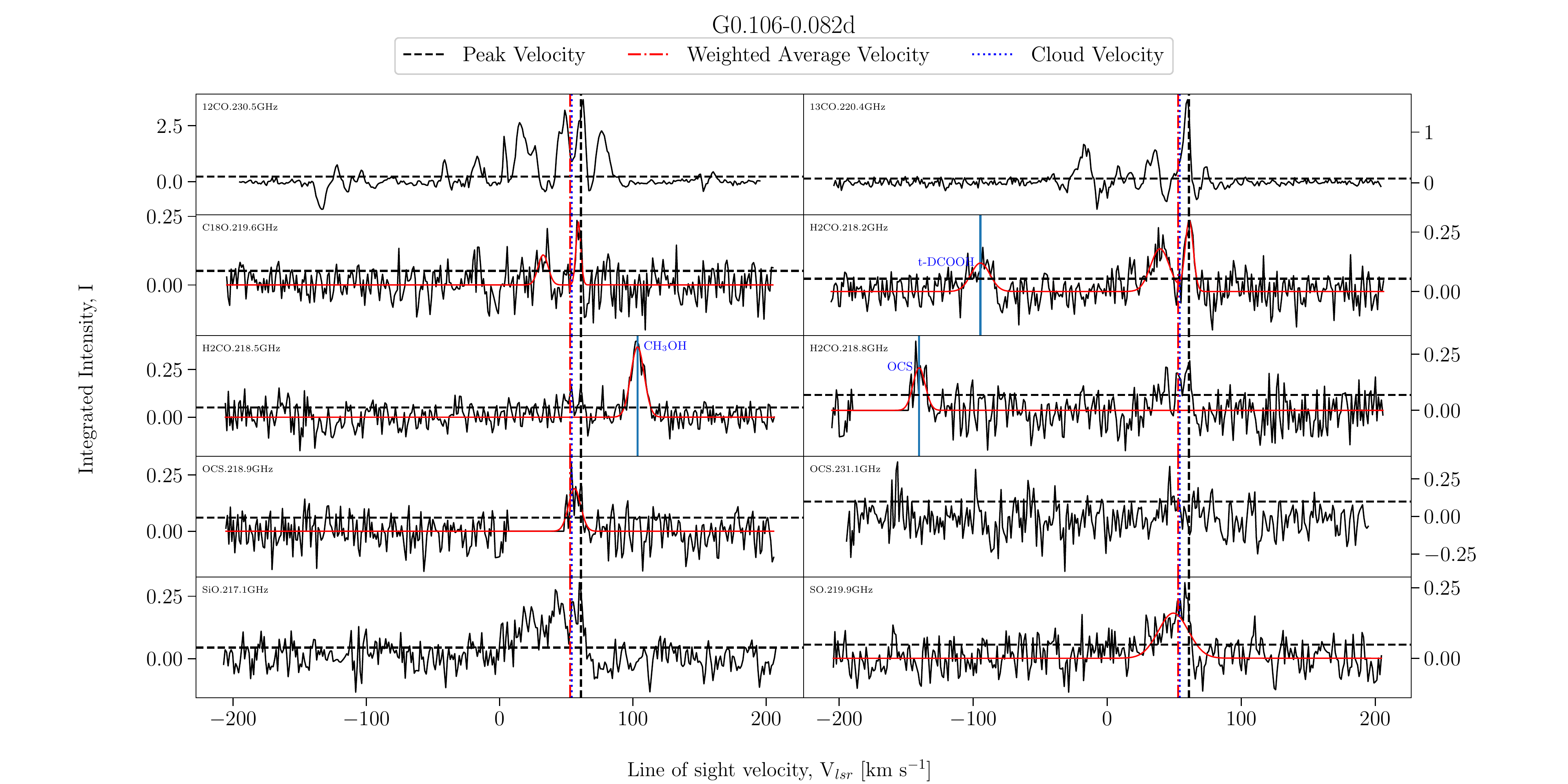}
    \caption{Fitted spectra for dendrogram leaf G0.106-0.082d, with scouse fits overlaid in red.}
    \label{fig:0106d}
\end{figure*}
\clearpage

\section{G0.068-0.075b Moment Maps and Spectral Fits}

\begin{figure*}
    \centering
    \includegraphics[angle=-90,width=0.83\textwidth]{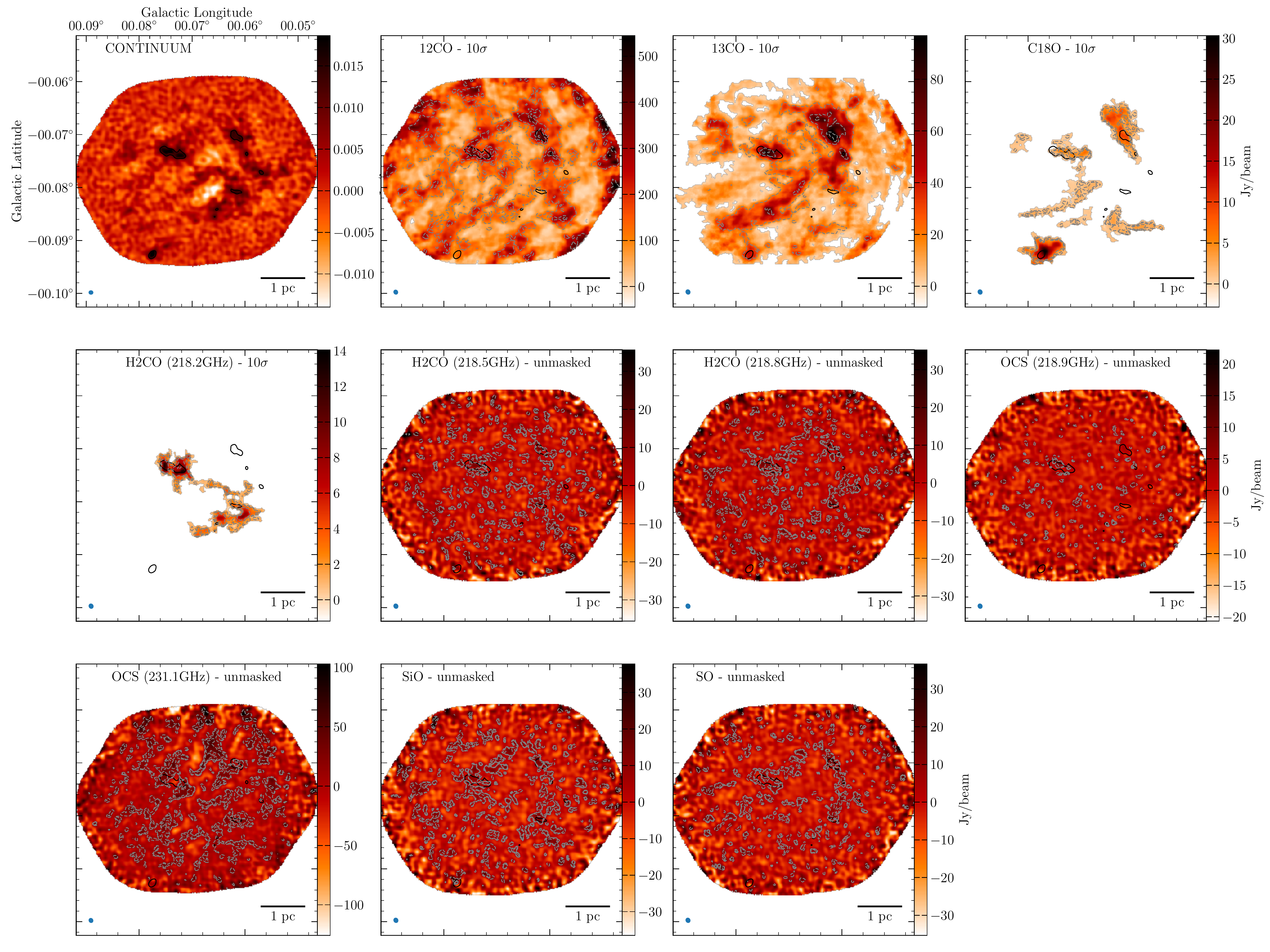}
    \caption{G0.068-0.075 integrated intensity moment maps}
    \label{fig:0068int}
\end{figure*}

\begin{figure*}
    \centering
    \includegraphics[angle=-90,width=0.83\textwidth]{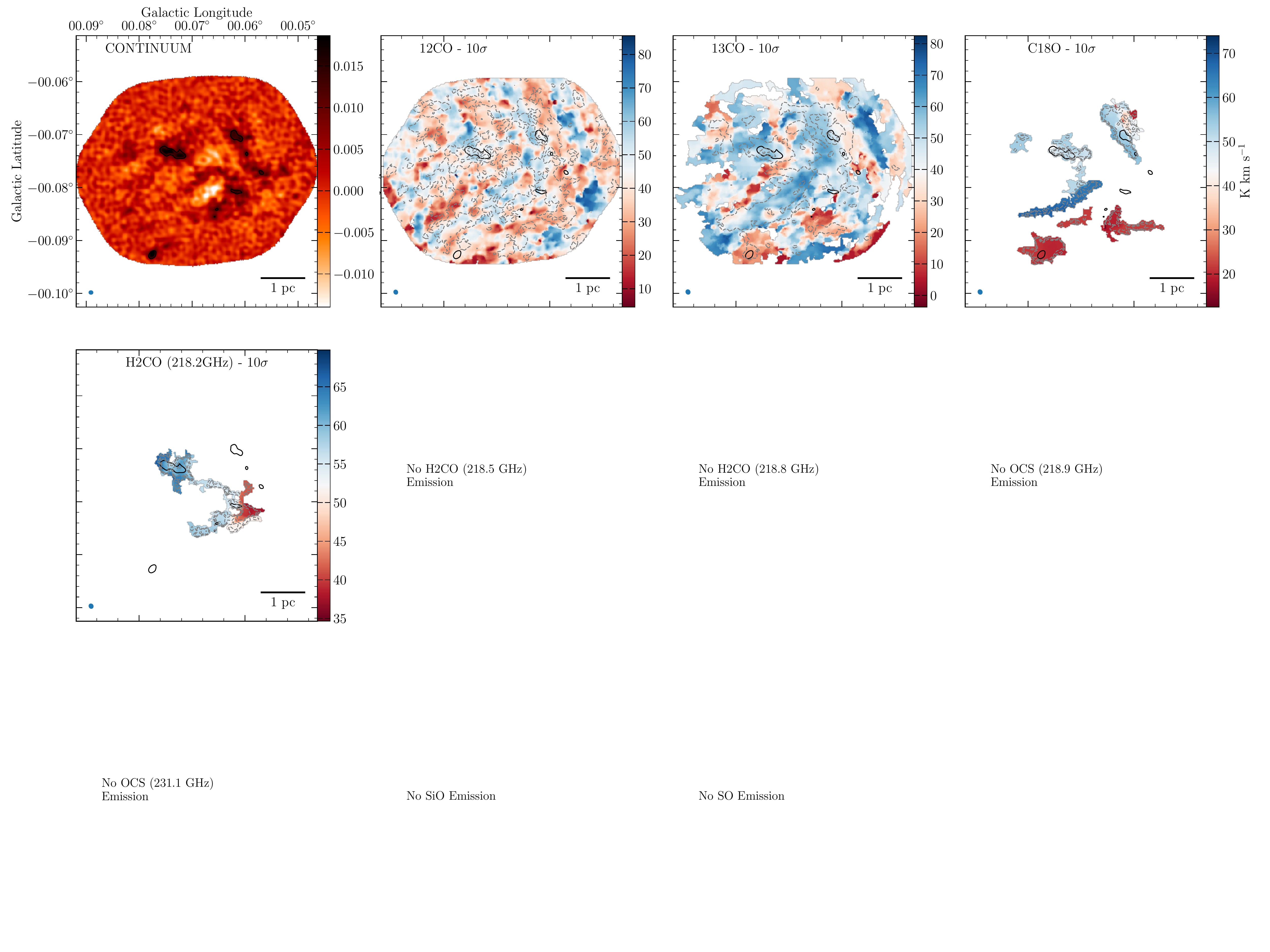}
    \caption{G0.068-0.075 \Vlsr moment maps}
    \label{fig:0068vel}
\end{figure*}

\begin{figure*}
    \centering
    \includegraphics[angle=-90,width=0.83\textwidth]{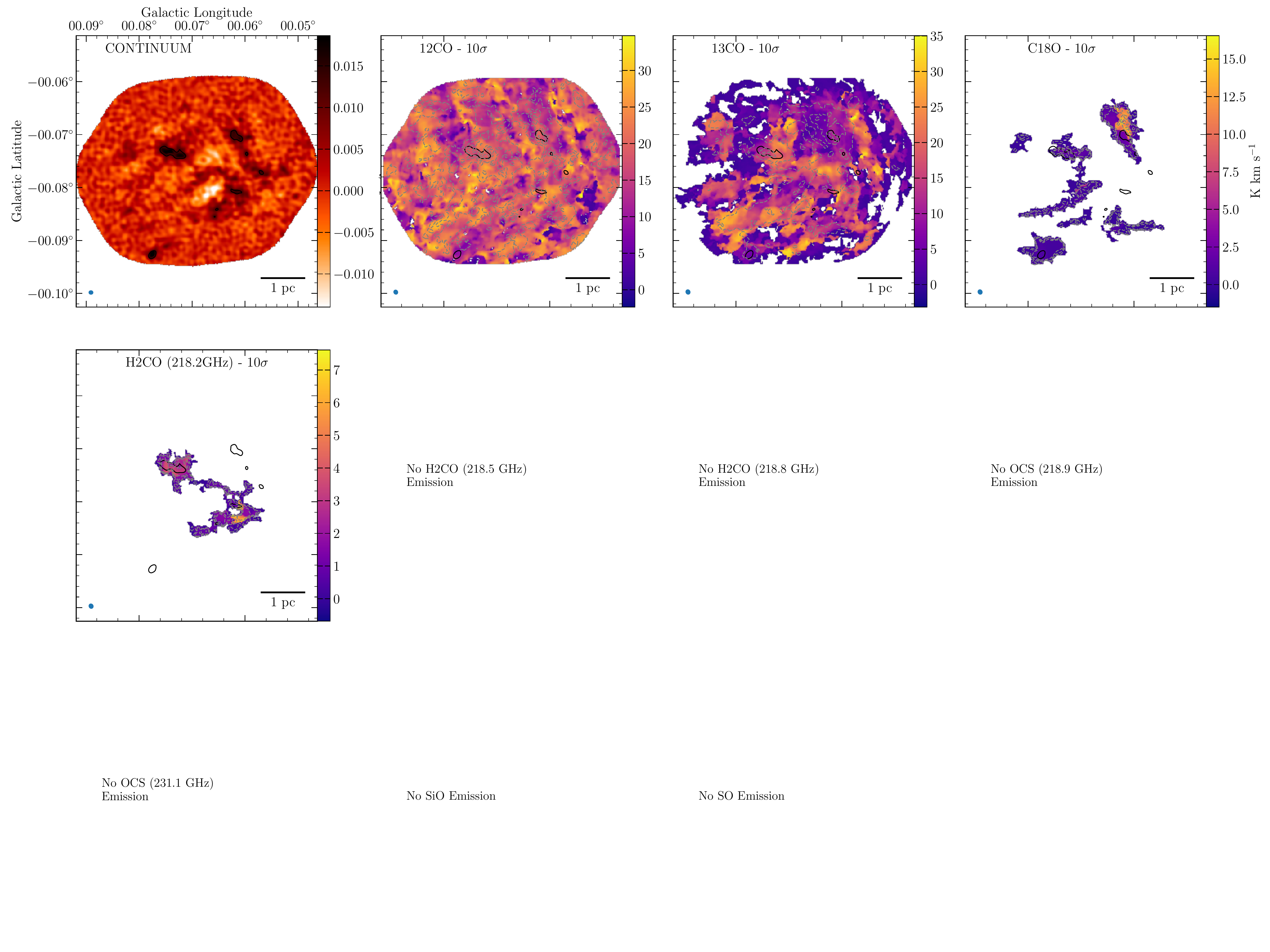}
    \caption{G0.068-0.075 velocity dispersion moment maps}
    \label{fig:0068fwhm}
\end{figure*}

\begin{figure*}
    \includegraphics[width=1\textwidth]{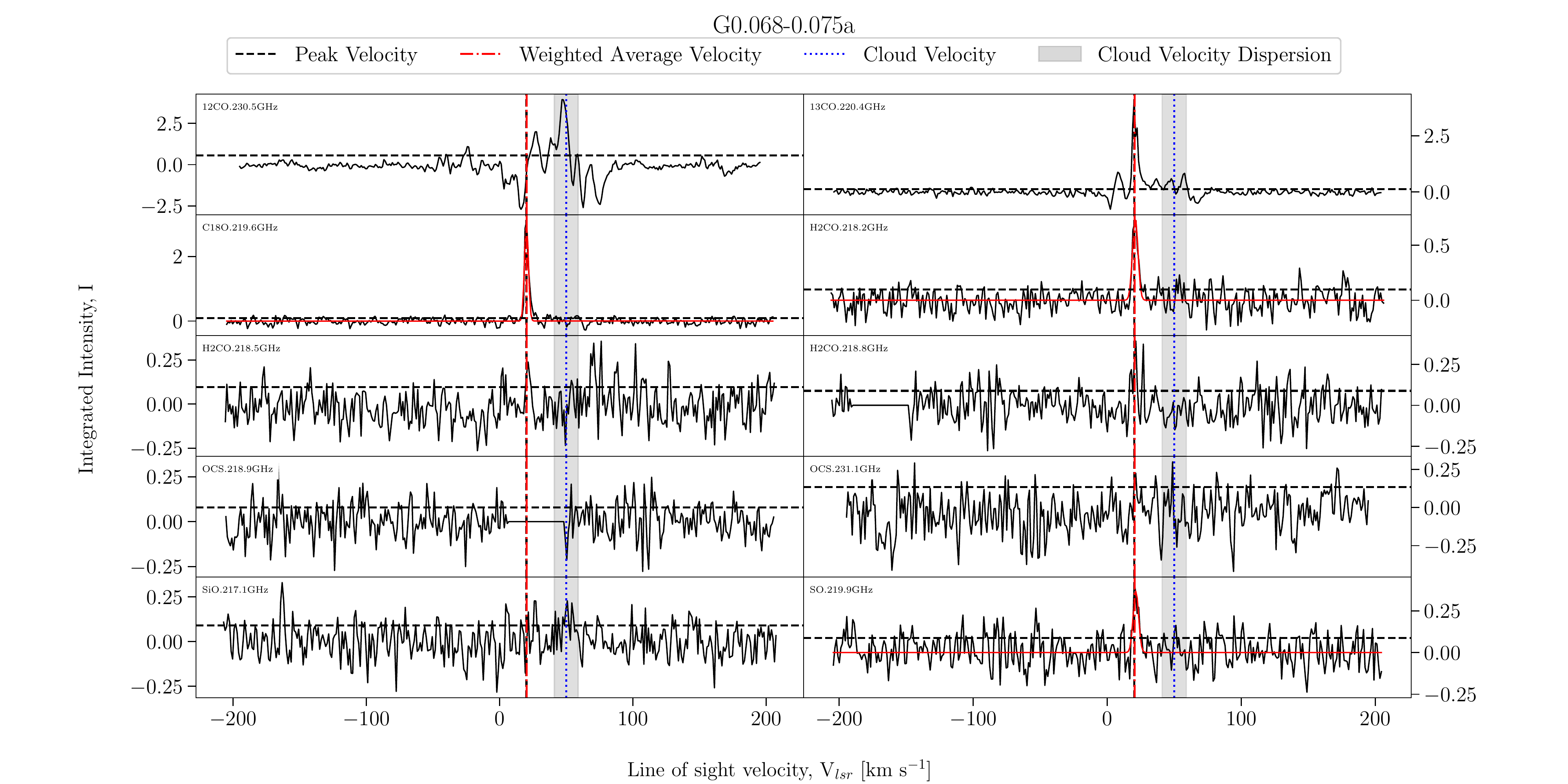}
    \caption{Fitted spectra for dendrogram leaf G0.068-0.075a, with scouse fits overlaid in red, cloud velocity and velocity dispersions are indicated by the blue dashed line and grey shaded area, respectively.}
    \label{fig:0068a}
\end{figure*}

\begin{figure*}
    \includegraphics[width=1\textwidth]{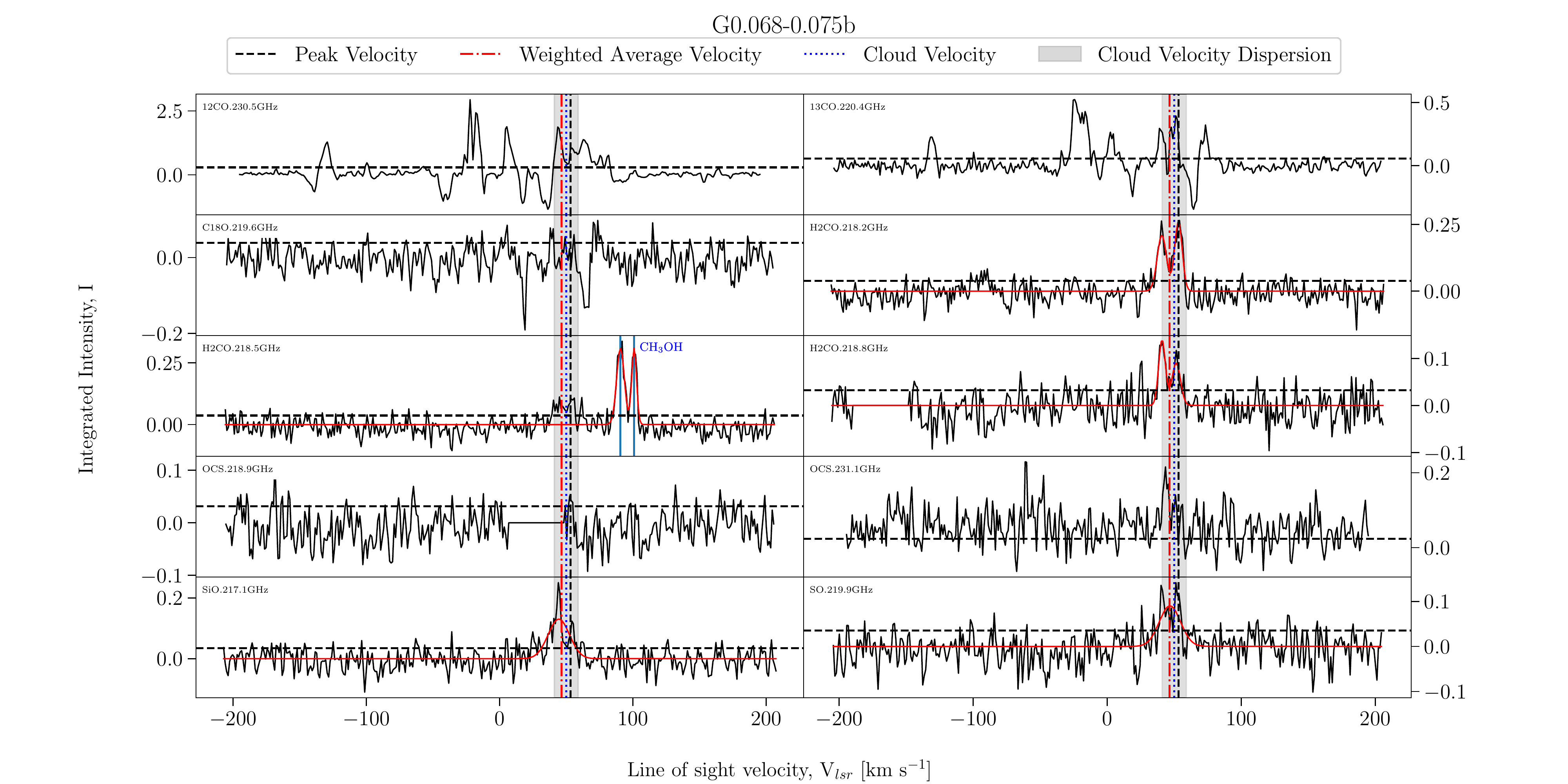}
    \caption{Fitted spectra for dendrogram leaf G0.068-0.075b, with scouse fits overlaid in red, cloud velocity and velocity dispersions are indicated by the blue dashed line and grey shaded area, respectively.}
    \label{fig:0068b}
\end{figure*}
\clearpage
\section{Outflow candidates \& position-velocity plots}
\label{sec:appendix_outflows}

\begin{figure*}
\centering
\begin{subfigure}[b]{1\textwidth}
    \centering
    \includegraphics[width=\linewidth]{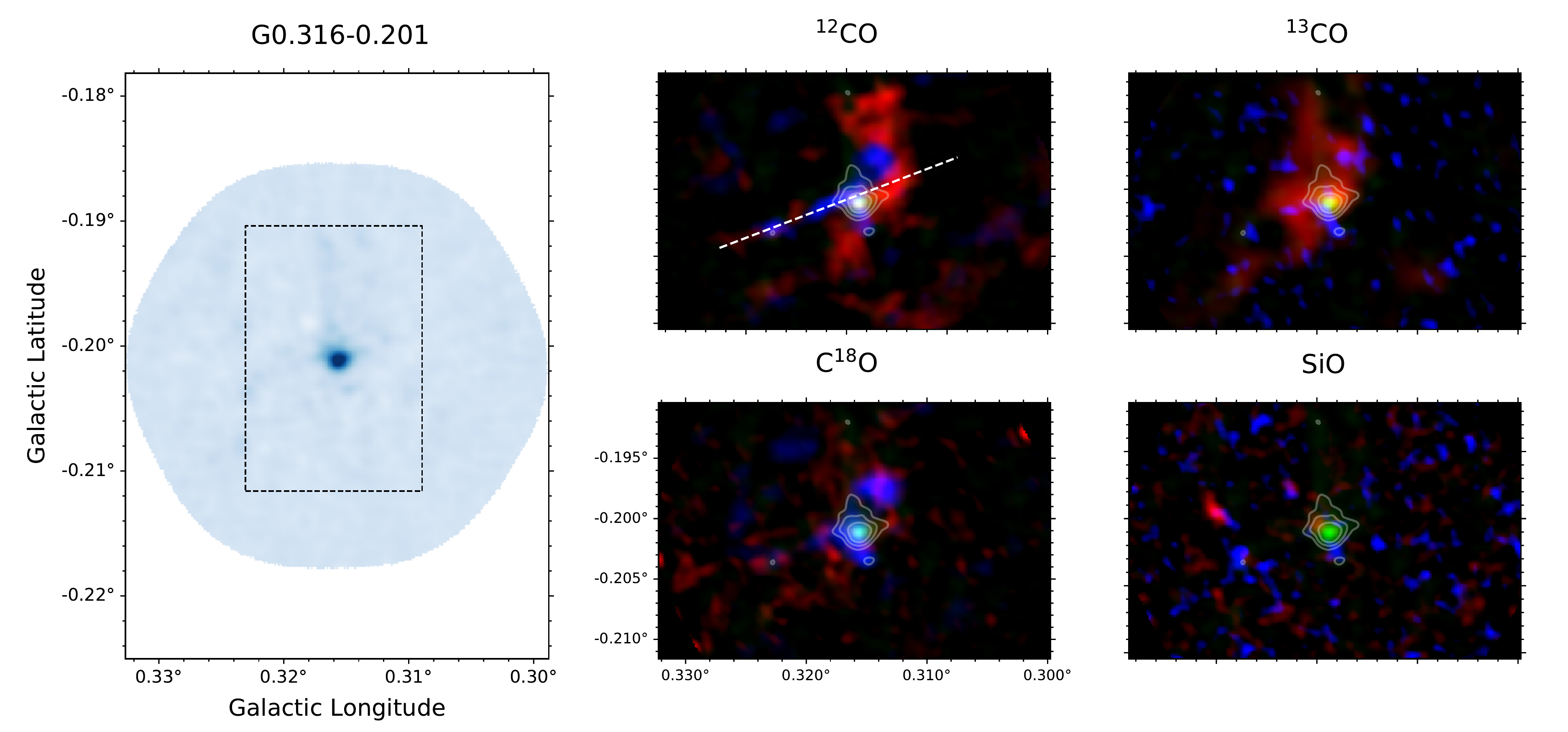}
    \caption{} 
\end{subfigure}%

\begin{subfigure}[b]{1\textwidth} 
    \centering
    \includegraphics[width=\linewidth]{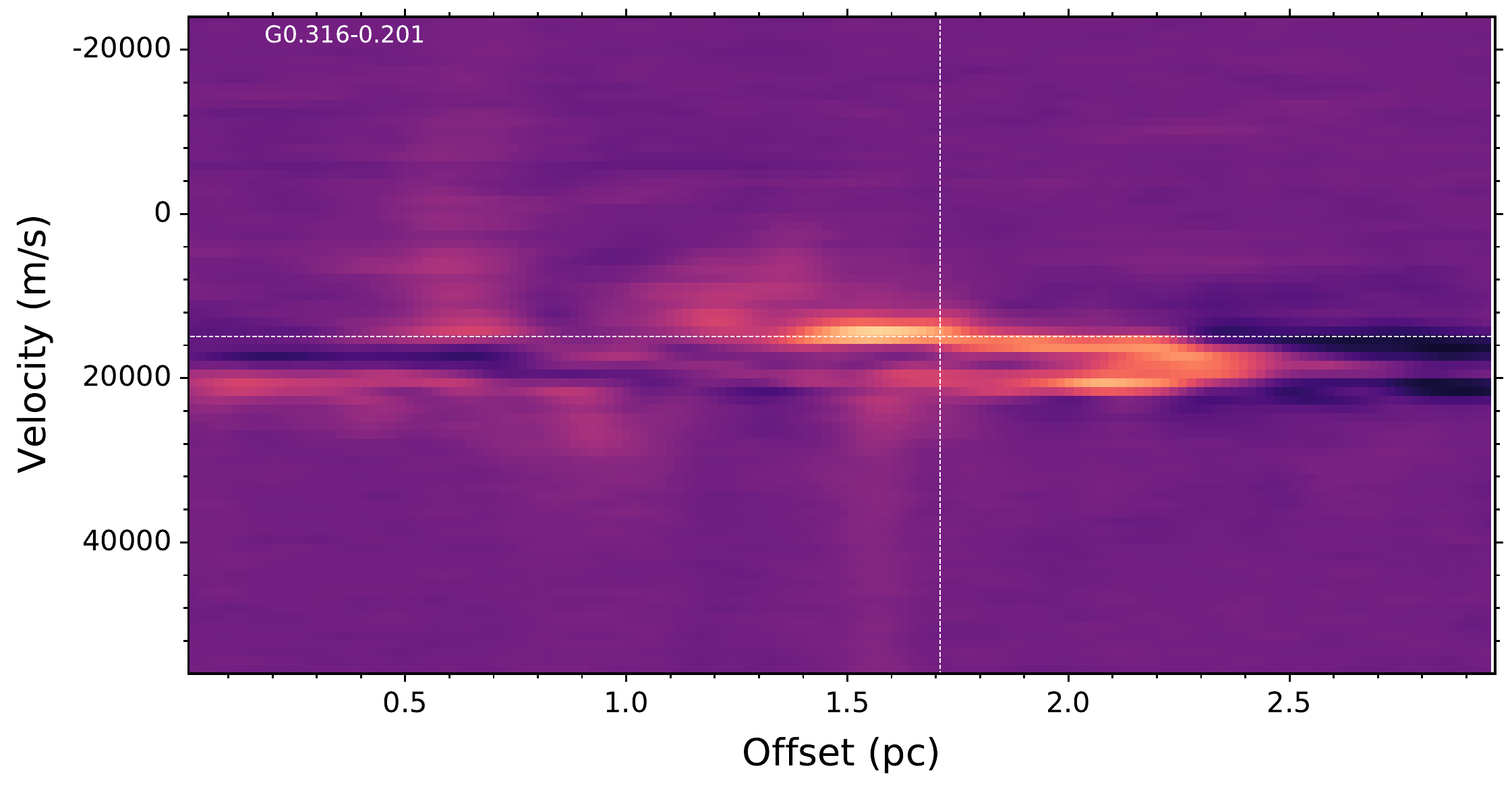}
    \caption{} 
\end{subfigure}%
    \caption{\textbf{(a)} \textit{Left}: SMA 1.3~mm dust continuum. The dotted black box indicates the region shown in the other panels. \textit{Right}: Four panels showing three-colour images for $^{12}$CO, $^{13}$CO, C$^{18}$O, and SiO. Red-shifted and blue-shifted integrated intensity (V$_{\textrm{lsr}}~\pm$~10~\kms) are shown in blue and red, respectively. Dust continuum is shown in green. The white dashed line overlaid on the $^{12}$CO emission indicates the region over which a PV-slice was taken. \textbf{(b)} PV-plot from the slice shown in \textbf{(a)}. The vertical dotted line denotes the central position of the continuum source across which the slice was taken. The horizontal dashed line denotes the assumed V$_{\textrm{lsr}}$ of the continuum source.} 
\label{fig:G0.316_outflows}
\end{figure*}

\begin{figure*}
\centering
\begin{subfigure}[b]{1\textwidth}
    \centering
    \includegraphics[width=\linewidth]{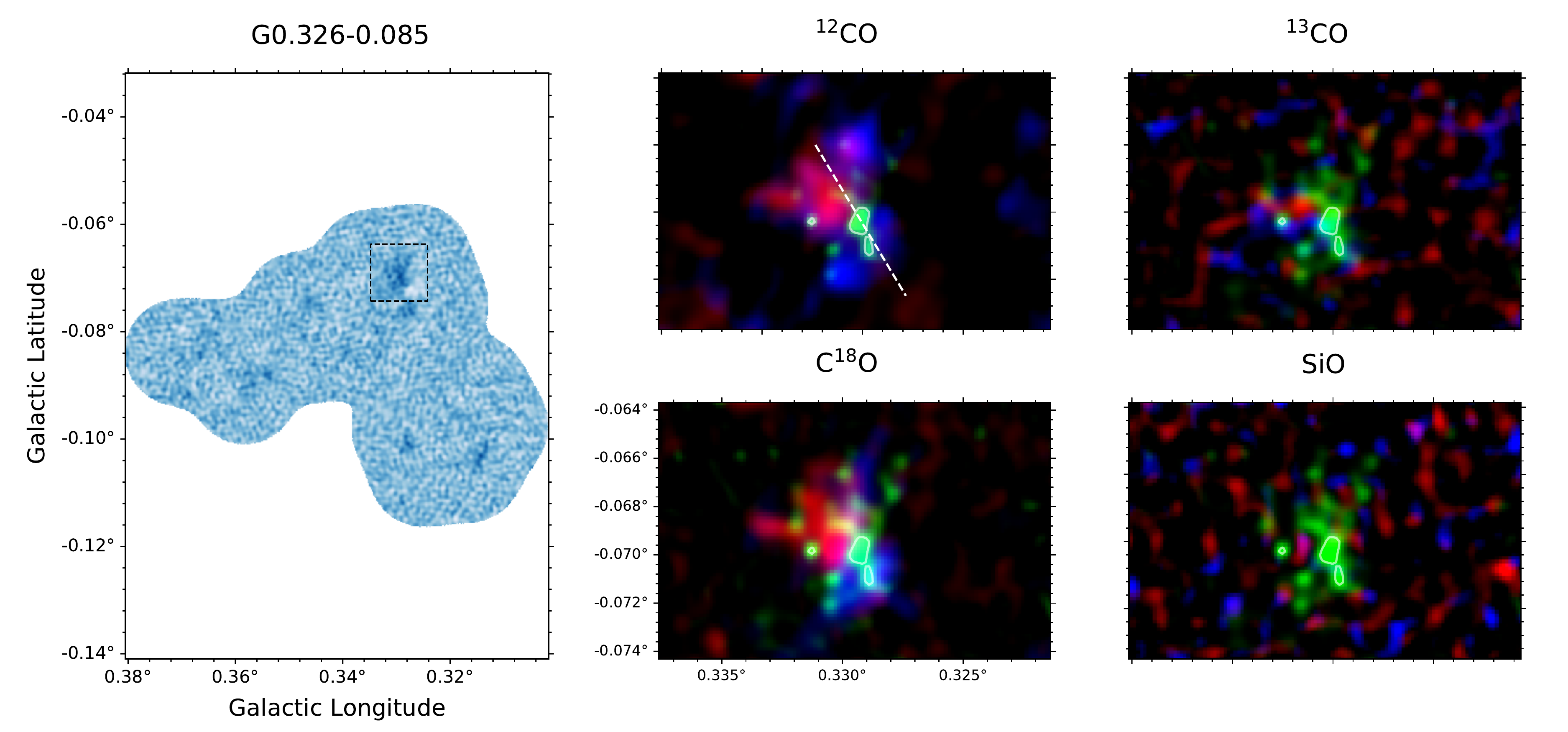}
    \caption{} 
\end{subfigure}%

\begin{subfigure}[b]{0.4\textwidth} 
    \centering
    \includegraphics[width=\linewidth]{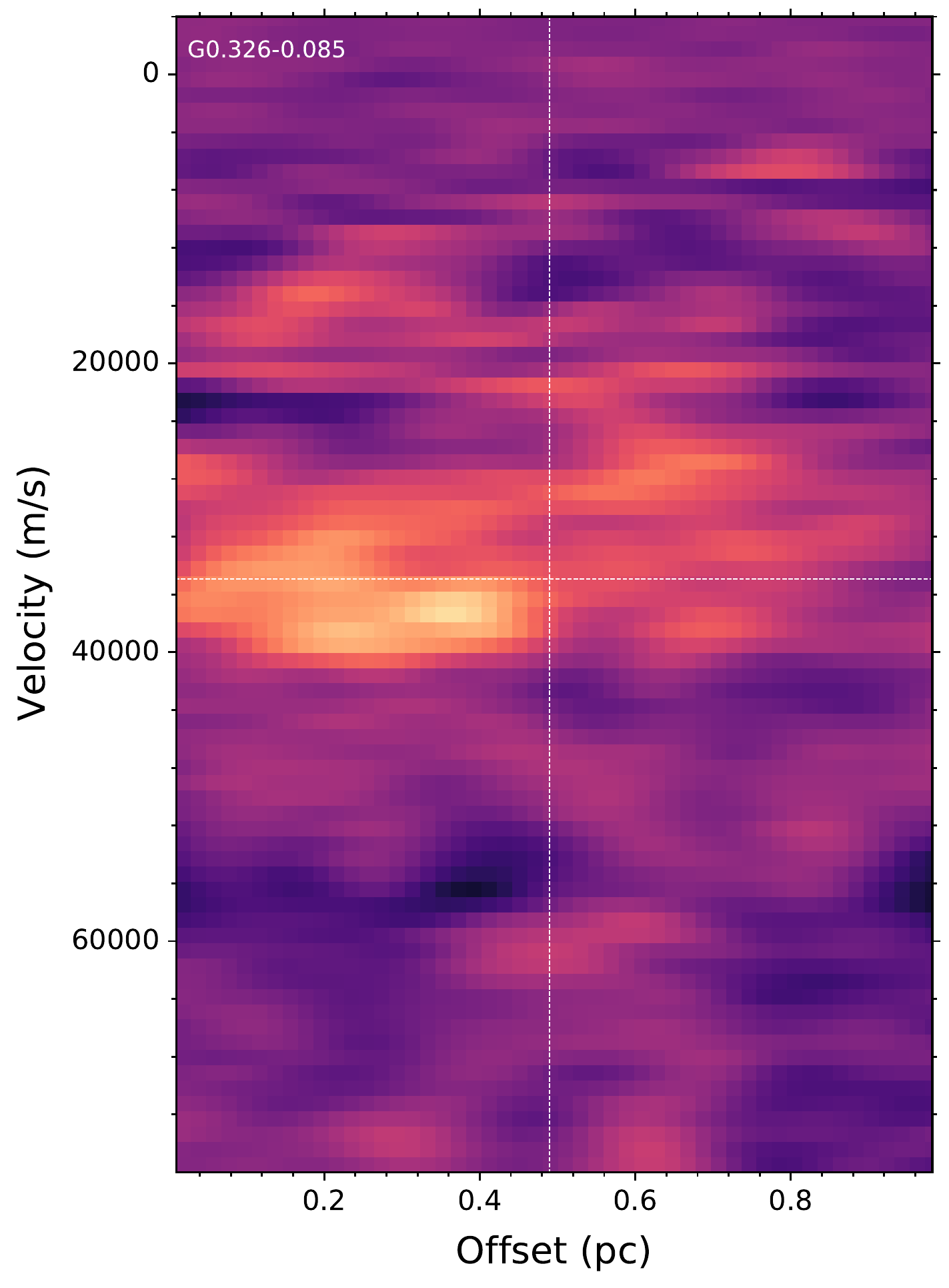}
    \caption{} 
\end{subfigure}%
    \caption{\textbf{(a)} \textit{Left}: SMA 1.3~mm dust continuum. The dotted black box indicates the region shown in the other panels. \textit{Right}: Four panels showing three-colour images for $^{12}$CO, $^{13}$CO, C$^{18}$O, and SiO. Red-shifted and blue-shifted integrated intensity (V$_{\textrm{lsr}}~\pm$~10~\kms) are shown in blue and red, respectively. Dust continuum is shown in green. The white dashed line overlaid on the $^{12}$CO emission indicates the region over which a PV-slice was taken. \textbf{(b)} PV-plot from the slice shown in \textbf{(a)}. The vertical dotted line denotes the central position of the continuum source across which the slice was taken. The horizontal dashed line denotes the assumed V$_{\textrm{lsr}}$ of the continuum source.} 
\label{fig:G0.326_outflows}
\end{figure*}

\begin{figure*}
\centering
\begin{subfigure}[b]{1\textwidth}
    \centering
    \includegraphics[width=\linewidth]{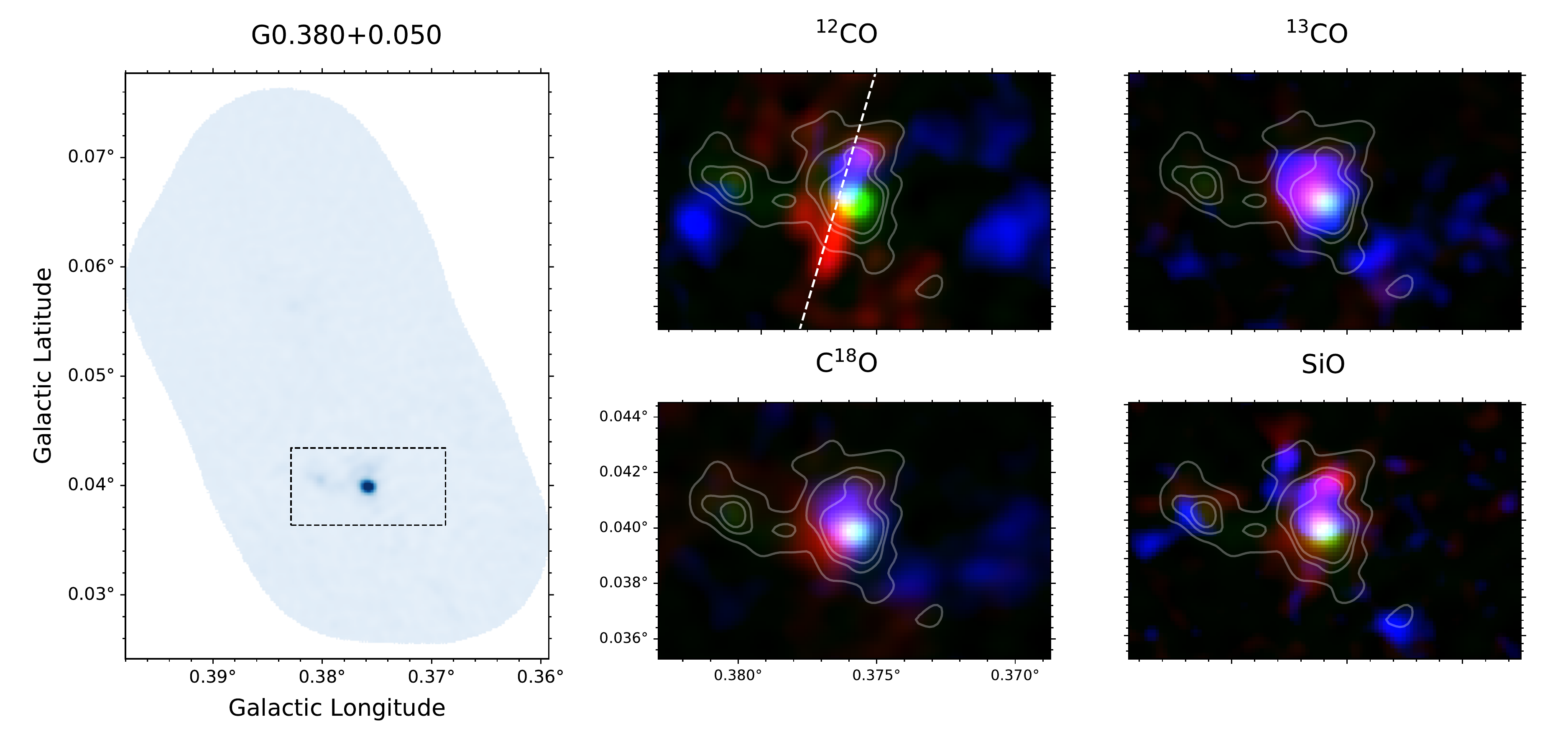}
    \caption{} 
\end{subfigure}%

\begin{subfigure}[b]{0.85\textwidth} 
    \centering
    \includegraphics[width=\linewidth]{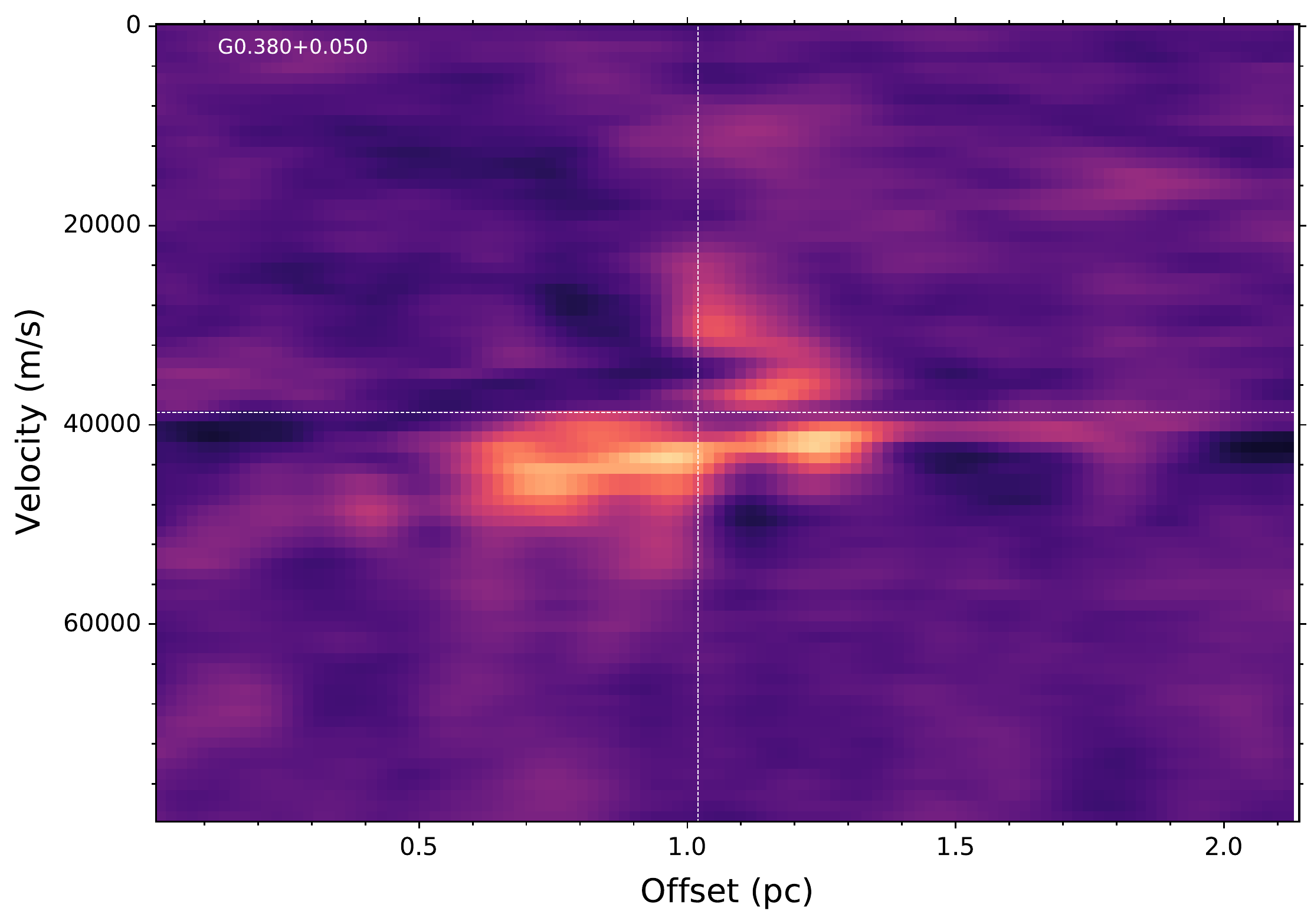}
    \caption{} 
\end{subfigure}%
    \caption{\textbf{(a)} \textit{Left}: SMA 1.3~mm dust continuum. The dotted black box indicates the region shown in the other panels. \textit{Right}: Four panels showing three-colour images for $^{12}$CO, $^{13}$CO, C$^{18}$O, and SiO. Red-shifted and blue-shifted integrated intensity (V$_{\textrm{lsr}}~\pm$~10~\kms) are shown in blue and red, respectively. Dust continuum is shown in green. The white dashed line overlaid on the $^{12}$CO emission indicates the region over which a PV-slice was taken. \textbf{(b)} PV-plot from the slice shown in \textbf{(a)}. The vertical dotted line denotes the central position of the continuum source across which the slice was taken. The horizontal dashed line denotes the assumed V$_{\textrm{lsr}}$ of the continuum source.} 
\label{fig:G0.380_outflows}
\end{figure*}

\begin{figure*}
\centering
\begin{subfigure}[b]{1\textwidth}
    \centering
    \includegraphics[width=\linewidth]{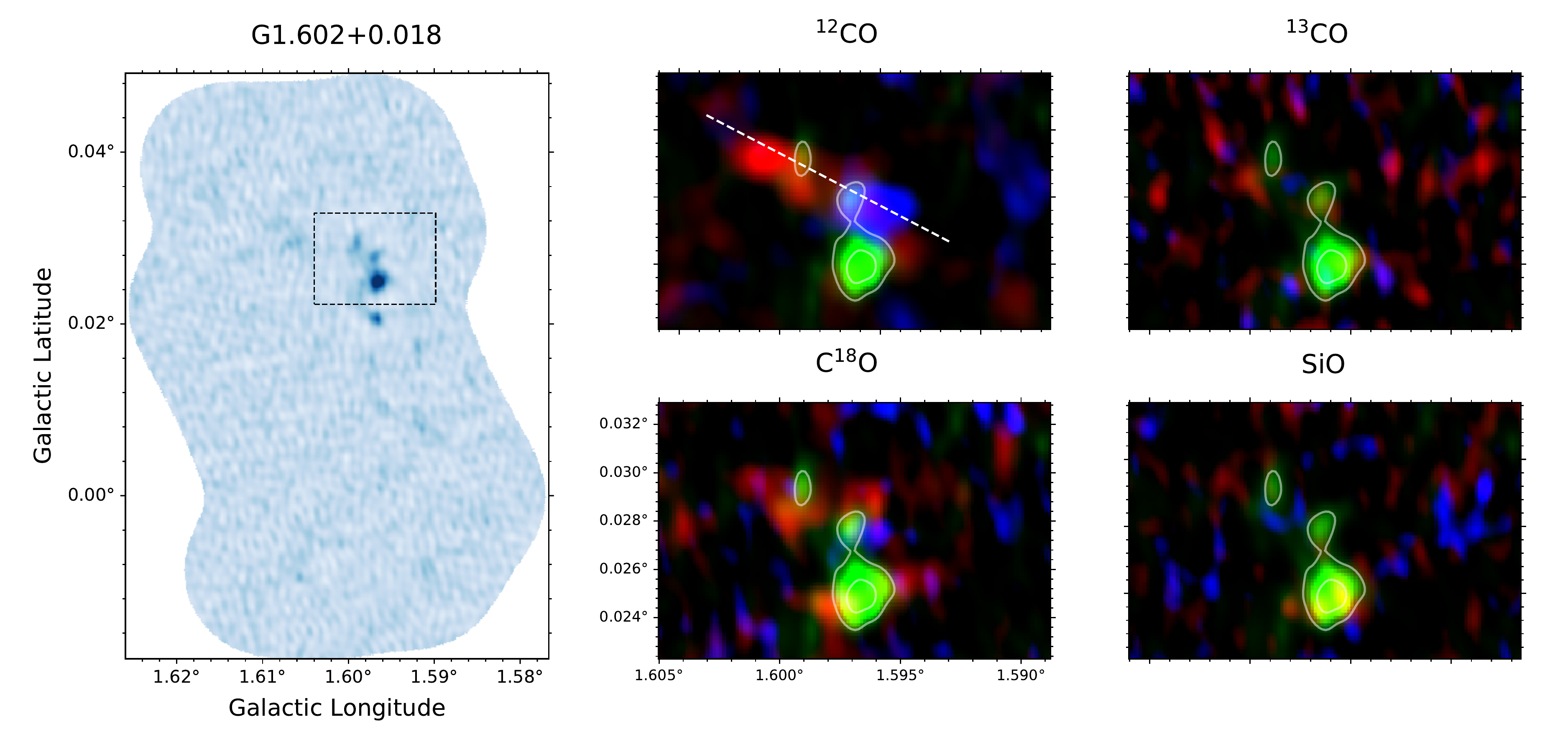}
    \caption{} 
\end{subfigure}%

\begin{subfigure}[b]{0.65\textwidth} 
    \centering
    \includegraphics[width=\linewidth]{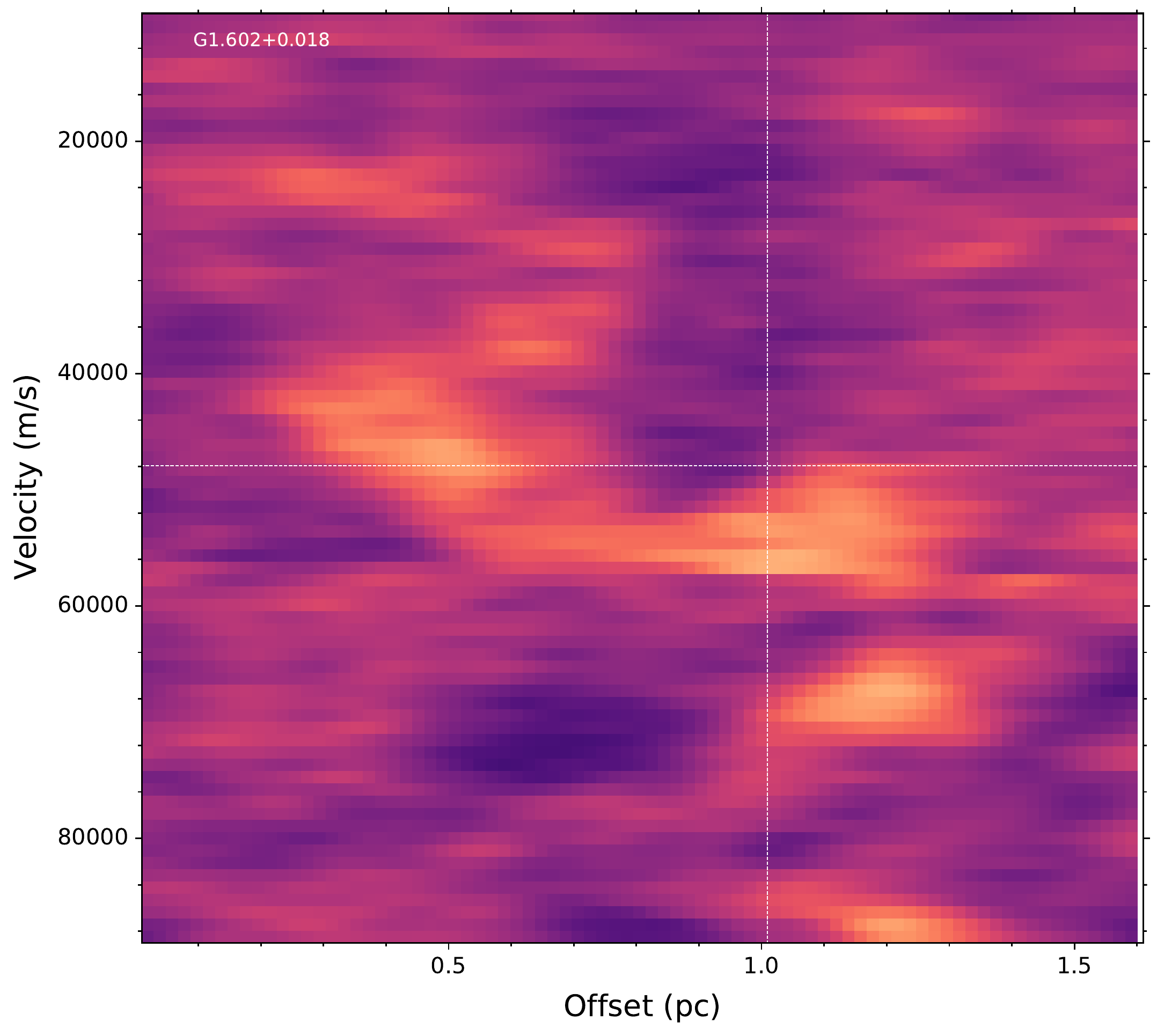}
    \caption{} 
\end{subfigure}%
    \caption{\textbf{(a)} \textit{Left}: SMA 1.3~mm dust continuum. The dotted black box indicates the region shown in the other panels. \textit{Right}: Four panels showing three-colour images for $^{12}$CO, $^{13}$CO, C$^{18}$O, and SiO. Red-shifted and blue-shifted integrated intensity (V$_{\textrm{lsr}}~\pm$~10~\kms) are shown in blue and red, respectively. Dust continuum is shown in green. The white dashed line overlaid on the $^{12}$CO emission indicates the region over which a PV-slice was taken. \textbf{(b)} PV-plot from the slice shown in \textbf{(a)}. The vertical dotted line denotes the central position of the continuum source across which the slice was taken. The horizontal dashed line denotes the assumed V$_{\textrm{lsr}}$ of the continuum source.} 
\label{fig:G1.602_outflows}
\end{figure*}

\begin{figure*}
\centering
\begin{subfigure}[b]{1\textwidth}
    \centering
    \includegraphics[width=\linewidth]{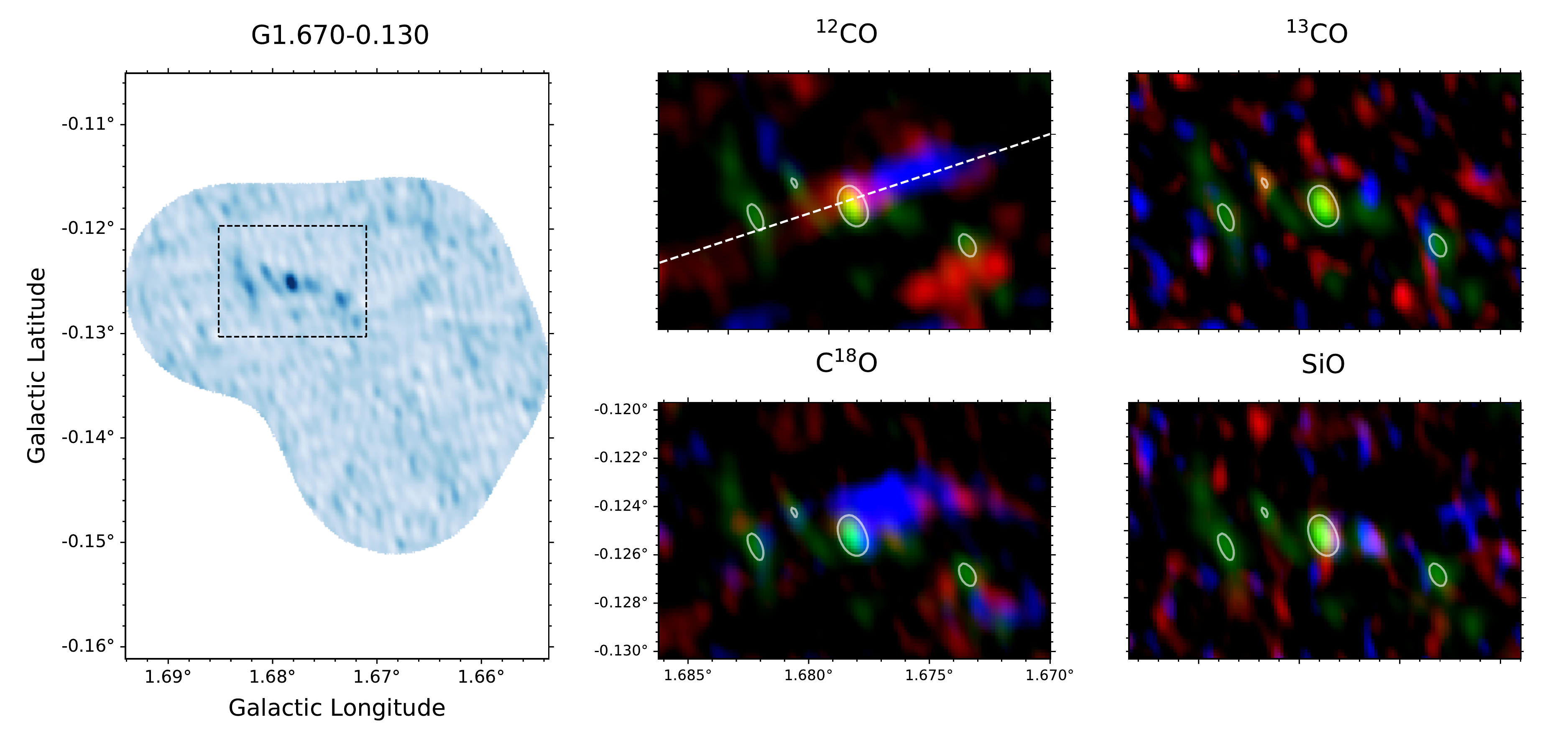}
    \caption{} 
\end{subfigure}%

\begin{subfigure}[b]{0.96\textwidth} 
    \centering
    \includegraphics[width=\linewidth]{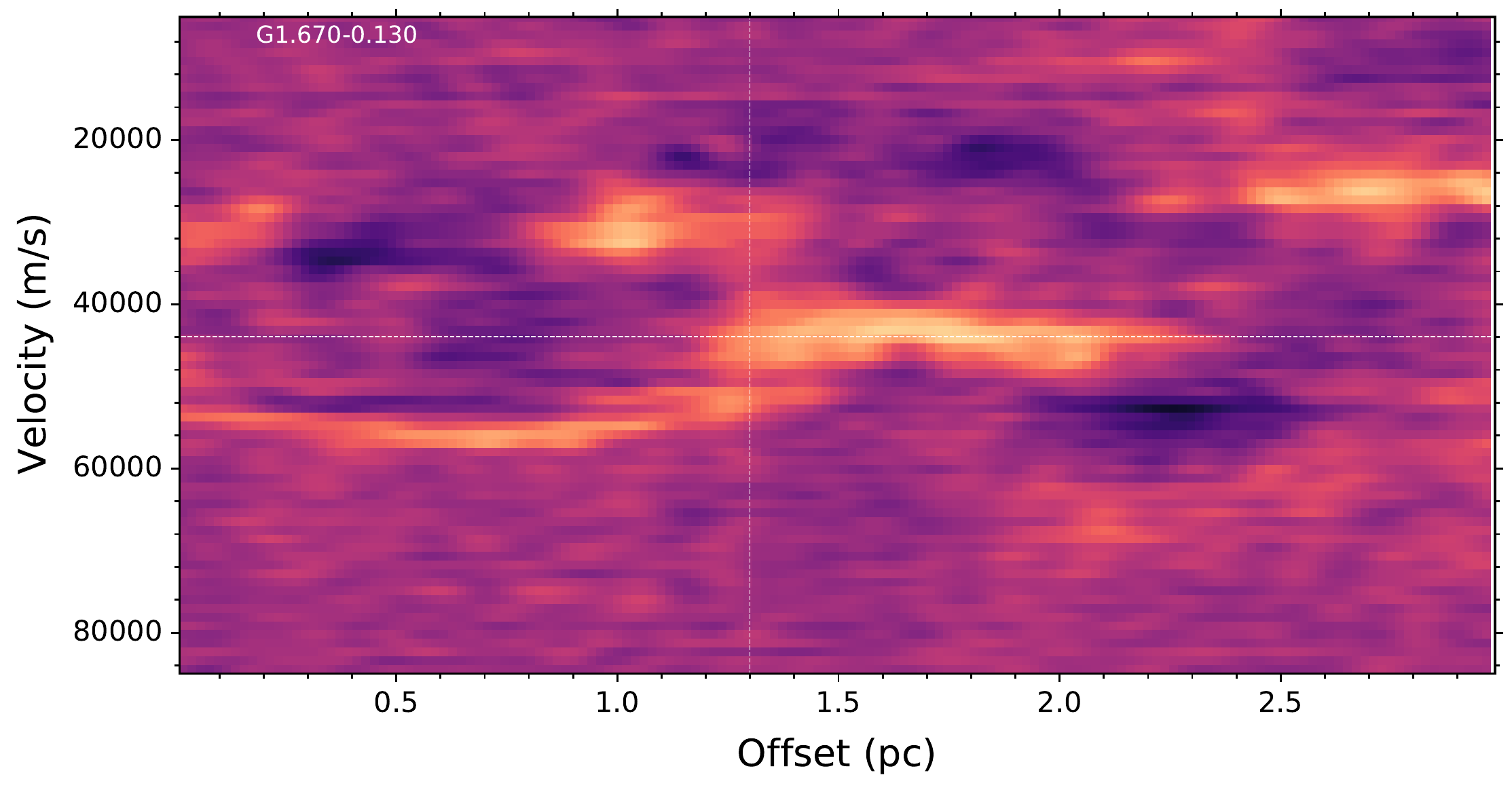}
    \caption{} 
\end{subfigure}%
    \caption{\textbf{(a)} \textit{Left}: SMA 1.3~mm dust continuum. The dotted black box indicates the region shown in the other panels. \textit{Right}: Four panels showing three-colour images for $^{12}$CO, $^{13}$CO, C$^{18}$O, and SiO. Red-shifted and blue-shifted integrated intensity (V$_{\textrm{lsr}}~\pm$~10~\kms) are shown in blue and red, respectively. Dust continuum is shown in green. The white dashed line overlaid on the $^{12}$CO emission indicates the region over which a PV-slice was taken. \textbf{(b)} PV-plot from the slice shown in \textbf{(a)}. The vertical dotted line denotes the central position of the continuum source across which the slice was taken. The horizontal dashed line denotes the assumed V$_{\textrm{lsr}}$ of the continuum source.} 
\label{fig:G1.670_outflows}
\end{figure*}

\begin{figure*}
\centering
\begin{subfigure}[b]{1\textwidth}
    \centering
    \includegraphics[width=\linewidth]{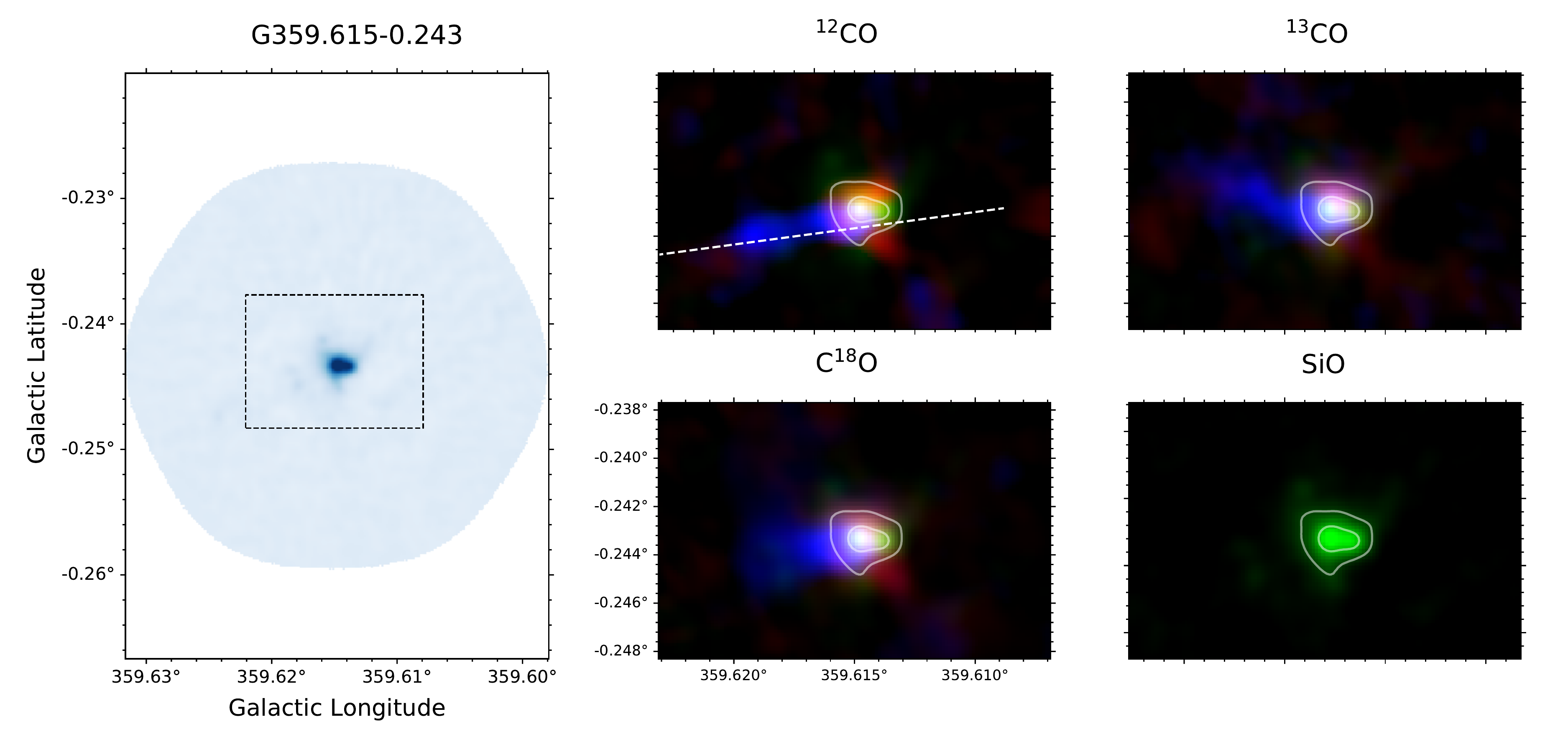}
    \caption{} 
\end{subfigure}%

\begin{subfigure}[b]{0.8\textwidth} 
    \centering
    \includegraphics[width=\linewidth]{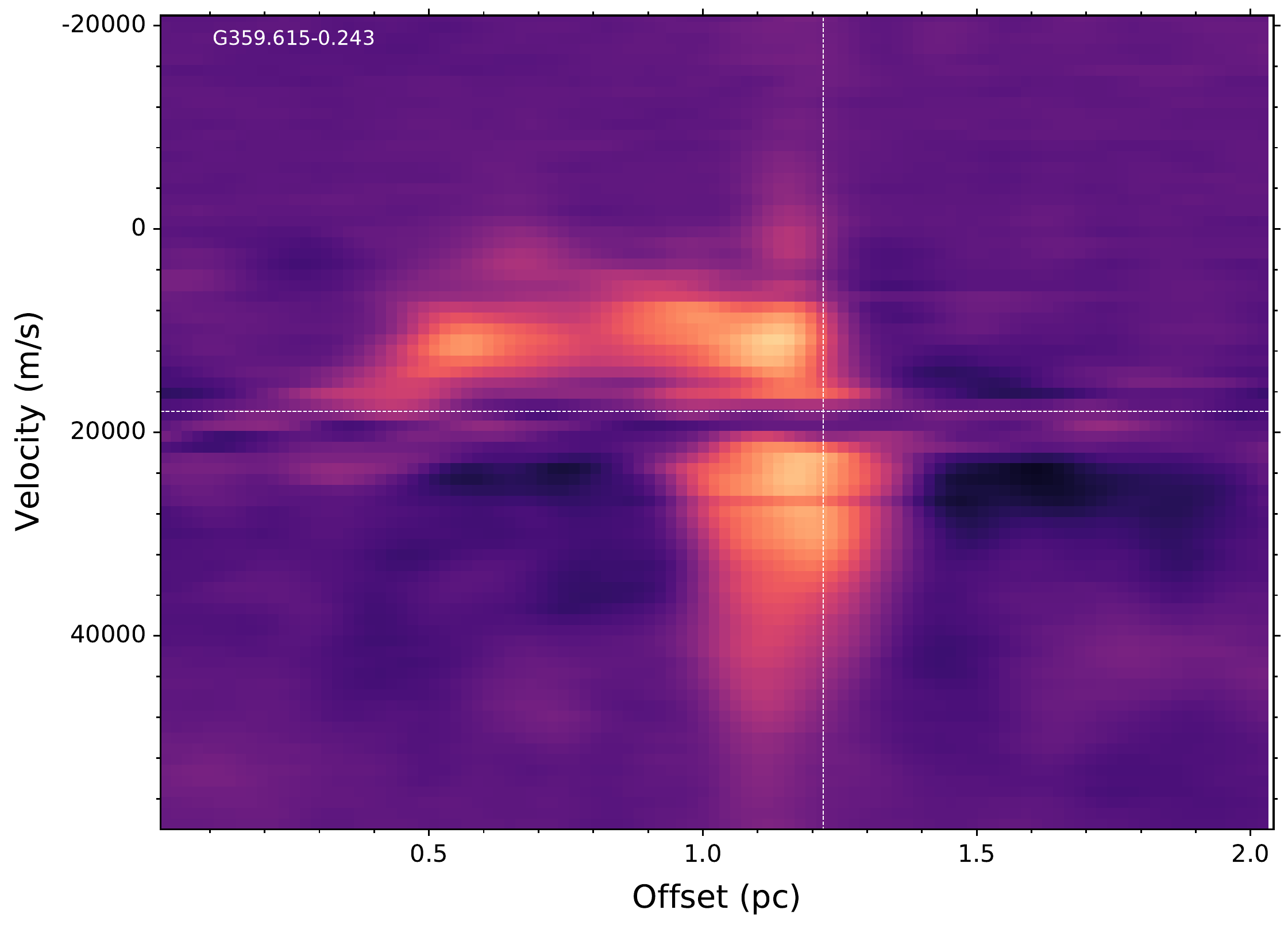}
    \caption{} 
\end{subfigure}%
    \caption{\textbf{(a)} \textit{Left}: SMA 1.3~mm dust continuum. The dotted black box indicates the region shown in the other panels. \textit{Right}: Four panels showing three-colour images for $^{12}$CO, $^{13}$CO, C$^{18}$O, and SiO. Red-shifted and blue-shifted integrated intensity (V$_{\textrm{lsr}}~\pm$~10~\kms) are shown in blue and red, respectively. Dust continuum is shown in green. The white dashed line overlaid on the $^{12}$CO emission indicates the region over which a PV-slice was taken. \textbf{(b)} PV-plot from the slice shown in \textbf{(a)}. The vertical dotted line denotes the central position of the continuum source across which the slice was taken. The horizontal dashed line denotes the assumed V$_{\textrm{lsr}}$ of the continuum source.} 
\label{fig:G359.615_outflows}
\end{figure*}
\end{document}